\documentclass[12pt,a4paper]{article}
\usepackage{graphics,graphicx}
\usepackage{amssymb,epsfig,amsmath,euscript,array,color}
\usepackage{cite}
\makeatletter \@addtoreset{equation}{section} \makeatother
\renewcommand{\theequation}{\thesection.\arabic{equation}}

\newcommand{\startappendix}{
\setcounter{section}{0}
\renewcommand{\thesection}{\Alph{section}}
\renewcommand{\theequation}{\Alph{section}.\arabic{equation}}}

\newcounter{multieqs}

\newenvironment{pretty}{}{}

\newcommand{\be}{\begin{equation}}
\newcommand{\ee}{\end{equation}}

\newcommand{\bm}[1]{\mbox{\boldmath $#1$}}

\newcommand{\partialslash}{\partial \!\!\! / }

\def\bd{\begin{document}}
\def\ed{\end{document}}
\def\nn{\nonumber}
\def\bea{\begin{eqnarray}}
\def\eea{\end{eqnarray}}
\let\bm=\bibitem
\let\la=\label

\def\npb#1#2#3{Nucl. Phys. {\bf{B#1}} #3 (#2)}
\def\plb#1#2#3{Phys. Lett. {\bf{#1B}} #3 (#2)}
\def\prl#1#2#3{Phys. Rev. Lett. {\bf{#1}} #3 (#2)}
\def\prd#1#2#3{Phys. Rev. {D \bf{#1}} #3 (#2)}
\def\cmp#1#2#3{Comm. Math. Phys. {\bf{#1}} #3 (#2)}
\def\cqg#1#2#3{Class. Quantum Grav. {\bf{#1}} #3 (#2)}
\def\nppsa#1#2#3{Nucl. Phys. B (Proc. Suppl.) {\bf{#1A}}#3 (#2)}
\def\ap#1#2#3{Ann. of Phys. {\bf{#1}} #3 (#2)}
\def\ijmp#1#2#3{Int. J. Mod. Phys. {\bf{A#1}} #3 (#2)}
\def\rmp#1#2#3{Rev. Mod. Phys. {\bf{#1}} #3 (#2)}
\def\mpla#1#2#3{Mod. Phys. Lett. {\bf A#1} #3 (#2)}
\def\jhep#1#2#3{J. High Energy Phys. {\bf #1} #3 (#2)}
\def\atmp#1#2#3{Adv. Theor. Math. Phys. {\bf #1} #3 (#2)}
\newcommand{\EQ}[1]{\begin{equation} #1 \end{equation}}
\newcommand{\AL}[1]{\begin{subequations}\begin{align} #1 \end{align}\end{subequations}}
\newcommand{\SP}[1]{\begin{equation}\begin{split} #1 \end{split}\end{equation}}
\newcommand{\ALAT}[2]{\begin{subequations}\begin{alignat}{#1} #2 \end{alignat}\end{subequations}}
\def\beqa{\begin{eqnarray}}
\def\eeqa{\end{eqnarray}}
\def\beq{\begin{equation}}
\def\eeq{\end{equation}}
\def\N{{\cal N}}
\def\sst{\scriptscriptstyle}
\def\thetabar{\bar\theta}
\def\Tr{{\rm Tr}}
\def\one{\mbox{1 \kern-.59em {\rm l}}}
 \def\Nh{\hat{N}}

\def\a{\alpha}      \def\da{{\dot\alpha}}
\def\b{\beta}       \def\db{{\dot\beta}}
\def\c{\gamma}  \def\G{\Gamma}  \def\cdt{\dot\gamma}
\def\d{\delta}  \def\D{\Delta}  \def\ddt{\dot\delta}
\def\e{\epsilon}        \def\vare{\varepsilon}
\def\f{\phi}    \def\F{\Phi}    \def\vvf{\f}
\def\h{\eta}
\def\k{\kappa}
\def\l{\lambda} \def\L{\Lambda}
\def\m{\mu} \def\n{\nu}
\def\o{\omega}
\def\p{\pi} \def\P{\Pi}
\def\r{\rho}
\def\s{\sigma}  \def\S{\Sigma}
\def\t{\tau}
\def\th{\theta} \def\Th{\Theta} \def\vth{\vartheta}
\def\X{\Xeta}
\def\z{\zeta}
\def\cA{{\cal A}} \def\cB{{\cal B}} \def\cC{{\cal C}}
\def\cD{{\cal D}} \def\cE{{\cal E}} \def\cF{{\cal F}}
\def\cG{{\cal G}} \def\cH{{\cal H}} \def\cI{{\cal I}}
\def\cJ{{\cal J}} \def\cK{{\cal K}} \def\cL{{\cal L}}
\def\cM{{\cal M}} \def\cN{{\cal N}} \def\cO{{\cal O}}
\def\cP{{\cal P}} \def\cQ{{\cal Q}} \def\cR{{\cal R}}
\def\cS{{\cal S}} \def\cT{{\cal T}} \def\cU{{\cal U}}
\def\cV{{\cal V}} \def\cW{{\cal W}} \def\cX{{\cal X}}
\def\cY{{\cal Y}} \def\cZ{{\cal Z}}
\def\ua{\underline{\alpha}}
\def\ub{\underline{\phantom{\alpha}}\!\!\!\beta}
\def\uc{\underline{\phantom{\alpha}}\!\!\!\gamma}
\def\um{\underline{\mu}}
\def\ud{\underline\delta}
\def\ue{\underline\epsilon}
\def\una{\underline a}\def\unA{\underline A}
\def\unb{\underline b}\def\unB{\underline B}
\def\unc{\underline c}\def\unC{\underline C}
\def\und{\underline d}\def\unD{\underline D}
\def\une{\underline e}\def\unE{\underline E}
\def\unf{\underline{\phantom{e}}\!\!\!\! f}\def\unF{\underline F}
\def\unm{\underline m}\def\unM{\underline M}
\def\unn{\underline n}\def\unN{\underline N}
\def\unp{\underline{\phantom{a}}\!\!\! p}\def\unP{\underline P}
\def\unq{\underline{\phantom{a}}\!\!\! q}
\def\unQ{\underline{\phantom{A}}\!\!\!\! Q}
\def\unH{\underline{H}}
\def\As {{A \hspace{-6.4pt} \slash}\;}
\def\bs {{b \hspace{-6.4pt} \slash}\;}
\def\Ds {{D \hspace{-6.4pt} \slash}\;}
\def\ds {{\del \hspace{-6.4pt} \slash}\;}
\def\ss {{\s \hspace{-6.4pt} \slash}\;}
\def\ks {{ k \hspace{-6.4pt} \slash}\;}
\def\ps {{p \hspace{-6.4pt} \slash}\;}
\def\pas {{{p_1} \hspace{-6.4pt} \slash}\;}
\def\pbs {{{p_2} \hspace{-6.4pt} \slash}\;}
\def\Fh{\hat{F}}
\def\Vh{\hat{V}}
\def\Xh{\hat{X}}
\def\ah{\hat{a}}
\def\xh{\hat{x}}
\def\yh{\hat{y}}
\def\ph{\hat{p}}
\def\xih{\hat{\xi}}
\def\psit{\tilde{\psi}}
\def\Psit{\tilde{\Psi}}
\def\tht{\tilde{\th}}
\def\lt{\tilde{\lambda}}
\def\llt{\tilde{l}}
\def\At{\tilde{A}}
\def\Qt{\tilde{Q}}
\def\Rt{\tilde{R}}
\def\Nt{\tilde{N}}

\def\at{\tilde{a}}
\def\st{\tilde{s}}
\def\ft{\tilde{f}}
\def\pt{\tilde{p}}
\def\qt{\tilde{q}}
\def\vt{\tilde{v}}
\def\nt{\tilde{n}}
\def\delb{\bar{\partial}}
\def\bz{\bar{z}}
\def\bD{\bar{D}}
\def\bB{\bar{B}}
\def\bk{{\bf k}}
\def\bl{{\bf l}}
\def\bp{{\bf p}}
\def\bq{{\bf q}}
\def\br{{\bf r}}
\def\bx{{\bf x}}
\def\by{{\bf y}}
\def\bR{{\bf R}}
\def\bV{{\bf V}}
\def\d{\delta}\def\D{\Delta}\def\ddt{\dot\delta}
\def\pa{\partial} \def\del{\partial}
\def\xx{\times}
\def\uno{\mbox{1 \kern-.59em {\rm l}}}
\def\trp{^{\top}}
\def\inv{^{-1}}
\def\dag{{^{\dagger}}}
\def\pr{^{\prime}}
\def\lan{\langle}
\def\ran{\rangle}
\def\rar{\rightarrow}
\def\lar{\leftarrow}
\def\lrar{\leftrightarrow}
\newcommand{\0}{\,\!}      %this is just NOTHING!
\def\one{1\!\!1\,\,}
\def\im{\imath}
\def\jm{\jmath}
\newcommand{\tr}{\mbox{tr}}
\newcommand{\slsh}[1]{/ \!\!\!\! #1}
\def\vac{|0\rangle}
\def\lvac{\langle 0|}
\def\hlf{\frac{1}{2}}
\def\ove#1{\frac{1}{#1}}
\def\Box{\square}
\def\ZZ{\mathbb{Z}}
\def\CC#1{({\bf #1})}
\def\bcomment#1{}
\def\bfhat#1{{\bf \hat{#1}}}
\def\VEV#1{\left\langle #1\right\rangle}
\newcommand{\ex}[1]{{\rm e}^{#1}} \def\ii{{\rm i}}
\def\rr{{\rm r}} \def\rs{{\rm s}}\def\rv{{\rm v}}
\def\ri{{\rm i}}\def\rj{{\rm j}}
\newcommand{\lrbrk}[1]{\left(#1\right)}
\newcommand{\sfrac}[2]{{\textstyle\frac{#1}{#2}}}

\def\Li2{{\rm Li}_2}

\font\mybb=msbm10 at 12pt
\def\bb#1{\hbox{\mybb#1}}

\font\myBB=msbm10 at 18pt
\def\BB#1{\hbox{\myBB#1}}

\begin{document}
\begin{flushright} Imperial-TP-JAPB-01-2011\end{flushright}
\begin{center}

{\Large \bf An Introduction to  } \\
\vspace{12pt}
{\Large \bf  String Theory
  \\}
\vspace{33pt}

{\bf James Bedford}\begin{pretty}\footnote{j.bedford@imperial.ac.uk}\end{pretty}

{\em The Blackett Laboratory\\
Imperial College\\
London SW7 2AZ\\
United Kingdom
 }

\vspace{40pt} {\bf Abstract}
\end{center}

\noindent
These notes are based on lectures given by Michael Green during Part III of the Mathematics
Tripos (the Certificate for Advanced Study in Mathematics) in the Spring of
2003. The course provided an introduction to
string theory, focussing on the Bosonic string, but treating the superstring as well. A background in quantum field theory
and general relativity is assumed. Some background in particle physics, group theory
and conformal field theory is useful, though not essential. 
A number of appendices on more advanced topics are also provided, including an introduction to orientifolds in various 
brane configurations which helps to populate a relatively sparse part of the literature.

\vspace{0.5cm}

\setcounter{page}{0}
\setcounter{tocdepth}{1}
\thispagestyle{empty}

\newpage\

\tableofcontents

\newpage\

\section{Introduction and Acknowledgements}
The raw form of these notes took shape during Michael Green's 2003 lectures on string theory at the 
Department of Applied Mathematics and Theoretical Physics of the University of Cambridge. 
The content and exposition thereof are broadly as set out by Professor Green, while the notes themselves represent the author's 
understanding of the material presented. Parts of the notes
follow the thread of the books by Green, Schwarz and Witten quite closely, as did the original course, although a number of 
additions and embellishments have been made over the years. In particular, Appendix \ref{orientifoldappendix} 
was originally compiled by the author in the early stages of research for \cite{BPZ}, and is produced here with the kind consent of the 
author's collaborators. I hope that these notes may help future Master's students and 
early-stage PhD students to get to grips with some of the basics of string theory, as they helped me. They may, in particular, be useful to those 
working their way through \cite{GSW1,GSW2} who wish to check some of the calculational details. While 
every effort has been made to excise mistakes and misprints, it is inevitable that some will still remain and the author takes full 
responsibility for these.

I am deeply indebted to Michael Green, not only for being an inspirational teacher, but also for his kind support in 
the preparation and publication of these notes. I am similarly grateful to Ruben Portugues, Aninda Sinha, Christian Stahn and Nemani Suryanarayana for 
all the insights they shared during problem classes. Further thanks go to Jacob Bourjaily, Lucy Byrne (n\'{e}e Wilson), Joseph Conlon, 
Dario Duo, Simon McNamara, Constantinos Papageorgakis, John Ward, Chris White and Konstantinos Zoubos for past and ongoing support and 
for many enlightening discussions on various points covered in the text. Finally, I would like to acknowledge a Queen Mary Studentship, a 
Marie Curie Early Stage Training Fellowship and an STFC Research Fellowship which have all provided support over the evolution of this work.

In keeping with the original lecture course, original papers have not generally been cited in the text. 
There is, however, a short and non-exhaustive list of other review articles and books 
in the bibliography which may be of some use to the reader. For more complete references the reader is referred to the original work cited in those publications, 
especially in \cite{GSW1,GSW2}.

\newpage

\section{Overview}
Quantum field theory deals with fundamental objects which
are taken to be point particles. String theory is, in a sense, a
natural extension of this. The fundamental units are now objects
extended in one dimension -- pieces of string. These strings can be
either open or closed. One of the main differences then, is
that these pieces of string can vibrate in space-time and the
vibrational modes of the strings lead to a very rich structure.

As will be seen later in these notes, the spectrum of the
\emph{closed} string necessarily contains a massless spin-2 particle --
the graviton. There is thus evidence here of a possible
unification of general relativity and quantum mechanics
(assuming that we are talking about a quantized string). Similarly, the
spectrum of the \emph{open} string necessarily contains a massless spin-1
particle -- \emph{c.f.} the photon, gluons (and $W^{\pm}, Z^{0}$ 
via the Higgs mechanism), so there is
evidence that string theory could contain the standard model of
particle physics. 

Open strings can always join to form closed strings (as
long as string interactions are permitted), but closed strings can
in principle exist without open strings. Thus a closed string theory need not
contain open string states, but any realistic open string theory
would contain closed string states. As we can't have open strings without
closed strings in an interacting theory, there is an indication of
a possible unification of Yang-Mills theories (the gauge theories
that make up the standard model) with general relativity. This is the basis for the 
claim that string theory could be a `theory of everything'. Furthermore, at low energies,
string theories reduce to specific conventional field
theories: Yang-Mills and general relativity. If supersymmetry (SUSY) is present -- as
for the superstring -- then the low energy limit is supergravity
(SUGRA).

Another attractive feature of string theory is that the divergences that particle theories 
and which are ultimately dealt with there using renormalisation are typically tamer in 
string theory. Na\"{i}vely, one can think of the fact that a string has a dimension -- a length -- compared 
with point particles which have dimension 0 as being the reason for this. As a simple example, the 
charge density of an electron (considered as a truly point-like object) is infinite as it has zero volume. 
However, in string theory the electron would arise as a particular vibrational mode of a string. The string has 
a given length and so there should be a finite charge density -- a charge per unit length. More 
concretely, when a loop expansion in stringy Feynman diagrams is considered, string theory is 
believed to be finite in the ultra-violet (UV). It has been shown to be finite to one- and two-loops, and 
in the most optimistic scenario, this may continue to hold to all loops.

In contrast with the standard model, no dimensionless parameters are needed
to formulate string theory, but there is one dimensionally meaningful parameter -- 
the string length $l_s$. The string tension, $T$, is then
given by:
\begin{eqnarray}
T = \frac{1}{2\pi{\alpha}'},
\end{eqnarray}
where ${\alpha}' = l_{s}^{2}$.

In these notes, we shall generally use `natural' units, in which
$\hbar = c = G = 1$\footnote{$\hbar$ is Planck's constant ($h$) divided
by $2\pi$, $c$ is the speed of light in vacuo, $G$ is Newton's
constant and $k_{B}$ is Boltzmann's constant.}. $\hbar = c = 1$ relates the
fundamental dimensions: $[M] = [L]^{-1} = [T]^{-1}$, and $G = 1$
then sets the length/mass/time scale. We can (and often do) set
$k_{B} = 1$ as well. This is a consistent choice and sets a
temperature scale. With $\hbar=c=G=1$ we have the Planck scale:
\begin{eqnarray*} t_{p} = \textrm{Planck Time} =
{\left(\frac{\hbar G}{c^5}\right)}^{\frac{1}{2}} \approx
5.391\times 10^{-44}\, \textrm{s}
\\ l_{p} = \textrm{Planck Length} = \left(\frac{\hbar
G}{c^3}\right)^{\frac{1}{2}} \approx 1.616\times10^{-33}\, \textrm{cm} \\ m_{p}
= \textrm{Planck Mass} = \left(\frac{\hbar
c}{G}\right)^{\frac{1}{2}} \approx 2.177\times10^{-5}\, \textrm{g}\ .
\end{eqnarray*}

A number of more recent topics in string theory are not covered in these notes, including 
D-branes (except in Appendix \ref{orientifoldappendix}), the AdS/CFT correspondence, M-theory 
and topological string theory to name but a few.

\section{Review of Point Particles}
The trajectories of point particles are described by their
world-lines. The position of a particle anywhere along its
world-line can be described by a $d$-dimensional vector
$X^{\mu}(\tau)$, where $\mu \in [0,d-1]$, and which depends on one
parameter, $\tau$, which we often choose to be the proper time.

\subsection{Action for Massive Particles}
If we are in Minkowski space, with metric $g_{\mu\nu} =
{\eta}_{\mu\nu} = \textrm{diag}(-,+,+,+ \ldots)$, then we can
describe this trajectory using the action:
\begin{eqnarray}
A =
-m\int_{{\tau}_{1}}^{{\tau}_{2}}{d{\tau}\sqrt{-{\eta}_{\mu\nu}\dot{X}^{\mu}\dot{X}^{\nu}}},
\end{eqnarray}
where $m$ is the mass of the particle and
$\dot{}\equiv\partial/{\partial\tau}$.

\subsubsection{Equation of Motion}
To get the equation of motion, we vary this which gives:
\begin{eqnarray}
\delta{A} =
m\int_{{\tau}_{1}}^{{\tau}_{2}}{d{\tau}\frac{\partial}{\partial\tau}\left[\frac{-\dot{X}_{\mu}}{\sqrt{-{\eta}_{\mu\nu}\dot{X}^{\mu}\dot{X}^{\nu}}}\right]\delta{X^{\mu}}}+{m\left[\frac{\dot{X}_{\mu}\delta{X^{\mu}}}{\sqrt{-{\eta}_{\mu\nu}\dot{X}^{\mu}\dot{X}^{\nu}}}\right]_{{\tau_{1}}}^{{\tau}_{2}}}.
\end{eqnarray}
The vanishing of the second (`surface') term, gives us our
boundary conditions. We can either require $\delta{X^{\mu}}$ to be
zero at $\tau_{1}$ and $\tau_{2}$, or we can require
$\dot{X}^{\mu}$ to be zero there. Clearly in this case the two are
essentially the same -- what else is there to vary $X^{\mu}$ with
respect to? The boundary conditions here are automatically
satisfied by conservation of momentum. On the other hand,
requiring that $\delta{A}$ vanishes for arbitrary
$\delta{X^{\mu}}$ gives the equation of motion:
\begin{eqnarray}
\frac{\partial}{\partial\tau}\left[\frac{-m\dot{X}_{\mu}}{\sqrt{-{\eta}_{\mu\nu}\dot{X}^{\mu}\dot{X}^{\nu}}}\right]
= 0.
\end{eqnarray}

\subsubsection{Symmetries}
The action is, as we might expect, invariant under global
Poincar\'e transformations:
\begin{eqnarray}
\label{poincarexfm}
X^{\mu} \rightarrow {{\Lambda}^{\mu}}_{\nu}X^{\nu}+a^{\mu},
\end{eqnarray}
where $\Lambda$ and $a$ describe constant transformations. In
addition there is world-line re-parametrization invariance, $\tau
\rightarrow \tilde{\tau}$, which is most easily seen by noticing
that the action can be written as:
\begin{eqnarray*}
A =
-m\int_{\tau_{1}}^{\tau_2}\sqrt{-\eta_{\mu\nu}d{X}^{\mu}d{X}^{\nu}}.
\end{eqnarray*}
The action cannot, however, describe massless particles which have
$m=0$ and hence $A\equiv0$.

\subsection{General Action}
We can approach the problem in a different way by using the
action:
\begin{eqnarray}
\hat{A}=\frac{1}{2}\int_{\tau_{1}}^{\tau_{2}}d{\tau}\left(e^{-1}\eta_{\mu\nu}\dot{X}^{\mu}\dot{X}^{\nu}-em^{2}\right),
\end{eqnarray}
where $e=e(\tau)$.

\subsubsection{Equations of Motion}
The equations of motion can be obtained as before. Varying with respect to (w.r.t.)
$X^{\mu}$ gives:
\begin{eqnarray}
\label{geneom}
\frac{\partial}{\partial\tau}\left(-e^{-1}\dot{X}^{\mu}\right)=0,
\end{eqnarray}
with similar boundary conditions\footnote{At least when the equation of
motion for $e$ has been taken into account.}, while varying
w.r.t. $e$ gives:
\begin{eqnarray}
\frac{{\eta}_{\mu\nu}\dot{X}^{\mu}\dot{X}^{\nu}}{e^{2}}+m^{2}=0.
\end{eqnarray}
Solving for $e$ and substituting back both into equation (\ref{geneom}) and
into $\hat{A}$ shows us that the equation of motion for the
particle is the same as before, and that the two actions are
equivalent.\footnote{$e^{-1}(\tau)$ is the one-dimensional
equivalent of $\sqrt{-\det g}g^{\alpha\beta}$ (see Section \ref{bdhaction}). \emph{i.e.}
$e^{-1}=\sqrt{g_{\tau\tau}}g^{\tau\tau}=\sqrt{g^{\tau\tau}}$. $g$
here is the `metric' along the world-line. In fact $e$ can be
thought of as an einbein on the world-line.} We can now describe
massless particles too.
\subsubsection{Symmetries} $\hat{A}$ has
the same global symmetries as before, and the re-parametrization
invariance is still here, but it is less obvious. Under $\tau
\rightarrow \tilde{\tau}$, we have $d\tau \rightarrow
d\tilde{\tau}$. Now:
\begin{eqnarray*}
e(\tau) = \sqrt{g_{\tau\tau}} \Rightarrow e(\tau)d\tau =
\sqrt{g_{\tau\tau}{d\tau}^2}
\end{eqnarray*}
and
\begin{eqnarray*}
g_{\tau\tau}d{\tau}d{\tau} \equiv
g_{\tilde{\tau}\tilde{\tau}}d{\tilde{\tau}}d{\tilde{\tau}},
\end{eqnarray*}
so
\begin{eqnarray*}
e(\tau)d\tau = \sqrt{g_{\tilde{\tau}\tilde{\tau}}}{d\tilde{\tau}}
= e(\tilde{\tau})d\tilde{\tau}.
\end{eqnarray*}
Seeing how this works in the action, we have finally that\footnote{To come into line with future
notation, we can alternatively write the infinitesimal field
transformations as: ${\delta}_{\eta}X^{\mu} =
\eta{\partial}_{\tau}X^{\mu}$ ; ${\delta}_{\eta}e =
{\partial}_{\tau}({\eta}e)$ to first order in $\eta$, and where
$\eta \sim {\delta}{\tau}$.}:
\begin{eqnarray}
\hat{A} \rightarrow \tilde{\hat{A}} =
\frac{1}{2}\int_{\tilde{{\tau}_{1}}}^{\tilde{{\tau}_{2}}}d\tilde{\tau}\left({\tilde{e}}^{-1}\left({\partial}_{\tilde{\tau}}X\right)^2
- \tilde{e}m^2\right).
\end{eqnarray}
\section{Bosonic Strings}
To generalize to strings, we now have a world-sheet which has the
form of a curved sheet or a curved cylinder depending on whether
the string is open or closed. There are coordinates
$X^{\mu}({\xi}^{\alpha}) \equiv
X^{\mu}(\tau,\sigma)$\footnote{\emph{i.e.} ${\xi}^{\alpha}$ has $\alpha =
0,1 : {\xi}^{0} = \tau, {\xi}^1 = \sigma$.}, which denote the
position of the string in space-time as a function of the
world-sheet parameters $\tau$ and $\sigma$. We will see that like
the point-particle case, there is re-parametrization invariance on
the world-sheet and we will generally use this to choose the
parameters $\tau$ and $\sigma$ to be orthogonal.

$\tau$ has the usual time-like range: $-\infty < \tau < \infty$,
and $\sigma$ can be chosen fairly arbitrarily for convenience. $0
\leq \sigma \leq 2\pi$ is clearly some sort of `sensible' choice
for closed strings. We could then set $0 \leq \sigma \leq \pi$ for
open strings, reflecting the `doubling up' of an open string to
form a closed string.
Alternatively we might wish to set $0 \leq \sigma \leq \pi$ for
both open \textbf{and} closed strings. This latter case is the one
that we'll generally use.
\subsection{Nambu-Goto Action}
The generalization from a point-particle to a string, of the
massive action $A$, is the Nambu-Goto action:
\begin{eqnarray}
A_{NG} =
-T\int{d^{2}\xi\sqrt{-\det\left(\eta_{\mu\nu}{\partial}_{\alpha}X^{\mu}{\partial}_{\beta}X^{\nu}\right)}},
\end{eqnarray}
where in obvious notation, $d^{2}\xi \equiv d\sigma{d\tau}$, and
we will often denote
$\eta_{\mu\nu}{\partial}_{\alpha}X^{\mu}{\partial}_{\beta}X^{\nu}$
by ${\hat{\gamma}}_{\alpha\beta}$ for convenience.
\subsection{Brink-Di Vecchia-Howe Action}\label{bdhaction}
As before, we can write down an equivalent action that is easier
to deal with by introducing a metric on the world-sheet,
${\gamma}_{\alpha\beta}$. This is the Brink-Di Vecchia-Howe action
(often referred to in the literature as the Polyakov action):
\begin{eqnarray}
A_{BDH} =
-\frac{T}{2}\int{d^{2}\xi\sqrt{-\det{\gamma}}\ {\gamma}^{\alpha\beta}\left({\eta_{\mu\nu}{\partial}_{\alpha}X^{\mu}{\partial}_{\beta}X^{\nu}}\right)}.
\end{eqnarray}
As before we can vary this action to find the equation of motion
for ${\gamma}_{\alpha\beta}$\footnote{To do this we must use the
well known result for the variation of a matrix, $M$:
$\frac{\delta |\det M|}{|\det M|}={\rm tr} (M^{-1}\delta
M)$.}, finding that ${\gamma}_{\alpha\beta} =
f(\sigma,\tau)\left(\eta_{\mu\nu}{\partial}_{\alpha}X^{\mu}{\partial}_{\beta}X^{\nu}\right)
\equiv f(\sigma,\tau){\hat{\gamma}}_{\alpha\beta}$.\footnote{As before we can also vary w.r.t. $X^{\mu}$ to find
the equation of motion satisfied by the position of the string in
space-time.} Substituting
back into the action shows us directly that $A_{NG} \equiv
A_{BDH}$. We will usually deal with the action in the form of
$A_{BDH}$.\footnote{We could in fact include other terms in
$A_{BDH}$ such as $A_{1} =
\lambda\int{d^{2}\xi\sqrt{-\det{\gamma}}}$, and $A_{2} =
\frac{1}{2\pi}\int{d^{2}\xi\sqrt{-\det{\gamma}}\ R^{\left(2\right)}\!\left(\gamma\right)}$,
but we leave them out here (see \emph{e.g.} \cite{GSW1}, Chapter 2, for a
discussion).}
\subsubsection{Symmetries}
$A_{BDH}$ has symmetries of both the world-sheet, and of the
background space-time. We again have global Poincar\'e invariance
(see (\ref{poincarexfm})) of the background space-time, from which we can compute
the conserved currents $P_{\alpha}^{\mu}$ and
$J_{\alpha}^{\mu\nu}$, and we have the world-sheet symmetries which amount to
re-parametrization invariance:
\begin{eqnarray}
{\delta}X^{\mu} = {\zeta}^{\alpha}{\partial}_{\alpha}X^{\mu},
\end{eqnarray}
where ${\zeta}^{\alpha} \sim \delta{\xi}^{\alpha}$, and under
which:
\begin{eqnarray}
\delta{\gamma}_{\alpha\beta} = 2\nabla_{(\alpha}{\zeta}_{\beta)}
&=& {\zeta}^{\rho}{\partial}_{\rho}{\gamma}_{\alpha\beta} +
2{\gamma}_{\rho(\beta}{\partial}_{\alpha)}{\zeta}^{\rho} \\
\delta\left(\sqrt{-\det\gamma}\right) &=&
{\partial}_{\alpha}\left({\zeta}^{\alpha}\sqrt{-\det\gamma}\right),
\end{eqnarray}
and Weyl invariance:
\begin{eqnarray}
\delta{\gamma}_{\alpha\beta} = \Lambda{\gamma}_{\alpha\beta}\ ,
\end{eqnarray}
where $\Lambda = \Lambda(\sigma,\tau)$.
\subsubsection{Gauge Fixing}
In order to quantize our particle theory, it is necessary to make a gauge
choice. Considering again the point-particle case, we can do this
in two obvious ways. We could solve for $e$ in $\hat{A}$,
substitute this back into $\hat{A}$ to give us $A$, and then make
some convenient choice such as $X^{0}(\tau) = \tau$.
Alternatively, we can fix the gauge by using the world-line
re-parametrization invariance to fix $e$. For instance we could
set $e = 1$. In this second case, the equation of motion for $e$
must then be imposed as a constraint.

Analogously, in the string case we could eliminate
${\gamma}_{\alpha\beta}$ to give us $A_{NG}$, and then set
$X^{0}(\sigma,\tau) = \tau$ and $X^{1}(\sigma,\tau) = \sigma$.
Alternatively we can use the world-sheet re-parametrization
invariance to fix ${\gamma}_{\alpha\beta}$. Again, the equation of
motion for ${\gamma}_{\alpha\beta}$ (vanishing of the world-sheet
energy-momentum tensor - see below) must then be imposed as a
constraint. This second route is the one that we'll follow.

Before we do this, it is convenient to note that the variation of
$A_{BDH}$ w.r.t. ${\gamma}_{\alpha\beta}$ defines a world-sheet
energy-momentum tensor (\emph{c.f.} GR):
\begin{eqnarray}
\label{enmom}
T_{\alpha\beta} =
-\frac{2}{T}\frac{1}{\sqrt{-\det\gamma}}\frac{\delta{A}_{BDH}}{\delta{\gamma}^{\alpha\beta}},
\end{eqnarray}
and its vanishing is the equation of motion for
${\gamma}_{\alpha\beta}$. We find that
\begin{eqnarray}
T_{\alpha\beta} = {\hat{\gamma}}_{\alpha\beta} -
\frac{1}{2}{\gamma}_{\alpha\beta}{\gamma}^{\epsilon\rho}{\hat{\gamma}}_{\epsilon\rho},
\end{eqnarray}
with field equation
\begin{eqnarray}
\frac{\delta{A}_{BDH}}{\delta{\gamma}^{\alpha\beta}} = 0
\Rightarrow T_{\alpha\beta} = 0.
\end{eqnarray}

Back to gauge fixing, and we can choose a form for
${\gamma}_{\alpha\beta}$ which looks like a `usual' flat metric:
\begin{eqnarray}
{\gamma}_{\alpha\beta} = e^{\phi(\sigma,\tau)}{\eta}_{\alpha\beta}
= e^{\phi(\sigma,\tau)}\left(\begin{array}{cc} -1 & 0 \\
0 & 1 \end{array}\right).
\end{eqnarray}
This is the so-called conformal gauge. In fact we can use Weyl
invariance to gauge away the $e^{\phi}$ dependence to give
\begin{eqnarray}
{\gamma}_{\alpha\beta} = {\eta}_{\alpha\beta}.
\end{eqnarray}
In conformal gauge, the Brink-Di Vecchia-Howe action then takes
the form:
\begin{eqnarray}
A_{BDH} &=&
-\frac{T}{2}\int{d^{2}\xi{\eta}^{\alpha\beta}{\eta_{\mu\nu}{\partial}_{\alpha}X^{\mu}{\partial}_{\beta}X^{\nu}}}
\\
\label{bdhconf} &=& \frac{T}{2}\int{d^{2}\xi\left({\dot{X}}^{2} -
{\acute{X}}^2\right)},
\end{eqnarray}
where again $\dot{} \equiv {\partial}/{\partial\tau}$, and now
$\acute{} \equiv {\partial}/{\partial\sigma}$, ${\dot{X}}^{2} =
{\eta}_{\mu\nu}\dot{X}^{\mu}\dot{X}^{\nu}$ \emph{etc.}

\subsection{Equations of Motion} In this gauge we can determine
the equations of motion much more easily. We have:
\begin{eqnarray}
\delta{A_{BDH}} \ &=& \
T\int{d^{2}\xi\left[-{\partial}_{\tau}^{2}{X}^{\mu} +
{\partial}_{\sigma}^{2}{X}^{\nu}\right]\delta{X_{\mu}}} \nonumber \\
\ &+& \
T\int{d\sigma\dot{X}^{\mu}\delta{X}_{\mu}}\Bigg{.}{\Bigg|}_{\tau =
-\infty}^{\tau = \infty} -
T\int{d\tau}\acute{X}^{\mu}\delta{X}_{\mu}\Bigg{.}{\Bigg|}_{\sigma}.
\end{eqnarray}
The string equation of motion is thus the 2-dimensional wave
equation\footnote{Here $\Box:=-\partial_{\tau}^2+\partial_{\sigma}^2$.}:
\begin{eqnarray}
\Box{X^{\mu}}= 0,
\end{eqnarray}
with boundary conditions given by the final two terms. The first
of these should vanish by conservation of momentum (being ensured
by the translational invariance of the action):
$\dot{X}^{\mu}\bigg{.}{\bigg|}_{-\infty} =
\dot{X}^{\mu}\bigg{.}{\bigg|}_{\infty}$. The conditions imposed in
order to ensure vanishing of the second term will depend on
whether the string we're considering is open or closed.

For closed strings, we require periodicity in $\sigma$ for
$X^{\mu}$ and its derivatives:
\begin{eqnarray*}
X^{\mu}(\sigma,\tau) &=& X^{\mu}(\sigma + 2\pi,\tau) \\
{\acute{X}}^{\mu}(\sigma,\tau) &=& {\acute{X}}^{\mu}(\sigma +
2\pi,\tau) \end{eqnarray*}
(for a closed string with $\sigma$
running between $0$ and $2\pi$ for example).

For open strings, the story is slightly different. There is no
`natural' periodicity and we wish to ensure that
${\acute{X}}^{\mu}(0,\tau)\delta{X}_{\mu}(0,\tau) =
{\acute{X}}^{\mu}(\pi,\tau)\delta{X}_{\mu}(\pi,\tau)$. Thus at one
end we can impose one of two conditions:
\begin{eqnarray*}
\textrm{Neumann} &\Rightarrow& {\acute{X}}^{\mu} = 0 \\
\textrm{Dirichlet} &\Rightarrow& \delta{X}^{\mu} = 0 \quad
(\textrm{\emph{i.e.}}\,\, {\dot{X}}^{\mu}) = 0\, ,
\end{eqnarray*}
which leaves three different overall possibilities.
Neumann-Neumann (NN), Dirichlet-Dirichlet (DD) and mixed
Neumann-Dirichlet (ND) boundary conditions.

\subsubsection{Solutions}
We can solve the wave equation by writing the solution as a sum of
two arbitrary functions:
\begin{eqnarray}
X^{\mu}(\sigma,\tau) &=& X_{R}^{\mu}(\tau - \sigma) +
X_{L}^{\mu}(\tau + \sigma) \nonumber \\
&=& X_{R}^{\mu}({\xi}^{-}) + X_{L}^{\mu}({\xi}^{+}),
\end{eqnarray}
where $X_{R}^{\mu}$ describes `right-moving' modes and
$X_{L}^{\mu}$ describes `left-moving' modes. At this point it is
useful to introduce `light-cone' coordinates. Define\footnote{In
fact the definition of ${\partial}_{\pm}$ follows naturally from
the definition of ${\xi}^{\pm}$ if one considers the differential of
some arbitrary $f(\sigma,\tau)\equiv f({\xi}^{+},{\xi}^{-})$ in terms
of both sets of variables.}:
\begin{eqnarray*}
{\xi}^{\pm} &=& \tau \pm \sigma ;\\
{\partial}_{\pm} &=& \frac{1}{2}\left({\partial}_{\tau} \pm
{\partial}_{\sigma}\right).
\end{eqnarray*}
In conformal gauge (with $e^{\phi}$ gauged away), the metric
tensor then becomes:
\begin{eqnarray}
{\gamma}_{\pm\pm} = {\gamma}^{\pm\pm} &=& 0; \nonumber \\
{\gamma}_{+-} = {\gamma}_{-+} &=& -\frac{1}{2}; \nonumber \\
{\gamma}^{+-} = {\gamma}^{-+} &=& -2,
\end{eqnarray}
so that \emph{e.g.}
\begin{eqnarray*}
U^{+} = {\eta}^{+-}U_{-} &=& -2U_{-}; \\
U_{+} = {\eta}_{+-}U^{-} &=& -\frac{1}{2}U^{-}
\end{eqnarray*}
\emph{etc.}

\noindent Using a mode expansion, we have:

For the \textbf{closed} string:
\begin{itemize}
\item With $0 \leq \sigma \leq 2\pi$
\begin{eqnarray}
X_{R}^{\mu} &=& \frac{1}{2}x^{\mu} +
\frac{1}{2}{\alpha}'p_{R}^{\mu}{\xi}^{-} +
i\sqrt{\frac{\alpha'}{2}}\sum_{n \neq
0}{\frac{{\alpha}_{n}^{\mu}}{n}e^{-in{\xi}^{-}}} \\
X_{L}^{\mu} &=& \frac{1}{2}x^{\mu} +
\frac{1}{2}{\alpha}'p_{L}^{\mu}{\xi}^{+} +
i\sqrt{\frac{\alpha'}{2}}\sum_{n \neq
0}{\frac{{\tilde{\alpha}}_{n}^{\mu}}{n}e^{-in{\xi}^{+}}},
\end{eqnarray}
so that
\begin{eqnarray}
X^{\mu} = x^{\mu} + {\alpha}'p^{\mu}\tau +
i\sqrt{\frac{\alpha'}{2}}\sum_{n \neq
0}{\frac{1}{n}\left({\alpha}_{n}^{\mu}e^{-in{\xi}^{-}} +
{\tilde{\alpha}}_{n}^{\mu}e^{-in{\xi}^{+}}\right)},
\end{eqnarray}
and where $p_{R}^{\mu} = p_{L}^{\mu} = p^{\mu}$ in Minkowski
space. \item With $0 \leq \sigma \leq \pi$
\begin{eqnarray}
X_{R}^{\mu} &=& \frac{1}{2}x^{\mu} + {\alpha}'p_{R}^{\mu}{\xi}^{-}
+ i\sqrt{\frac{\alpha'}{2}}\sum_{n \neq
0}{\frac{{\alpha}_{n}^{\mu}}{n}e^{-2in{\xi}^{-}}} \\
X_{L}^{\mu} &=& \frac{1}{2}x^{\mu} + {\alpha}'p_{L}^{\mu}{\xi}^{+}
+ i\sqrt{\frac{\alpha'}{2}}\sum_{n \neq
0}{\frac{{\tilde{\alpha}}_{n}^{\mu}}{n}e^{-2in{\xi}^{+}}},
\end{eqnarray}
so that
\begin{eqnarray}
X^{\mu} = x^{\mu} + 2{\alpha}'p^{\mu}\tau +
i\sqrt{\frac{\alpha'}{2}}\sum_{n \neq
0}{\frac{1}{n}\left({\alpha}_{n}^{\mu}e^{-2in{\xi}^{-}} +
{\tilde{\alpha}}_{n}^{\mu}e^{-2in{\xi}^{+}}\right)},
\end{eqnarray}
where again the same relations hold between $p_{R}^{\mu}$,
$p_{L}^{\mu}$ and $p^{\mu}$.
\end{itemize}
For the \textbf{open} string (recall $0 \leq \sigma \leq
\pi$)\footnote{In each of these open-string solutions, we have
assumed that \textit{all} of the space-time indices have the
stated boundary conditions, though it is not hard to generalize to
hybrid cases. In $d$ dimensions (with $d-1$ spatial directions) we
can have $p$ Neumann and $d-1-p$ Dirichlet boundary conditions for
any particular string endpoint. This defines a $(p+1)$-dimensional
hypersurface known as a D$p$-brane or D-brane for short.}:
\begin{itemize}
\item With NN boundary conditions
\begin{eqnarray}
\label{neumann}
X^{\mu} = x^{\mu} + 2{\alpha}'p^{\mu}\tau +
i\sqrt{2{\alpha}'}\sum_{n \neq
0}{\frac{{\alpha}_{n}^{\mu}}{n}e^{-in\tau}\cos{n\sigma}}.
\end{eqnarray}
\item With DD boundary conditions
\begin{eqnarray}
X^{\mu} = a^{\mu} + \frac{1}{\pi}(b^{\mu} - a^{\mu})\sigma +
\sqrt{2{\alpha}'}\sum_{n \neq
0}{\frac{{\alpha}_{n}^{\mu}}{n}e^{-in\tau}\sin{n\sigma}}.
\end{eqnarray}
\item With ND boundary conditions
\begin{eqnarray}
X^{\mu} = b^{\mu} + i\sqrt{2{\alpha}'}\sum_{r \in \,\mathbb{Z} +
\frac{1}{2}}{\frac{{\alpha}_{r}^{\mu}}{r}e^{-ir\tau}\cos{r\sigma}}.
\end{eqnarray}
\item With DN boundary conditions
\begin{eqnarray}
X^{\mu} = a^{\mu} + \sqrt{2{\alpha}'}\sum_{r \in \,\mathbb{Z} +
\frac{1}{2}}{\frac{{\alpha}_{r}^{\mu}}{r}e^{-ir\tau}\sin{r\sigma}}.
\end{eqnarray}
\end{itemize}
In these solutions, the requirement that $X^{\mu}(\sigma,\tau)$ is
a real function means that:
\begin{eqnarray}
X^{\mu} = ({X^{\mu}})^{\dagger} \Rightarrow
({\alpha}_{n}^{\mu})^{\dagger} &=& {\alpha}_{-n}^{\mu} \nonumber \\
({\tilde{\alpha}}_{n}^{\mu})^{\dagger} &=&
{\tilde{\alpha}}_{-n}^{\mu},
\end{eqnarray}
where $\dagger$ is understood to act as a complex conjugation
operation on the unquantized mode coefficients and as a hermitian
conjugation operation on the quantized mode operators. $x^{\mu}$
is like a centre of mass for the string, and $p^{\mu}$ is like its
overall momentum, as can be verified by explicitly calculating the
total momentum of a string. $a^{\mu}$ and $b^{\mu}$ are constant vectors.

\subsubsection{Poisson Brackets}
From the form of the action in conformal gauge
(\ref{bdhconf}), we can identify the momentum conjugate to
$X^{\mu}$ as
\begin{eqnarray}
P^{\mu} \equiv \frac{\delta{A_{BDH}}}{\delta{\dot{X}}_{\mu}} =
T{\dot{X}}^{\mu}.
\end{eqnarray}
We thus have:
\begin{itemize}
\item For the \textbf{closed} string $(0 \leq \sigma \leq \pi)$:
\begin{eqnarray}
P^{\mu} &=& T\left\{2{\alpha}'p^{\mu} +
2\sqrt{\frac{{\alpha}'}{2}}\sum_{n \neq
0}\left({\alpha}_n^{\mu}e^{-2in{\xi}^{-}} +
{\tilde{\alpha}}_n^{\mu}e^{-2in{\xi}^{+}}\right)\right\} \nonumber \\
&=& T\sqrt{2{\alpha}'}\sum_{n =
-\infty}^{\infty}\left({\alpha}_n^{\mu}e^{-2in{\xi}^{-}} +
{\tilde{\alpha}}_n^{\mu}e^{-2in{\xi}^{+}}\right),
\end{eqnarray}
where we have set ${\alpha}_{0}^{\mu} = {\tilde{\alpha}}_{0}^{\mu}
= \sqrt{{\alpha}'/{2}}\,p^{\mu}$. \item For the \textbf{open} string (with NN boundary conditions):
\begin{eqnarray}
P^{\mu} = T\sqrt{2{\alpha}'}\sum_{n =
-\infty}^{\infty}{\alpha}_{n}^{\mu}e^{-in\tau}\cos{n\sigma},
\end{eqnarray}
where we have set ${\alpha}_{0}^{\mu} = \sqrt{2{\alpha}'}\,p^{\mu}$.
\end{itemize}
This means that we are now in a position to write down some
Poisson brackets\footnote{Recall that if $f$ and $g$ are two
functions of generalized coordinates $q_{i}$, their conjugate
momenta $p_{i}$ and time $t$, then their Poisson bracket is
defined as: $[f,g]_{P.B.} \equiv
\frac{\partial{f}}{\partial{q_{i}}}\frac{\partial{g}}{\partial{p^{i}}}
-
\frac{\partial{f}}{\partial{p_{i}}}\frac{\partial{g}}{\partial{q^{i}}}$.
This yields, for instance, $[q,p]_{P.B.} = 1$, $[q,q]_{P.B.} = 0$,
$[p,p]_{P.B.} = 0$ \emph{etc.}}.
\begin{eqnarray}
\left[X^{\mu}(\sigma),X^{\nu}(\sigma')\right]_{P.B.} &=&
\left[P^{\mu}(\sigma),P^{\nu}(\sigma')\right]_{P.B.} = 0
\nonumber \\
\label{xppb}
\left[X^{\mu}(\sigma),P^{\nu}(\sigma')\right]_{P.B.} &=&
\delta(\sigma - \sigma'){\eta}^{\mu\nu},
\end{eqnarray}
which means that the mode coefficients must obey:
\begin{eqnarray}
\label{modepb}
\left[{\alpha}_m^{\mu},{\alpha}_n^{\nu}\right]_{P.B.} =
\left[{\tilde{\alpha}}_m^{\mu},{\tilde{\alpha}}_n^{\nu}\right]_{P.B.}
&=& -im{\delta}_{m+n}{\eta}^{\mu\nu} \nonumber \\
\left[{\alpha}_m^{\mu},{\tilde{\alpha}}_n^{\nu}\right]_{P.B.} &=&
0 \nonumber \\
\left[x^{\mu},p^{\nu}\right]_{P.B.} &=& {\eta}^{\mu\nu},
\end{eqnarray}
where we use the conventions of \cite{GSW1,GSW2} so that
\begin{eqnarray*}
{\delta}_{m+n} = \left\{\begin{array}{ll} 1 & m+n=0 \\ 0 & m+n
\neq 0 \end{array}\right..
\end{eqnarray*}
We can then go on to calculate such things as the Hamiltonian and
the world-sheet energy-momentum tensor. The Hamiltonian is:
\begin{eqnarray}
H &=& \int{d\sigma\left({\dot{X}}^{\mu}P_{\mu} -
\mathcal{L}\right)}
\nonumber \\
&=& \frac{T}{2}\int{d\sigma\left({\dot{X}}^2 + {X'}^2\right)}.
\end{eqnarray}
In order to calculate this in terms of modes, it is useful to
remember that as $X^{\mu}$ is real (and hence so are
${\partial}_{\tau}X^{\mu}$ and ${\partial}_{\sigma}X^{\mu}$), we
can write ${\dot{X}}^{2}$ as
${\dot{X}}^{\mu}{{\dot{X}}_{\mu}}^{\dagger}$ \emph{etc.} We end up with:
\begin{eqnarray}
H &=& {\alpha'}T\left\{\frac{1}{2}\pi\sum_{n \neq
0}{\alpha}_{n}\cdot{\alpha}_{-n} + \frac{1}{2}\pi\sum_{n \neq
0}{\alpha}_{n}\cdot{\alpha}_{-n} +
\pi{\alpha}_0\cdot{\alpha}_0\right\} \nonumber \\
&=& \frac{1}{2}\sum_{n =
-\infty}^{\infty}{\alpha}_{-n}\cdot{\alpha}_{n}
\end{eqnarray}
for open strings, and
\begin{eqnarray}
H = \frac{1}{2}\sum_{n =
-\infty}^{\infty}\left({\alpha}_{-n}\cdot{\alpha}_{n} +
{\tilde{\alpha}}_{-n}\cdot{\tilde{\alpha}}_{n}\right)
\end{eqnarray}
for closed strings. Here ${\alpha}_{n}\cdot{\alpha}_{m} \equiv
{\alpha}_{n}^{\mu}{\alpha}_{m}^{\nu}{\eta}_{\mu\nu}$.

It is easiest to deal with the energy-momentum tensor in
light-cone coordinates. Recall that in conformal gauge, the
energy-momentum tensor takes the form:
\begin{eqnarray}
T_{\alpha\beta} =
{\partial}_{\alpha}X^{\mu}{\partial}_{\beta}X_{\mu} -
\frac{1}{2}{\eta}_{\alpha\beta}{\eta}^{\epsilon\rho}{\partial}_{\epsilon}X^{\mu}{\partial}_{\rho}X_{\mu},
\end{eqnarray}
which we can write as
\begin{eqnarray}
T_{++} &=& {\partial}_{+}X^{\mu}{\partial}_{+}X_{\mu} \nonumber \\
T_{--} &=& {\partial}_{-}X^{\mu}{\partial}_{-}X_{\mu} \nonumber \\
T_{+-} &=& T_{-+} \equiv 0
\end{eqnarray}
in light-cone coordinates. The constraint equation
$T_{\alpha\beta} = 0$ then takes the form:
\begin{eqnarray}
{\partial}_{+}X^{\mu}{\partial}_{+}X_{\mu} =
{\partial}_{-}X^{\mu}{\partial}_{-}X_{\mu} = 0,
\end{eqnarray}
with $T_{+-} = T_{-+} = 0$ being satisfied identically. This last
statement just encapsulates the tracelessness of the
energy-momentum tensor. By Noether's theorem the energy-momentum
tensor is also conserved:
\begin{eqnarray}
{\partial}^{\alpha}T_{\alpha\beta} &=& 0 \nonumber \\ \Rightarrow
{\partial}_{-}T_{++} = {\partial}_{+}T_{--} &=& 0.
\end{eqnarray}
So, in terms of modes we can now write:
\begin{itemize}
\item For \textbf{closed} strings ($0 \leq \sigma \leq \pi$):
\begin{eqnarray}
T_{++} = 2{\alpha'}\sum_{n,m \in
\,\mathbb{Z}}{\tilde{\alpha}}_{n}\cdot{{\tilde{\alpha}}_{-m}}e^{-2in{\xi}^{+}}e^{2im{\xi}^{+}}
\nonumber \\
T_{--} = 2{\alpha'}\sum_{n,m \in
\,\mathbb{Z}}{\alpha}_{n}\cdot{{\alpha}_{-m}}e^{-2in{\xi}^{-}}e^{2im{\xi}^{-}}.
\end{eqnarray}
\item For \textbf{open} strings (NN boundary conditions):
\begin{eqnarray}
T_{\pm\pm} = \frac{\alpha'}{2}\sum_{n,m \in
\mathbb{Z}}{\alpha}_{n}\cdot{{\alpha}_{-m}}e^{-in{\xi}^{\pm}}e^{im{\xi}^{\pm}}.
\end{eqnarray}
\end{itemize}
Let us also expand the energy-momentum tensor in terms of its
modes $\tilde{L}_{m}$ and $L_{m}$:
\begin{itemize}
\item \textbf{Closed} strings:
\begin{eqnarray}
T_{++} = 4\alpha'\sum_{n =
-\infty}^{\infty}{\tilde{L}}_{n}e^{-2in{\xi}^{+}} \nonumber \\
T_{--} = 4\alpha'\sum_{n =
-\infty}^{\infty}{L}_{n}e^{-2in{\xi}^{-}}.
\end{eqnarray}
\item \textbf{Open} strings:
\begin{eqnarray}
T_{\pm\pm} = \alpha'\sum_{n =
-\infty}^{\infty}{L}_{n}e^{-in{\xi}^{\pm}}.
\end{eqnarray}
\end{itemize}
Inverting these expressions, we can work out the modes as:
\begin{itemize}
\item \textbf{Closed} strings:
\begin{eqnarray}
\label{modes}
{\tilde{L}}_{m} =
\frac{1}{4\pi\alpha'}\int_{0}^{\pi}d\sigma{e^{2im\sigma}}T_{++}\Bigg.\Bigg|_{\tau
= 0} &=& \frac{1}{2}\sum_{n = -\infty}^{\infty}{\tilde{\alpha}}_{m-n}\cdot{{\tilde{\alpha}}_{n}} \nonumber \\
{L}_{m} =
\frac{1}{4\pi\alpha'}\int_{0}^{\pi}d\sigma{e^{-2im\sigma}}T_{--}\Bigg.\Bigg|_{\tau
= 0} &=& \frac{1}{2}\sum_{n =
-\infty}^{\infty}{\alpha}_{m-n}\cdot{{\alpha}_{n}}.
\end{eqnarray}
\item \textbf{Open} strings\footnote{In this case, we regard these
definitions as purely mathematical means of extracting the mode
coefficients because the natural range of $\sigma$ is not $0 \leq
\sigma \leq 2\pi$. There are other means of combating this -- see \emph{e.g.} \cite{GSW1},
Chapter 2, for more.}:
\begin{eqnarray}
\label{modes2}
L_{m} &=&
\frac{1}{2\pi\alpha'}\int_{0}^{2\pi}d\sigma{e^{im\sigma}}T_{++}\Bigg.\Bigg|_{\tau
= 0} \nonumber \nonumber \\ &\equiv&
\frac{1}{2\pi\alpha'}\int_{0}^{2\pi}d\sigma{e^{-im\sigma}}T_{--}\Bigg.\Bigg|_{\tau
= 0} \nonumber \nonumber \\ &=& \frac{1}{2}\sum_{n =
-\infty}^{\infty}{\alpha}_{m-n}\cdot{{\alpha}_{n}}.
\end{eqnarray}
\end{itemize}
As a result of all this, we can finally write the Virasoro
constraints (vanishing of the world-sheet energy-momentum tensor)
in the simple form:
\begin{eqnarray}
\label{virasoroconstraints}
{\tilde{L}}_{m} = L_{m} = 0 \quad \forall \quad m.
\end{eqnarray}
\section{Quantization}
\subsection{Canonical Covariant Quantization}
In order to quantize the Bosonic string we use the Poisson
brackets we have found and make the substitution:
\begin{eqnarray}
\left[f,g\right]_{P.B.} \rightarrow -i[\hat{f},\hat{g}],
\end{eqnarray}
where $f$ and $g$ now become operators $\hat{f}$ and $\hat{g}$.
The poisson brackets found in (\ref{xppb}) and (\ref{modepb}) thus become:
\begin{eqnarray}
\left[X^{\mu}(\sigma),X^{\nu}(\sigma')\right] &=&
\left[P^{\mu}(\sigma),P^{\nu}(\sigma')\right] = 0
\nonumber \\
\left[X^{\mu}(\sigma),P^{\nu}(\sigma')\right] &=& i\delta(\sigma -
\sigma'){\eta}^{\mu\nu},
\end{eqnarray}
and
\begin{eqnarray}
\left[{\alpha}_m^{\mu},{\alpha}_n^{\nu}\right] =
\left[{\tilde{\alpha}}_m^{\mu},{\tilde{\alpha}}_n^{\nu}\right]
&=& m{\delta}_{m+n}{\eta}^{\mu\nu} \nonumber \\
\left[{\alpha}_m^{\mu},{\tilde{\alpha}}_n^{\nu}\right] &=&
0 \nonumber \\
\left[x^{\mu},p^{\nu}\right] &=& i{\eta}^{\mu\nu}.
\end{eqnarray}
By comparison with a Simple Harmonic Oscillator, we can identify
the ${\alpha}_{m}^{\mu}$ and ${\tilde{\alpha}}_{m}^{\mu}$ as
raising and lowering operators for negative or positive $m$
respectively. \emph{i.e.}

\noindent Raising operators:
\begin{eqnarray*}
\left({\alpha}_{m}^{\mu}\right)^{\dagger} \equiv {{\alpha}_{-m}^{\mu}} \quad m > 0 \\
\left({\tilde{\alpha}}_{m}^{\mu}\right)^{\dagger} \equiv
{{\tilde{\alpha}}_{-m}^{\mu}} \quad m > 0.
\end{eqnarray*}
Lowering operators:
\begin{eqnarray*}
{\alpha}_{m}^{\mu} \quad m > 0 \\
{\tilde{\alpha}}_{m}^{\mu} \quad m > 0.
\end{eqnarray*}
We therefore define states (with specific centre-of-mass momentum
$p^{\mu}$) by:
\begin{eqnarray*}
\left|\phi;p^{\mu}\right>
\end{eqnarray*}
such that
\begin{eqnarray*}
{\alpha}_{m}^{\mu}\left|\phi;p^{\mu}\right> =
{\tilde{\alpha}}_{m}^{\mu}\left|\phi;p^{\mu}\right> = 0 \quad m >
0,
\end{eqnarray*}
and
\begin{eqnarray*}
{\alpha}_{-m}^{\mu}\left|\phi;p^{\mu}\right> \quad ; \quad
{\tilde{\alpha}}_{-m}^{\mu}\left|\phi;p^{\mu}\right>
\end{eqnarray*}
denote excited states. We usually denote the ground state by
$\left|0;p^{\mu}\right>$. With this viewpoint, the
${\alpha}^{\mu}$ are creating oscillations on a string of overall
momentum $p^{\mu}$. The conditions (\ref{virasoroconstraints}) are then
translated into operator equations. Comparing with the
Gupta-Bleuler treatment of QED, it is clear that the naive
conditions
\begin{eqnarray*}
L_{m}\left|\phi\right> = \tilde{L}_{m}\left|\phi\right> = 0 \quad
\forall \quad m
\end{eqnarray*}
are inconsistent\footnote{See Section \ref{stringstatessection} for
more on why.}, and that we can actually only require the weaker
conditions of:
\begin{eqnarray}
\left<\phi\right|L_{m}\left|\phi\right> =
\left<\phi\right|\tilde{L}_{m}\left|\phi\right> = 0 \quad \forall
\quad m,
\end{eqnarray}
\emph{i.e.} that
\begin{eqnarray}
L_{m}\left|\phi\right> = \tilde{L}_{m}\left|\phi\right> = 0 \quad
\forall \quad m > 0,
\end{eqnarray}
or equivalently
\begin{eqnarray}
\left<\phi\right|L_{-m} = \left<\phi\right|\tilde{L}_{-m} = 0
\quad \forall \quad m > 0,
\end{eqnarray}
where $L_{-m} = {L_{m}}^{\dagger}$ and $\left|\phi\right>$ is a
physical state.

We should also be clear that special care is needed if $m=0$. We
have:
\begin{eqnarray*}
L_{m} = \frac{1}{2}\sum_{n =
-\infty}^{\infty}{\alpha}_{m-n}\cdot{{\alpha}_{n}},
\end{eqnarray*}
with a similar expression for ${\tilde{L}}_{m}$. However, for
$L_{0}$ (and ${\tilde{L}}_{0}$), ${\alpha}_{-n}$ does not commute with
${\alpha}_{n}$. There is a normal ordering ambiguity, so we
\textit{define} $L_{0}$ by\footnote{See also equation (\ref{howtonormalorder}).}:
\begin{eqnarray}
L_{0} &=& \frac{1}{2}{\alpha}_{0}^{2} + \frac{1}{2}\sum_{n \neq
0}{:{\alpha}_{-n}\cdot{\alpha}_{n}:} \\
&=& \frac{1}{2}{\alpha}_{0}^{2} +
\sum_{n=1}^{\infty}{\alpha}_{-n}\cdot{\alpha}_{n},
\end{eqnarray}
with similar expressions for ${\tilde{L}}_{0}$. Because of this
normal ordering ambiguity, we should then include an undetermined
constant in the quantum constraints provided by $L_{0}$
$({\tilde{L}}_{0})$. We thus modify these conditions to:
\begin{eqnarray}
(L_{0} - a)\left|\phi\right> &=& 0, \\
({\tilde{L}}_{0} - \tilde{a})\left|\phi\right> &=& 0,
\end{eqnarray}
with $a$ and $\tilde{a}$ undetermined constants. The $L_{0}$
$({\tilde{L}}_{0})$ conditions are important both classically and
quantum mechanically as they give rise to the mass of the string.

\subsubsection{String Mass}
\underline{Classically}:
\begin{itemize}
\item For the \textbf{open} string, ${\alpha}_{0}^{\mu}=
\sqrt{2\alpha'}\,p^{\mu}$, so:
\begin{eqnarray*}
L_{0} = {\alpha'}p^2 +
\sum_{n=1}^{\infty}{\alpha}_{-n}\cdot{\alpha}_{n} = 0,
\end{eqnarray*}
and with $p^2 = -M^2$ as usual gives:
\begin{eqnarray}
M^2 =
\frac{1}{\alpha'}\sum_{n=1}^{\infty}{\alpha}_{-n}\cdot{\alpha}_{n}
= \frac{1}{\alpha'}N,
\end{eqnarray}
where we have defined
\begin{eqnarray}
N = \sum_{n=1}^{\infty}{\alpha}_{-n}\cdot{\alpha}_{n}.
\end{eqnarray}
\item For the \textbf{closed} string, we usually combine the $L_{0}$ and
${\tilde{L}}_{0}$ conditions into:
\begin{eqnarray}
L_{0} - {\tilde{L}}_{0} &=& 0 \Rightarrow N=\tilde{N}, \\
L_{0} + {\tilde{L}}_{0} &=& 0 \Rightarrow
\frac{1}{2}\left({\alpha}_{0}^2 + {\tilde{\alpha}}_{0}^2\right) =
-(N + \tilde{N}),
\end{eqnarray}
the first of which is known as \textit{level-matching}. The closed string
has
${\alpha}_{0}^{\mu}={\tilde{\alpha}}_{0}^{\mu}=\sqrt{{\alpha'}/2}\,p^{\mu}$,
so we end up with
\begin{eqnarray}
M^2 = \frac{2}{\alpha'}\sum_{n =
1}^{\infty}\left({\alpha}_{-n}\cdot{\alpha}_{n} +
{\tilde{\alpha}}_{-n}\cdot{\tilde{\alpha}}_{n}\right) =
\frac{2}{\alpha'}(N+\tilde{N}).
\end{eqnarray}
\end{itemize}

\noindent\underline{Quantum Mechanically}: \begin{itemize} \item
For the \textbf{open} string:
\begin{eqnarray}
\left(L_{0} - a\right)\left|\phi\right> = 0 &\Rightarrow&
\frac{1}{2}{\alpha}_{0}^{2}\left|\phi\right> = \left(a -
\sum_{n=1}^{\infty}{\alpha}_{-n}\cdot{\alpha}_{n}\right)\left|\phi\right>
\nonumber \\
&\Rightarrow& M^2\left|\phi\right> = -\frac{1}{\alpha'}\left(a -
\sum_{n=1}^{\infty}{\alpha}_{-n}\cdot{\alpha}_{n}\right)\left|\phi\right>,
\nonumber
\end{eqnarray}
and since $\left|\phi\right>$ is any arbitrary physical state, we
must have that
\begin{eqnarray}
M^2 = -\frac{1}{\alpha'}\left(a -
\sum_{n=1}^{\infty}{\alpha}_{-n}\cdot{\alpha}_{n}\right)\ ,
\end{eqnarray}
\emph{i.e.} $M^2=(N-a)/\alpha'$.
\item Similar considerations for the \textbf{closed} string lead to level-matching:
\begin{eqnarray}
N-a = \tilde{N}-\tilde{a},
\end{eqnarray}
and mass
\begin{eqnarray}
M^2 = \frac{2}{\alpha'}\left(N+\tilde{N}-a-\tilde{a}\right).
\end{eqnarray}
\end{itemize}
\subsubsection{The Virasoro Algebra}\label{virasalgebra}
At this point it is very useful for us to know something about the
algebra of the $L_{m}$ and ${\tilde{L}}_{m}$. This is known as the
Virasoro algebra. Classically, the algebra of the $L$s is easily found. One can start off by 
showing that
\begin{eqnarray}
\label{usefulpb}
\left[L_{m},\alpha_{n}^{\mu}\right]_{P.B.}=in\alpha_{m+n}^{\mu}\ ,
\end{eqnarray}
using the definition of $L_m$ as
\begin{eqnarray}
L_m=\frac{1}{2}\sum_{n=-\infty}^{\infty}\alpha_{m-n}\cdot\alpha_{n}\nonumber\ ,
\end{eqnarray}
and the algebras of the $\alpha$'s which is
\begin{eqnarray}
\left[{\alpha}_m^{\mu},{\alpha}_n^{\nu}\right]_{P.B.} 
= -im{\delta}_{m+n}{\eta}^{\mu\nu} \nonumber\ .
\end{eqnarray}
We can then calculate the required commutator:
\begin{eqnarray}
\left[L_{m},L_{n}\right]_{P.B.} &=& \frac{1}{2}\eta_{\mu\nu}\left[\sum_{p=-\infty}^{\infty}\alpha_{m-p}^{\mu}\alpha_p^\nu,L_n\right]_{P.B.}\nonumber\\
&=& \frac{1}{2}\eta_{\mu\nu}\sum_{p=-\infty}^{\infty}\left(\alpha_{m-p}^{\mu}\left[\alpha_p^\nu,L_n\right]_{P.B.}+
\left[\alpha_{m-p}^{\mu},L_n\right]_{P.B.}\alpha_p^\nu\right)\nonumber\\
&=& \frac{-i}{2}\eta_{\mu\nu}\sum_{p=-\infty}^{\infty}\left(p\alpha_{m-p}^{\mu}\alpha_{p+n}^{\nu}+
(m-p)\alpha_{m+n-p}^{\mu}\alpha_p^\nu\right)\nonumber\\
&=& -i\left(m\frac{1}{2}\sum_{p=-\infty}^{\infty}\alpha_{m+n-p}\cdot\alpha_{p}-\frac{1}{2}\sum_{p=-\infty}^{\infty}p\alpha_{m+n-p}\cdot\alpha_{p}
+\frac{1}{2}\sum_{p=-\infty}^{\infty}p\alpha_{m-p}\cdot\alpha_{p+n}\right)\nonumber\\
&=&  -i\left(m\frac{1}{2}\sum_{p=-\infty}^{\infty}\alpha_{m+n-p}\cdot\alpha_{p}-\frac{1}{2}\sum_{p=-\infty}^{\infty}p\alpha_{m+n-p}\cdot\alpha_{p}
+\frac{1}{2}\sum_{q=-\infty}^{\infty}(q-n)\alpha_{m+n-q}\cdot\alpha_{q}\right)\nonumber\\
&=&  -i\left(m\frac{1}{2}\sum_{p=-\infty}^{\infty}\alpha_{m+n-p}\cdot\alpha_{p}-\frac{1}{2}\sum_{p=-\infty}^{\infty}p\alpha_{m+n-p}\cdot\alpha_{p}
+\frac{1}{2}\sum_{p=-\infty}^{\infty}(p-n)\alpha_{m+n-p}\cdot\alpha_{p}\right)\nonumber\\
&=& -i(m-n)\underbrace{\frac{1}{2}\sum_{p=-\infty}^{\infty}\alpha_{m+n-p}\cdot\alpha_{p}}_{L_{m+n}}\ .\nonumber
\end{eqnarray}
From the first to the second line a standard commutator result from Appendix \ref{mathappendix} has been used. From the 
second to the third we have used (\ref{usefulpb}). The third to the fourth is a simple rearrangement, while from the fourth 
to the fifth the variable of summation in the last term has been changed to $q=p+n$. In the following line it has simply 
been relabelled as $p$ since it is a dummy variable. Therefore we finally find:
\begin{eqnarray}
\left[L_{m},L_{n}\right]_{P.B.} = -i(m-n)L_{m+n}\ .
\end{eqnarray}
However, when we try to compute
the quantum analogue of this, we run into ordering issues. For
$m+n \neq 0$ it is similarly easy to show that
\begin{eqnarray}
\left[L_{m},L_{n}\right] = (m-n)L_{m+n}, \nonumber
\end{eqnarray}
but more care is needed if $m+n = 0$. Considering this case, we
find that there is an anomaly which modifies the algebra by the
addition of a term to the commutator. This is called the central
extension, and there are many different ways to calculate it\footnote{See
Appendix \ref{VirasoroAppendix}, or \cite{GSW1,Polchinski1,Lust,Zwiebach} for example.},
but the end result is to provide an extra constant term
when $m+n=0$, so that the full Virasoro algebra is:
\begin{eqnarray}
\left[L_{m},L_{n}\right] = (m-n)L_{m+n} +
\frac{d}{12}(m^3-m){\delta}_{m+n},
\end{eqnarray}
with $d$ the dimension of spacetime.
\subsubsection{String States}\label{stringstatessection}
We can now think about constructing string states.

For the \textbf{open} string, the physical state conditions are:
\begin{eqnarray}
L_{m}\left|\phi;p\right> &=& 0 \quad \forall \quad m > 0 \nonumber \\
\left(L_{0}-a\right)\left|\phi;p\right> &=& 0.
\end{eqnarray}
If we think about these acting on the ground state
$\left|0;p\right>$ for a minute, we see that as ${\alpha}_{n}$ is
an annihilation operator for $n > 0$, it will kill the ground
state, and thus ensure the $L_{m}$ conditions automatically.
Similarly it will ensure that $N$ annihilates the ground state, so
that the mass condition is:
\begin{eqnarray}
{\alpha}'p^2 = a \Rightarrow M^2 = -\frac{a}{\alpha'}.
\end{eqnarray}

Consider now states at the first excited level:
\begin{eqnarray}
{\zeta}_{\mu}{\alpha}_{-1}^{\mu}\left|0;p\right>,
\end{eqnarray}
where $\zeta_{\mu}$ is a polarisation vector with $d$ independent
components before gauge constraints are taken into account. Now,
$N$ applied to this state has value 1 (we usually say that $N=1$
\emph{etc.}), and we have for the mass-shell condition:
\begin{eqnarray}
\left(L_{0}-a\right){\zeta}_{\mu}{\alpha}_{-1}^{\mu}\left|0;p\right>
= 0 \Rightarrow M^2 = -\frac{a}{\alpha'}(a-1).
\end{eqnarray}
The $L_{m}$ conditions can then be dealt with by using the
easily-proven identity
$\left[L_{m},{\alpha}_{n}^{\mu}\right]=-n{\alpha}_{m+n}^{\mu}$.
They are identically satisfied for $m>1$, but for $m=1$ we get:
\begin{eqnarray}
L_{1}{\zeta}_{\mu}{\alpha}_{-1}^{\mu}\left|0;p\right> =
{\zeta}_{\mu}{\alpha}_{0}^{\mu}\left|0;p\right> =
\sqrt{2\alpha'}\,{\zeta}_{\mu}p^{\mu}\left|0;p\right>. \nonumber
\end{eqnarray}
Requiring this to be zero then gives the condition:
\begin{eqnarray}
{\zeta}_{\mu}p^{\mu} = 0.
\end{eqnarray}

\noindent Clearly the value of $a$ is crucial here in determining
the masses of these states. We can consider there to be `3'
possible choices for $a$: \begin{enumerate} \item If we choose
$a>1$, then $M^2 <0$ and this first excited state is tachyonic (!)
$p^2>0$ and so we can choose some frame in which $p^{\mu} =
(0,\underline{0},p)$ ,\emph{i.e.} $p^{\mu}=0$ for $\mu=0 \ldots d-2$ and
$p^{d-1}=p$. Then ${\zeta}_{\mu}p^{\mu}=0\Rightarrow
{\zeta}_{d-1}=0$. However this does not provide any constraint on
${\zeta}_{0}$. The norm of this first excited state is:
\begin{eqnarray}
\left<0;p\right|{\zeta}_{\mu}{\alpha}_{1}^{\mu}{\zeta}_{\nu}{\alpha}_{-1}^{\nu}\left|0;p\right>
&=&
{\zeta}_{\mu}{\zeta}_{\nu}\left\{\left<0;p\right|{\alpha}_{-1}^{\mu}{\alpha}_{1}^{\nu}\left|0;p\right>+{\eta}^{\mu\nu}\left<0;p\right.\left|0;p\right>\right\}
\nonumber \\
&=&{\zeta}_{\mu}{\zeta}_{\nu}{\eta}^{\mu\nu}\left<0;p\right.\left|0;p\right>
\nonumber \\
&=&{\zeta}\cdot{\zeta},
\end{eqnarray}
if the ground state is normalized to 1. So for this first excited
state to have positive-definite norm, we require that
${\zeta}\cdot\zeta \geq 0$. There is no constraint on
${\zeta}_{0}$, and as this entry will pick up a minus when squared
due to the signature of space-time, we cannot ensure that
$\zeta\cdot\zeta \geq 0$. \item If we choose $a=1$ then
$M^2=-p^2=0$ and $p^{\mu}$ is a null vector. We can choose a frame
in which $p^{\mu}=(\omega,\underline{0},\omega)$ and the
constraint is
${\zeta}_{\mu}p^{\mu}=\omega({\zeta}_0+{\zeta}_{d-1})=0\Rightarrow{\zeta}_0=-{\zeta}_{d-1}$.
Using this we have that
${\zeta}\cdot{\zeta}=-{\zeta}_{0}^2+{\zeta}_{i}^2+{\zeta}_{d-1}^2={\zeta}_{i}^2
\geq 0$\footnote{Here $i=1\ldots{d-2}$.}.\item If we choose $a<1$
then we have $M^2>0\Rightarrow{p}^2<0$ and we can choose a frame
in which $p^{\mu}=(m,\underline{0})$ so that
${\zeta}\cdot{p}=0\Rightarrow{\zeta}_{0}=0$. This means that we
can again have ${\zeta}\cdot{\zeta} \geq 0$.
\end{enumerate}
Clearly, for states to be physically reasonable we must have
$a\leq1$. \newline

This is the first condition for the absence of negative norm
states (ghosts\footnote{Note, these are not ghosts in the BRST
sense.}) in string theory. The absence of these states in general
is known as the `No-Ghost Theorem'. Following \cite{GSW1},
we'll take a quick look at this. \newline

We define \underline{physical} states by
$L_{m}\left|\phi\right>=0$ for $m>0$ and
$\left(L_{0}-a\right)\left|\phi\right>=0$. A state is called
\underline{spurious} if it obeys
$\left(L_{0}-a\right)\left|\psi\right>=0$, but is orthogonal to
\textbf{all} physical states:
\begin{eqnarray}
\left<\phi\right.\left|\psi\right>=0.
\end{eqnarray}
As $\left<\phi\right|L_{-n}=0$ for $n>0$, we can always write
spurious states in the form
\begin{eqnarray}
\label{spurious2}
\left|\psi\right>=\sum_{n>0}L_{-n}\left|{\chi}_{n}\right>,
\end{eqnarray}
where the $\left|{\chi}_{n}\right>$ are states that obey
$\left(L_{0}-a+n\right)\left|{\chi}_{n}\right>=0$. In fact we can
truncate this infinite sum, since for $n\geq3$ the $L_{-n}$ can be
represented as iterated commutators of $L_{-1}$ and $L_{-2}$, so
we can simply write a spurious state as:
\begin{eqnarray}
\left|\psi\right>=L_{-1}\left|{\chi}_{1}\right>+L_{-2}\left|{\chi}_{2}\right>.
\end{eqnarray}

We get something special if a state is both spurious
\underline{and} physical. That is, if we have a state that
satisfies:
\begin{eqnarray}
\left<\phi\right.\left|\psi\right>=0 \quad ; \quad
L_{m}\left|\psi\right>=0,\quad \forall \quad m>0 \quad ; \quad
\left(L_{0}-a\right)\left|\psi\right>=0.
\end{eqnarray}
As we can write these states in the form of (\ref{spurious2}),
we have that
\begin{eqnarray}
\left<\psi\right.\left|\psi\right>=\sum_{m>0}\left<{\chi}_{m}\right|L_{m}\left|\psi\right>=0,
\end{eqnarray}
so these states have zero norm. These states are orthogonal to all
physical states (including themselves) and are sometimes called
`null' states.

We can construct states of this type by considering spurious
states of the form
\begin{eqnarray}
\left|\psi\right>=L_{-1}\left|\tilde{\chi}\right>,
\end{eqnarray}
where $\left|\tilde{\chi}\right>$ is an arbitrary state satisfying
$L_{m}\left|\tilde{\chi}\right>=0$ for $m>0$ and
$\left(L_{0}-a+1\right)\left|\tilde{\chi}\right>=0$. Here
$\left|\tilde{\chi}\right>$ could be the zero momentum state
$\left|0;0\right>$ or really any physical state with suitably
shifted $p^{\mu}$. In addition to being spurious,
$\left|\psi\right>$ also satisfies all the conditions for being
physical apart from the $L_1$ condition:
\begin{eqnarray*}
L_{1}\left|\psi\right>=L_{1}L_{-1}\left|\tilde{\chi}\right>=2L_{0}\left|\tilde{\chi}\right>,
\end{eqnarray*}
which vanishes if $a=1$. In this case the $\left|\psi\right>$
states are both spurious and physical and hence have zero norm.
Clearly by applying $L_{-1}$ to an arbitrary state
$\left|\tilde{\chi}\right>$, an infinite number of states with
zero norm can be made. If $a=1$, the first excited state of the
open string is massless and is the simplest example where
$\left|\tilde{\chi}\right>=\left|0;k\right>$.

In fact, if we consider spurious states with the structure:
\begin{eqnarray}
\left|\psi\right>=\left(L_{-2}+{\gamma}L_{-1}^2\right)\left|\tilde{\chi}\right>,
\end{eqnarray}
then we have an infinite set of zero norm states if $a=1$, $d=26$
and $\gamma=3/2$. The existence of this infinite extra set of null
states is essential for the decoupling of all the time-like modes
in the critical dimension. The general rule (No-Ghost Theorem) is
that the spectrum is ghost-free provided that $a=1$ and $d=26$ or
$a\leq1$ and $d\leq25$.

As a general statement in $d=26$ with $a=1$, we can write any
physical state as:
\begin{eqnarray}
\left|\phi\right>=\left|{\phi}_{S}\right>+\left|{\phi}_{T}\right>,
\end{eqnarray}
where $\left|{\phi}_{S}\right>$ is null (spurious and physical)
and $\left|{\phi}_{T}\right>$ has positive norm.\newline

Summarizing, for the \textbf{open} string we have (we'll take
$a=1$ and $d=26$, and see even more evidence that this should be
the case later):

\begin{enumerate}
\item A tachyonic ground state:
\begin{eqnarray}
\left|0;p\right> \qquad M^2=-\frac{1}{\alpha'}.
\end{eqnarray}
\item With $a=1$, the first excited state is massless . What's
more, the $L_{1}$ condition $p^{\mu}{\zeta}_{\mu}=0$ therefore
allows for transformations of the form ${\zeta}_{\mu} \rightarrow
{\zeta}_{\mu}+{p_{\mu}\chi}$. But this is just the fourier
transform of the $U(1)$ gauge equivalence $A_{\mu} \rightarrow
A_{\mu} + {\partial}_{\mu}\chi$. That is to say, this state is a
massless photon!
\begin{eqnarray}
{\zeta}_{\mu}{\alpha}_{-1}^{\mu}\left|0;p\right> \qquad M^2=0.
\end{eqnarray}
\item The second excited state has the form:
\begin{eqnarray}
\left({\zeta}_{\mu\nu}^{(2)}{\alpha}_{-1}^{\mu}{\alpha}_{-1}^{\nu}+{\zeta}_{\mu}^{(1)}{\alpha}_{-2}^{\mu}\right)\left|0;p\right>
\qquad M^2=1/{\alpha'},
\end{eqnarray}
with ${\zeta}_{\mu\nu}^{(2)}$ a symmetric $2^{\textrm{nd}}$ rank tensor. The
$L_{1}$ and $L_{2}$ (`gauge') conditions are:
\begin{eqnarray}
\sqrt{2\alpha'}\,p^{\mu}{\zeta}_{\mu\nu}^{(2)}=-{\zeta}_{\nu}^{(1)}\\
{\eta}^{\mu\nu}{\zeta}_{\mu\nu}^{(2)}=-2\sqrt{2\alpha'}\,{\zeta}_{\mu}^{(1)}p^{\mu}=4\alpha'p^{\mu}p^{\nu}{\zeta}_{\mu\nu}^{(2)}\ ,
\end{eqnarray}
which gives a non-null, massive spin-2 particle. These are the lowest
lying states.
\end{enumerate}

For the \textbf{closed} string we have:

\begin{eqnarray*}
L_{m}&=&\frac{1}{2}\sum_{n}{\alpha}_{m-n}\cdot{\alpha}_{n} \quad
m\neq 0 \quad ; \quad
L_{0} = N+\frac{{\alpha}_{0}^2}{2} \\
\tilde{L}_{m}&=&\frac{1}{2}\sum_{n}{\tilde{\alpha}}_{m-n}\cdot{\tilde{\alpha}}_{n}
\quad m\neq 0 \quad ; \quad L_{0} =
\tilde{N}+\frac{{\tilde{\alpha}}_{0}^2}{2},
\end{eqnarray*}
where the physical state conditions are:
\begin{eqnarray*}
L_{m}\left|\phi\right>&=&\tilde{L}_{m}\left|\phi\right>=0 \quad
m\neq
0 \\
(L_{0}-a)\left|\phi\right>&=&(\tilde{L}_{0}-\tilde{a})\left|\phi\right>=0.
\end{eqnarray*}
Here we essentially have a doubling of the physical space of
states:
\begin{eqnarray*}
|\phi,\tilde{\phi}\rangle_{closed}
\equiv|\phi\rangle_{open}\otimes|\tilde{\phi}\rangle_{open},
\end{eqnarray*}
where the states $|\phi\rangle$ and $|\tilde{\phi}\rangle$ are
created by acting on the vacuum with ${\alpha}_{-n}$ and
$\tilde{\alpha}_{-n}$ respectively.

In fact we can make linear combinations of the above $L_{0}$ and
$\tilde{L}_{0}$ conditions to give:
\begin{eqnarray}
\label{mass-shell}
\left(L_{0}+\tilde{L}_{0}-(a+\tilde{a})\right)\left|\phi\right>=0
\\
\label{level-matching}
\left(L_{0}-\tilde{L}_{0}-(a-\tilde{a})\right)\left|\phi\right>=0,
\end{eqnarray}
where (\ref{mass-shell}) is known as the `mass-shell'
condition, and (\ref{level-matching}) is known as `level-matching'.
In Minkowski space, where $p_{L}^{\mu}=p_{R}^{\mu}=p^{\mu}$, the
level-matching condition tells us that
$(N-\tilde{N})|\phi\rangle=0$, \emph{i.e.} $N=\tilde{N}$.

$N$ and $\tilde{N}$ both kill the ground state $(N=\tilde{N}=0)$,
so we have:
\begin{eqnarray*}
\left(L_{0}-a\right)|0;p\rangle=0 &\Rightarrow&
\left(N+\frac{{\alpha}'p^2}{4}-a\right)|0;p\rangle=0 \\
&\Rightarrow& M^2=-p^2=-\frac{4a}{\alpha'}.
\end{eqnarray*}
Similar considerations for $\tilde{L}_{0}$ lead to:
\begin{eqnarray*}
M^2=-\frac{4\tilde{a}}{\alpha'}.
\end{eqnarray*}
Thus if we require a consistent mass formula for the string, we
must have $\tilde{a}=a$, and then the open string results fix
$a=1$. Again we find that the ground state is tachyonic!

Level matching now tells us that there are no states with
$(N,\tilde{N})=(1,0)/(0,1)$, so the first excited state has
$N=\tilde{N}=1$:
\begin{eqnarray*}
{\zeta}_{\mu\nu}{\alpha}_{-1}^{\mu}{\tilde{\alpha}}_{-1}^{\nu}|0;p\rangle,
\end{eqnarray*}
with $\zeta$ a non-symmetric $2^{\textrm{nd}}$ rank tensor, and gauge conditions
from $L_1$ and $\tilde{L}_1$:
\begin{eqnarray*}
p^{\mu}{\zeta}_{\mu\nu}=p^{\nu}{\zeta}_{\mu\nu}=0.
\end{eqnarray*}
The mass is again zero as for the first excited state of the open
string, and as such we can find a frame in which
$p^{\mu}=\omega(1,\underline{0},1)$. In this case the constraints
become:
\begin{eqnarray*}
p^{\mu}{\zeta}_{\mu\nu}=0 &\Rightarrow&
{\zeta}_{0\nu}+{\zeta}_{(d-1)\nu}=0 \\
p^{\nu}{\zeta}_{\mu\nu} &\Rightarrow&
{\zeta}_{\mu0}+{\zeta}_{\mu(d-1)}=0,
\end{eqnarray*}
which means that $\mu,\nu$ effectively have transverse values
$i,j=1\ldots{d-2}$. Effectively
${\zeta}_{\mu\nu}\rightarrow{\zeta}_{ij}$, and we can decompose
this under the irreducible representations of $SO(d-2)$:
\begin{eqnarray}
{\zeta}_{ij}={\zeta}_{(ij)}+{\zeta}_{[ij]}+{\zeta}_{i}^{i}.
\end{eqnarray}
${\zeta}_{(ij)}$ is symmetric and trace-free - the graviton
$g_{ij}$. ${\zeta}_{[ij]}$ is antisymmetric - the antisymmetric
tensor $B_{ij}$ and ${\zeta}_{i}^{i}$ is scalar - the dilaton
$\Phi$. We see that the spectrum of the closed Bosonic string
contains a graviton!\newline

Summarizing for the \textbf{closed} string, with $a=\tilde{a}=1$
and $d=26$, we have:

\begin{enumerate}
\item A tachyonic ground state:
\begin{eqnarray}
|0;p\rangle \quad M^2=-\frac{4}{\alpha'}.
\end{eqnarray}
\item A massless first-excited state:
\begin{eqnarray}
{\zeta}_{\mu\nu}{\alpha}_{-1}^{\mu}{\tilde{\alpha}}_{-1}^{\nu}|0;p\rangle
\quad M^2=0,
\end{eqnarray}
which contains the graviton $g$, the antisymmetric tensor $B$ and
the dilaton $\Phi$. These are the lowest-lying states.
\end{enumerate}

This quantisation procedure that we have been pursuing shows us
that the quantisation is manifestly covariant, but not manifestly
free of ghosts. We also know that even when we have chosen our
gauge $({\gamma}_{\alpha\beta}={\eta}_{\alpha\beta})$, there is
still the freedom (residual gauge symmetry) to make
reparametrizations. We can use this freedom to make a specific
non-covariant choice (the so-called light-cone gauge) in which the
covariance is obscure, but the spectrum is manifestly free of
ghosts. It is also possible to show that the two views are
equivalent, which implies a covariant \textbf{and} ghost-free
theory. We shall now look at this light-cone gauge quantisation
procedure.

\subsection{Light-Cone Gauge Quantisation}

We begin by introducing light-cone coordinates in the space-time:
\begin{eqnarray}
X^{+}=\frac{X^{0}+X^{d-1}}{\sqrt{2}} \quad
X^{-}=\frac{X^{0}-X^{d-1}}{\sqrt{2}}.
\end{eqnarray}
This is similar to forming ${\xi}^{\pm}=\tau\pm\sigma$, but there
is a big difference in that we are now singling out two
coordinates (from $d$) in an arbitrary and non-covariant way.
Thus:
\begin{eqnarray*}
X^{\mu} \rightarrow X^{+},X^{-},X^{i} \quad i=1\ldots{d-2} \\
{\eta}_{ij}=1,\quad
{\eta}_{+-}={\eta}_{-+}={\eta}^{+-}={\eta}^{-+}=-1.
\end{eqnarray*}
The components of a vector $V^{\mu}$ are:
\begin{eqnarray*}
V^{\pm}=\frac{1}{\sqrt{2}}(V^{0}\pm V^{d-1}), \quad V^{i} \quad
i=1\ldots{d-2},
\end{eqnarray*}
and the inner product is:
\begin{eqnarray*}
V^{\mu}W_{\mu}=V^{i}W^{i}-V^{+}W^{-}-V^{-}W^{+}.
\end{eqnarray*}

In terms of the ${\xi}^{\pm}$, the residual gauge invariance
corresponds to the possibility of making arbitrary
reparametrizations:
\begin{eqnarray}
{\xi}^{\pm} \rightarrow
{\tilde{\xi}}^{\pm}\left({\xi}^{\pm}\right).
\end{eqnarray}
For closed strings we may reparametrize them independently, but
for open strings they are linked by the boundary conditions. They
thus transform $\tau=({\xi}^{+}+{\xi}^{-})/2$ and
$\sigma=({\xi}^{+}-{\xi}^{-})/2$ into
\begin{eqnarray}
\tilde{\tau}&=&\frac{1}{2}\left[{\tilde{\xi}}^{+}\left({\xi}^{+}\right)+{\tilde{\xi}}^{-}\left({\xi}^{-}\right)\right]
\\
\tilde{\sigma}&=&\frac{1}{2}\left[{\tilde{\xi}}^{+}\left({\xi}^{+}\right)-{\tilde{\xi}}^{-}\left({\xi}^{-}\right)\right].
\end{eqnarray}
This form for $\tilde{\tau}$ clearly asserts that it is a solution
to the massless wave equation:
\begin{eqnarray*}
\left({\partial}_{\sigma}^{2}-{\partial}_{\tau}^{2}\right)\tilde{\tau}=0.
\end{eqnarray*}
On the other hand, once we have chosen $\tilde{\tau}$,
$\tilde{\sigma}$ is automatically determined. We also know that
$X^{\mu}$ obeys the massless wave equation, so we can make a
reparametrization to choose $\tilde{\tau}$ to be equal to one of
the $X^{\mu}$. The usual light-cone gauge choice is:
\begin{eqnarray}
X^{+}(\sigma,\tau)=x^{+}+2\alpha'p^{+}\tau.
\end{eqnarray}
This is manifestly independent of $\sigma$, and in the classical
description corresponds to setting ${\alpha}_{n}^{+}=0$ for $n\neq
0$. The $X^{+}$ coordinate actually corresponds to the time
coordinate in a frame in which the string is travelling at
infinite momentum.

Having thus fixed $X^{+}$, the Virasoro constraints become:
\begin{eqnarray}
T_{\pm\pm}=\frac{1}{4}\left(\dot{X}\pm X'\right)^2=0 \nonumber \\
\Rightarrow \left(\dot{X}^{i}\pm
{{X}^{i}}'\right)-2\left(\dot{X}^{+}\pm
{{X}^{+}}'\right)\left(\dot{X}^{-}\pm {{X}^{-}}'\right)=0 \nonumber \\
\Rightarrow \left({\dot{X}^{-}}\pm
{X^{-}}'\right)=\frac{1}{4p^{+}\alpha'}\left({\dot{X}}^{i}\pm
{X^{i}}'\right)^2,
\end{eqnarray}
which can be solved for $X^{-}$ in terms of the $X^{i}$. Thus in
light-cone gauge we can actually eliminate both $X^{+}$ and
$X^{-}$, leaving only the transverse oscillators $X^{i}$.

If we now consider a solution to the equations of motion of the
open string with NN boundary conditions (see eqn.(\ref{neumann})), we can
explicitly solve to find:
\begin{eqnarray}
{\alpha}_{n}^{-}=\frac{1}{\sqrt{2\alpha'}\,p^{+}}
\left\{\frac{1}{2}\sum_{i=1}^{d-2}\sum_{m=-\infty}^{\infty}:{\alpha}_{n-m}^{i}{\alpha}_{m}^{i}:
-a{\delta}_{n}\right\}.
\end{eqnarray}
As in the covariant treatment, we have introduced an unknown
normal-ordering constant $a$ into ${\alpha}_{0}^{-}$. In
light-cone gauge, the identification of ${\alpha}_{0}^{-}$ with
$\sqrt{2\alpha'}\,p^{-}$ is the mass-shell condition. For the
\textbf{open} string we have:
\begin{eqnarray}
{\alpha}_{0}^{+}=\sqrt{2\alpha'}\,p^{+} \quad &;& \quad
{\alpha}_{0}^{-}=\sqrt{2\alpha'}\,p^{-} \nonumber \\
{\alpha}_{n}^{+}=0 \quad &\forall& n\neq 0,
\end{eqnarray}
while for the \textbf{closed} string the relations are:
\begin{eqnarray}
{\alpha}_{0}^{+}=\sqrt{\frac{\alpha'}{2}}\,p^{+} \quad &;& \quad
{\alpha}_{0}^{-}=\sqrt{\frac{\alpha'}{2}}\,p^{-} \nonumber \\
{\alpha}_{n}^{+}=0 \quad &\forall& n\neq 0,
\end{eqnarray}
with similar relations for ${\tilde{{\alpha}}}^{\pm}$. Then:
\begin{eqnarray}
L_{m} &=&
\frac{1}{2}\sum_{-\infty}^{\infty}{\alpha}_{m-n}\cdot{\alpha}_{n}
\nonumber \\
&=&
\frac{1}{2}\sum_{-\infty}^{\infty}\left\{{\alpha}_{m-n}^{i}{\alpha}_{n}^{i}
- {\alpha}_{m-n}^{+}{\alpha}_{n}^{-} -
{\alpha}_{m-n}^{-}{\alpha}_{n}^{+}\right\} \nonumber \\
&=& -{\alpha}_{0}^{+}{\alpha}_{m}^{-} +
\frac{1}{2}\sum_{-\infty}^{\infty}{\alpha}_{m-n}^{i}{\alpha}_{n}^{i},
\end{eqnarray}
and similarly for ${\tilde{L}}_{m}$. So:
\begin{eqnarray}
L_{0} = -2{\alpha}'p^{+}p^{-} + N + {\alpha}'p^{i}p^{i}
\end{eqnarray}
for the open string, where now $N =
\sum_{n=1}^{\infty}{\alpha}_{-n}^{i}{\alpha}_{n}^{i}$. The
mass-shell condition then becomes:
\begin{eqnarray}
\left[(-2p^{+}p^{-}+p^{i}p^{i}){\alpha}'+N-a\right]|\phi\rangle
&=&
0 \nonumber \\
\Rightarrow M^2 &=& \frac{1}{\alpha'}\left(N-a\right).
\end{eqnarray}
Clearly this is the same as before, except that now we only have
transverse $({\alpha}_{n}^{i})$ contributing.

The lowest \textbf{open} string states in light cone gauge are
thus:
\begin{enumerate}
\item Ground state
\begin{eqnarray}
|0;p\rangle \qquad M^2=-\frac{a}{\alpha'}
\end{eqnarray}
\item 1st excited state
\begin{eqnarray}
{\zeta}_{i}{\alpha}_{-1}^{i}|0;p\rangle \qquad
M^2=\frac{1-a}{\alpha'}
\end{eqnarray}
\item 2nd excited state
\begin{eqnarray}
\left({\zeta}_{ij}^{(2)}{\alpha}_{-1}^{i}{\alpha}_{-1}^{j}+{\zeta}_{i}^{(1)}{\alpha}_{-2}^{i}\right)|0;p\rangle
\qquad M^2=\frac{2-a}{\alpha'}.
\end{eqnarray}
\end{enumerate}
As only `Euclidean' indices are involved, we can see heuristically
that this gauge is good for the absence of ghosts: the `Euclidean'
metric has signature $(+,+,+,...) \Rightarrow$ positive definite
norms.

We also get a restriction on the parameter $a$. Consider the first
excited state. This is a transversely polarized vector, and when
subjected to a Lorentz transformation such vectors acquire a
longitudinal polarization, unless they are massless. Thus,
requiring the first excited state to be massless gives $a=1$.

We can also see how the dimension of space-time is restricted. The
constant $a$ arises from the need to normal order $L_{0}$. Let's
consider doing this explicitly:
\begin{eqnarray}
L_{0} =
\frac{1}{2}\sum_{-\infty}^{\infty}{\alpha}_{-n}^{i}{\alpha}_{n}^{i}
&=&
\frac{1}{2}\left\{\sum_{n=1}^{\infty}{\alpha}_{-n}^{i}{\alpha}_{n}^{i}+\sum_{n=-1}^{-\infty}{\alpha}_{-n}^{i}{\alpha}_{n}^{i}
+ {{\alpha}_{0}}^{2}\right\} \nonumber \\
&=&
\frac{1}{2}\left\{\sum_{n=1}^{\infty}{\alpha}_{-n}^{i}{\alpha}_{n}^{i}+\sum_{n=1}^{\infty}{\alpha}_{n}^{i}{\alpha}_{-n}^{i}
+ {{\alpha}_{0}}^{2}\right\} \nonumber \\
&=&
\frac{1}{2}\left\{\sum_{n=1}^{\infty}{\alpha}_{-n}^{i}{\alpha}_{n}^{i}+\sum_{n=1}^{\infty}{\alpha}_{-n}^{i}{\alpha}_{n}^{i}
+ {\eta}_{ij}{\eta}^{ij}\sum_{n=1}^{\infty}n + {{\alpha}_{0}}^{2}\right\} \nonumber \\
&=&
\frac{1}{2}\left\{\sum_{n=1}^{\infty}{\alpha}_{-n}^{i}{\alpha}_{n}^{i}+\sum_{n=-1}^{-\infty}{\alpha}_{n}^{i}{\alpha}_{-n}^{i}
+ {{\alpha}_{0}}^{2} + (d-2)\sum_{n=1}^{\infty}n\right\} \nonumber \\
&=&
\frac{1}{2}\sum_{-\infty}^{\infty}:{\alpha}_{-n}^{i}{\alpha}_{n}^{i}:
+ \frac{d-2}{2}\sum_{n=1}^{\infty}n.
\end{eqnarray}
We can thus identify $-a = \frac{d-2}{2}\sum_{n=1}^{\infty}n$,
which is ostensibly divergent. However, we can deal with it using
`zeta function regularization'. The Riemann zeta function can be
defined by:
\begin{eqnarray}
\zeta(s) = \sum_{n=1}^{\infty}n^{-s},
\end{eqnarray}
for $\Re(s)>1$. But the zeta function also has a unique analytic
continuation to the point $s=-1$, where $\zeta(-1)=-1/12$. Thus
\begin{eqnarray}
\sum_{n=1}^{\infty}n = -\frac{1}{12} \quad (+\infty).
\end{eqnarray}
The requirement that $a=1$ then tells us that $d=26$. Critical
Bosonic string theory lives in 26 dimensions!

\subsection{Path Integral Quantization}

We know from quantum field theory that we can use the modern
covariant formalism of path integrals to quantise a theory and do
calculations. For instance, the two-point correlation function
(propagator) for a scalar field $\phi$ can be written as:
\begin{eqnarray*}
{\tau}_{2}(x_{1},x_{2})=\langle\phi(x_{1})\phi(x_{2})\rangle=\frac{1}{N}\int{\mathcal{D}
\phi}\,\phi(x_{1})\phi(x_{2})e^{iS[\phi]},
\end{eqnarray*}
where $N=\int\mathcal{D}\phi e^{iS[\phi]}$. Alternatively we could
write down a partition function:
\begin{eqnarray}
Z=\int\mathcal{D}X^{\mu}(\tau)\mathcal{D}e\,e^{iS[X,e]},
\end{eqnarray} \emph{i.e.} a path integral over the embedding
and the metric on the world-line, and use this to calculate things
with. The generalisation of this in string theory is:
\begin{eqnarray}
Z=\int\mathcal{D}X^{\mu}(\sigma,\tau)\mathcal{D}{\gamma}_{\alpha\beta}(\sigma,\tau)e^{iS[X,\gamma]},
\end{eqnarray}
or if we Wick-rotate to Euclidean space:
\begin{eqnarray}
Z_{E}=\int\mathcal{D}X\mathcal{D}\gamma e^{-S_{E}[X,\gamma]}.
\end{eqnarray}
Recall that we still have a gauge group of
\begin{eqnarray}
G = \textrm{Diffeomorphisms}\otimes\textrm{Weyl Transformations}
\end{eqnarray}
under which the theory is invariant. Eventually we wish to fix the
gauge to be the conformal one:
\begin{eqnarray}
{\gamma}_{\alpha\beta}\rightarrow {\tilde{\gamma}}_{\alpha\beta}
=e^{\phi}\left(\begin{array}{cc} \mp1 & 0 \\
0 & 1 \end{array}\right),
\end{eqnarray}
where we must use a minus if we are in Minkowski space and a plus
if we are in Euclidean space.

Under reparametrizations we have:
\begin{eqnarray}
\delta X^{\mu}&=&{\zeta}^{\alpha}{\partial}_{\alpha}X^{\mu} \\
\delta{\gamma}_{\alpha\beta}&=&2{\nabla}_{(\alpha}{\zeta}_{\beta)},
\end{eqnarray}
and
\begin{eqnarray}
\delta{\gamma}_{\alpha\beta}=2\Lambda{\gamma}_{\alpha\beta}
\end{eqnarray}
under Weyl transformations. So, under a general transformation we
have:
\begin{eqnarray}
\delta{\gamma}_{\alpha\beta}&=&
2{\nabla}_{(\alpha}{\zeta}_{\beta)}+2\Lambda{\gamma}_{\alpha\beta}
\\
&=&
(\mathcal{P}\zeta)_{\alpha\beta}+2{\tilde{\Lambda}}{\gamma}_{\alpha\beta},
\end{eqnarray}
where
\begin{eqnarray}
(\mathcal{P}\zeta)_{\alpha\beta}&=&2{\nabla}_{(\alpha}{\zeta}_{\beta)}-({\nabla}_{\gamma}{\zeta}^{\gamma}){\gamma}_{\alpha\beta}
\\
\& \quad
2\tilde{\Lambda}&=&2\Lambda+{\nabla}_{\gamma}{\zeta}^{\gamma}.
\end{eqnarray}
$\mathcal{D}{\gamma}_{\alpha\beta}\equiv\mathcal{D}(\mathcal{P}\zeta)\mathcal{D}\tilde{\Lambda}$,
and we wish to fix the gauge by transforming
\begin{eqnarray*}
\mathcal{D}{\gamma}_{\alpha\beta}&\rightarrow&\mathcal{D}\zeta\mathcal{D}\Lambda
\\
\textrm{\emph{i.e.}}\quad\mathcal{D}(\mathcal{P}\zeta)\mathcal{D}\tilde{\Lambda}&\rightarrow&\mathcal{D}\zeta\mathcal{D}\Lambda.
\end{eqnarray*}
So, we seek a Jacobian satisfying:
\begin{eqnarray}
\mathcal{D}(\mathcal{P}\zeta)\mathcal{D}\tilde{\Lambda}=\mathcal{D}\zeta\mathcal{D}\Lambda\left|\frac{\partial(\mathcal{P}\zeta,\tilde{\Lambda})}{\partial(\zeta,\Lambda)}\right|.
\end{eqnarray}
In fact it is easily checked that the Jacobian can be written as:
\begin{eqnarray}
\left|\begin{array}{cc} \mathcal{P} & 0 \\ \ast &
1\end{array}\right|,
\end{eqnarray}
where the precise form of $\ast$ is unimportant due to the
presence of the 0, and we have that
$\mathcal{D}\zeta\mathcal{D}\Lambda\equiv\mathcal{D}G$. Thus:
\begin{eqnarray}
Z_{E}=\underbrace{\int\mathcal{D}G}_{{\textrm{Vol}}_{G}}\int\mathcal{D}Xe^{-S_{E}[X,\tilde{\gamma}]}\det{\mathcal{P}},
\end{eqnarray}
where ${\textrm{Vol}}_{G}$ is the volume of the symmetry group and
$\tilde{\gamma}$ is the gauge-fixed metric on the world-sheet. Now
a standard finite-dimensional result is:
\begin{eqnarray}
\det{M}=\int d{\theta}_{1}\ldots d{\theta}_{n}\int
d{\overline{\theta}}_{1}\ldots
d{\overline{\theta}}_{n}e^{-{\overline{\theta}}_{i}M_{ij}{\theta}_{j}},
\end{eqnarray}
for ${\theta}_{i}$ complex Grassmann variables.
The corresponding infinite-dimensional generalization is:
\begin{eqnarray}
\det{\mathcal{P}}=\int\mathcal{D}b\int\mathcal{D}c\,e^{-\int
b\mathcal{P}c},
\end{eqnarray}
for operator $\mathcal{P}$. In our case, $\mathcal{P}$ is the
operator:
\begin{eqnarray}
{\mathcal{P}}_{\alpha\beta}^{\rho}={\delta}_{\alpha}^{\rho}{\nabla}_{\beta}
+{\delta}_{\beta}^{\rho}{\nabla}_{\alpha}
-{\tilde{\gamma}}_{\alpha\beta}{\nabla}^{\rho}
\end{eqnarray}
and so we finally have the Euclidean partition function:
\begin{eqnarray}
Z_{E}={\textrm{Vol}}_{G}\int\mathcal{D}X\,\mathcal{D}b\,\,\mathcal{D}c\,\exp\left\{{-\frac{T}{2}\int
d^{2}\xi\sqrt{\det{\tilde{\gamma}}}\,b_{\alpha\beta}(2{\tilde{\gamma}}^{\beta(\alpha}{\nabla}^{\rho)}
-{\tilde{\gamma}}^{\alpha\rho}{\nabla}^{\beta})c_{\rho}}\right\},
\end{eqnarray}
where the $c^{\alpha}$ and $b^{\alpha\beta}$ are anticommuting
ghosts and antighosts\footnote{For further discussion see
\cite{GSW1}, Chapter 3.}. In fact, $b^{\alpha\beta}$ is traceless
and $b_{\alpha\beta}{\nabla}^{\alpha}c^{\beta}$ vanishes on-shell
(\emph{i.e.} when the equations of motion are used).

\subsubsection{The Polyakov Path Integral}
Following Polyakov, we now write:
\begin{eqnarray}
Z={\textrm{Vol}}_{G}\int\mathcal{D}X\,\mathcal{D}b\,\mathcal{D}c\,e^{-A_{q}}
\end{eqnarray}
where:
\begin{eqnarray}
A_{\textrm{q}}&=&A+A^{(bc)} \\
A&=&-\frac{1}{2\pi}\int
d^{2}\xi\,{\partial}^{\alpha}X^{\mu}{\partial}_{\alpha}X_{\mu} \\
A^{(bc)}&=&-\frac{i}{2\pi}\int
d^{2}\xi\sqrt{-\det{\tilde{\gamma}}}\ {\tilde{\gamma}}^{\alpha\beta}c^{\gamma}{\nabla}_{\alpha}b_{\beta\gamma}.
\end{eqnarray}
We have set $\alpha'=1/2$ in the above formul{\ae}, and unless it appears explicitly from now on
it can be assumed to take this value.

\subsubsection{(Faddeev-Popov) Ghosts}
As before (see eq. (\ref{enmom})), we can define an energy momentum tensor for
$A_{\textrm{q}}$, which yields a ghost contribution of \footnote{The convention
here is that $A_{(\a\b)}\equiv \frac{1}{2}\left(A_{\a\b}+A_{\b\a}\right)$.}
\begin{eqnarray}
\label{enmomghost}
T_{\alpha\beta}^{(bc)} \ = \ -i\left[\frac{1}{2}c^{\gamma}{\nabla}_{(\alpha}b_{\beta)\gamma}
\ + \ \left({\nabla}_{(\alpha}c^{\gamma}\right)b_{\beta)\gamma} \ - \
\frac{1}{2}\left(\frac{1}{2}c^{\gamma}{\nabla}^{\rho}b_{\rho\gamma}
+\left({\nabla}^{\rho}c^{\gamma}\right)
b_{\rho\gamma}\right)
{\tilde{\gamma}}_{\alpha\beta}\right]\ .
\end{eqnarray}
It is often easier to work in light-cone gauge, where it is easily checked that
\begin{eqnarray}
\label{abclight}
A^{(bc)} \ = \ \frac{i}{\pi}\int\,d^{2}\xi\left(c^{+}{\partial}_{-}b_{++} \ + \
c^{-}{\partial}_{+}b_{--}\right)\ ,
\end{eqnarray}
and
\begin{equation}
\label{enmomlight}
T_{\pm\pm}^{(bc)} \ = \ -i\left[\frac{1}{2}c^{\pm}{\partial}_{\pm}b_{\pm\pm} \ + \
\left({\partial}_{\pm}c^{\pm}\right)b_{\pm\pm}\right]\ ,
\end{equation}
and from which the equations of motion straightforwardly follow as:
\begin{eqnarray}
\label{ghosteom}
{\partial}_{\pm}c^{\mp} \ &=& \ 0 \ , \\
{\partial}_{\pm}b_{\mp\mp} \ &=& \ 0 \ .
\end{eqnarray}

The conjugate momenta to $b_{++}$ and $b_{--}$ are
\begin{eqnarray}
\label{ghostmom}
{\Pi}^{(b_{\pm\pm})} \ = \
\frac{\delta A^{(bc)}}{\delta({\partial}_{\tau}b_{\pm\pm})} \ = \
\frac{ic^{\pm}}{2\pi}\ ,
\end{eqnarray}
so we should impose quantum anti-commutation relations of the form:
\begin{eqnarray}
\label{ghostcom}
\left\{b_{\pm\pm}(\sigma,\,\tau),\, {\Pi}^{(b_{\pm\pm})}({\sigma}',\,\tau)\right\}
\ &=& \ i\delta(\sigma  -  {\sigma}') \\
\label{ghostcom2}
\Rightarrow
\left\{b_{\pm\pm}(\sigma,\,\tau),\, c^{\pm}({\sigma}',\,\tau)\right\}
\ &=& \ 2\pi\delta(\sigma  -  {\sigma}')\ ,
\end{eqnarray}
from which we can see that $b_{++}$ is conjugate to $c^{+}$ and $b_{--}$
is conjugate to $c^{-}$.

The equations of motion imply that
\begin{eqnarray}
\label{ghostholomorphic}
c^{\pm} \ &=& \ c^{\pm}({\xi}^{\pm}) \\
b_{\pm\pm} \ &=& \ b_{\pm\pm}({\xi}^{\pm})\ ,
\end{eqnarray}
so as before we can write down solutions for the ghosts in terms of left and right movers.
For the \textbf{open} string $(0 \leq \sigma \leq \pi)$, the boundary conditions imply that $c^{+}=c^{-}$
at the ends of the string (\emph{i.e.} left-movers are reflected into right-movers), so the mode coefficients must
also be equal:
\begin{eqnarray}
\label{cexp}
c^{\pm} \ = \ \sum_{-\infty}^{\infty}c_{n}e^{-in{\xi}^{\pm}}\ .
\end{eqnarray}
Similarly, $b_{++} = b_{--}$ at the ends, so
\begin{equation}
\label{bexp}
b_{\pm\pm} \ = \ \sum_{-\infty}^{\infty}b_{n}e^{-in{\xi}^{\pm}}\ ,
\end{equation}
where the mode anti-commutation relations are:
\begin{eqnarray}
\label{modcom}
\left\{c_{m},\, b_{n}\right\} \ &=& \ {\delta}_{m+n} \nonumber \\
\left\{c_{m},\, c_{n}\right\} \ = \ \left\{b_{m},\, b_{n}\right\} \ &=& \ 0 \ .
\end{eqnarray}
For the \textbf{closed} string $(0 \leq \sigma \leq \pi)$ on the other hand, the boundary condition is
just periodicity in $\sigma$ so that the $c_{\pm}$ and $b_{\pm\pm}$ have mode expansions:
\begin{eqnarray}
\label{closedghostexp}
c^{+} \ &=& \ \sqrt{2}\sum_{-\infty}^{\infty}\tilde{c}_{n}e^{-2in{\xi}^{+}}\ , \nonumber \\
c^{-} \ &=& \ \sqrt{2}\sum_{-\infty}^{\infty}{c}_{n}e^{-2in{\xi}^{-}}\ , \nonumber \\
b_{++} \ &=& \ \sqrt{2}\sum_{-\infty}^{\infty}\tilde{b}_{n}e^{-2in{\xi}^{+}}\ , \nonumber \\
b_{--} \ &=& \ \sqrt{2}\sum_{-\infty}^{\infty}{b}_{n}e^{-2in{\xi}^{-}}\ ,
\end{eqnarray}
where similar mode commutation relations to (\ref{modcom}) apply and the extra factors of $\sqrt{2}$
ensure the normalization of (\ref{ghostcom2}).

Using these, we can again extract the ghost contribution to the Fourier modes of the
world-sheet energy-momentum tensor (see (\ref{modes}) and (\ref{modes2})). For the \textbf{closed} string:
\begin{eqnarray}
\label{ghostenmom}
\tilde{L}_{m}^{(bc)} \ &=& \ \frac{1}{2\pi}\int_{0}^{\pi}d\sigma e^{2im\sigma}T_{++}\Bigg.\Bigg|_{\tau = 0}\ , \nonumber \\
{L}_{m}^{(bc)} \ &=& \ \frac{1}{2\pi}\int_{0}^{\pi}d\sigma e^{-2im\sigma}T_{--}\Bigg.\Bigg|_{\tau = 0}\ .
\end{eqnarray}
While for the \textbf{open} string:
\begin{eqnarray}
\label{ghostenmom2}
{L}_{m}^{(bc)} \ &=& \ \frac{1}{\pi}\int_{0}^{2\pi}d\sigma e^{im\sigma}T_{++}\Bigg.\Bigg|_{\tau = 0}\ ,
\end{eqnarray}
which yields
\begin{eqnarray}
{L}_{m}^{(bc)} = \sum_{n=-\infty}^{\infty}(m-n):b_{m+n}c_{-n}:\ ,
\end{eqnarray}
where we have indicated that it must be normal ordered in the quantum theory.
Again, as in (\ref{modes2}), (\ref{ghostenmom2}) is regarded as a formal operation.

The algebra of the $L_{m}^{(bc)}$'s is then\footnote{See Appendix \ref{VirasoroAppendix} for further details.}:
\begin{eqnarray}
\label{Lb}
\left[L_{m}^{(bc)},\, b_{n}\right] \ &=& \ (m-n)b_{m+n}\ , \\
\label{Lc}
\left[L_{m}^{(bc)},\, c_{n}\right] \ &=& \ -(2m+n)c_{m+n}\ , \\
\label{LL}
\left[L_{m}^{(bc)},\, L_{n}^{(bc)}\right] \ &=& \
(m-n)L_{m+n}^{(bc)} \ + \ \frac{1}{6}(m-13m^3){\delta}_{m+n}\ ,
\end{eqnarray}
and we can therefore define the complete Virasoro generators by:
\begin{equation}
\label{Ltot}
L_{m}^{\textrm{(tot)}} \ = \ L_{m}^{(\alpha)} \ + \ L_{m}^{(bc)} \ - \ a{\delta}_{m}\ ,
\end{equation}
where we have shifted our earlier definition of $L_{0}$ so that the zeroth constraint is
$L_{0}^{\textrm{(tot)}}=0$. The complete Virasoro algebra is thus:
\begin{equation}
\label{vircomplete}
\left[L_{m}^{\textrm{(tot)}},\, L_{n}^{\textrm{(tot)}}\right] \ = \
(m-n)L_{m+n}^{\textrm{(tot)}} \ + \
\left[\frac{d}{12}(m^3-m) + \frac{1}{6}(m-13m^3) + 2am\right]{\delta}_{m+n}\ ,
\end{equation}
and one can see that in $d=26$ with $a=1$ there is no anomaly and we have:
\begin{equation}
\label{anomfree}
\left[L_{m}^{\textrm{(tot)}},\, L_{n}^{\textrm{(tot)}}\right] \ = \
(m-n)L_{m+n}^{\textrm{(tot)}}\ .
\end{equation}
Only for these values is the theory really conformally-invariant.

\subsubsection{String States (again)}

The presence of the Faddeev-Popov ghosts $(b,c)$ encodes
the gauge invariance of the theory. No gauge symmetry is left in
$A_{\textrm{q}}$, but there is a new global symmetry acting
on classical fields and ghosts. The quantum action of the Bosonic
string in a conformal gauge is invariant under the global Fermionic
transformations:
\begin{eqnarray}
\label{ghostxfms}
{\delta}_{\epsilon}X^{\mu} \ &=& \ \epsilon(c^{+}{\partial}_{+}+c^{-}{\partial}_{-})X^{\mu}\ , \nonumber \\
{\delta}_{\epsilon}b_{\pm\pm} \ &=& \ 2i\epsilon T_{\pm\pm}^{\textrm{(tot)}}\ , \nonumber \\
{\delta}_{\epsilon}c^{\pm} \ &=& \ \epsilon(c^{+}{\partial}_{+}+c^{-}{\partial}_{-})c^{\pm}\ ,
\end{eqnarray}
where $\epsilon$ is a constant Grassmann parameter and $T_{\pm\pm}^{\textrm{(tot)}}\equiv T_{\pm\pm}^{(\alpha)} +
T_{\pm\pm}^{(bc)}$\footnote{Note: The last transformation of (\ref{ghostxfms}) is really
${\delta}_{\epsilon}c^{\pm} \ = \ \epsilon c^{\pm}{\partial}_{\pm}c^{\pm}$,
but can be written in the stated form due to the equations of motion (\ref{ghosteom}).}. Note also that these
transformations imply that ${\delta}_{\epsilon}T_{\pm\pm}^{\textrm{(tot)}}=0$, and that they are nilpotent
on-shell\footnote{\emph{i.e.} ${\delta}_{{\epsilon}_{1}}{\delta}_{{\epsilon}_{2}}\phi=0$ for any world-sheet
field $\phi$, \textit{if} the equations of motion are obeyed.}.

Applying the Noether procedure then gives rise to a conserved current as usual. In this case,
this is in fact the BRST current, $J_{B}^{\alpha}$ such that ${\partial}_{\alpha}J_{B}^{\alpha}=0$:
\begin{equation}
\label{brstcurrent}
J_{B\, \pm} \ = \ 2c^{\pm}\left(T_{\pm\pm}^{(\alpha)}+\frac{1}{2}T_{\pm\pm}^{(bc)}\right)\ .
\end{equation}
By integrating these over $\sigma$, we can recover the BRST charge
\begin{equation}
\label{openbrstcharge}
Q_{B} \ = \ \frac{1}{2\pi}\int_{0}^{\pi}d\sigma(J_{B\, +}+J_{B\, -})\Bigg.\Bigg|_{\tau = 0}\ ,
\end{equation}
for \textbf{open} strings, and
\begin{eqnarray}
\label{closedbrstcharge}
{\tilde{Q}}_{B} \ &=& \ \frac{1}{2\pi}\int_{0}^{2\pi}d\sigma J_{B\, +}\Bigg.\Bigg|_{\tau = 0}\ , \nonumber \\
Q_{B} \ &=& \ \frac{1}{2\pi}\int_{0}^{2\pi}d\sigma J_{B\, -}\Bigg.\Bigg|_{\tau = 0}\ ,
\end{eqnarray}
for \textbf{closed} strings $(0\leq \sigma \leq 2\pi)$. In fact these exactly correspond to
the BRST operator defined by attempting to carry out the usual BRST quantization procedure\footnote{See
\cite{GSW1}, Chapter 3 for more on these.}:
\begin{equation}
\label{BRST}
Q_{B} \ \equiv \ Q \ = \ \sum_{-\infty}^{\infty}L_{-m}^{(\alpha)}c_{m} \ - \
\frac{1}{2}\sum_{-\infty}^{\infty}(m-n):c_{-m}c_{-n}b_{m+n}: \ - \ ac_{0}\ ,
\end{equation}
and when we compare with the $L_{m}$'s derived previously, we can write this as
\begin{equation}
\label{BRST2}
Q \ = \ \sum_{-\infty}^{\infty}:\left(L_{-m}^{(\alpha)} + \frac{1}{2}L_{-m}^{(bc)} - a{\delta}_{m}\right)c_{m}:\ .
\end{equation}

Finally, the ghost-number current is defined by:
\begin{equation}
\label{numbercurrent}
j_{\pm} \ = \ c^{\pm}b_{\pm\pm}\ ,
\end{equation}
which gives charges
\begin{equation}
\label{openumbercharge}
U \ = \ \frac{1}{2\pi}\int_{0}^{\pi}d\sigma(j_{+}+j_{-})\Bigg.\Bigg|_{\tau = 0}\ ,
\end{equation}
for the \textbf{open} string, and
\begin{equation}
\label{closednumbercharge}
U_{\pm} \ = \ \frac{1}{2\pi}\int_{0}^{2\pi}d\sigma j_{\pm}\Bigg.\Bigg|_{\tau = 0}
\end{equation}
for the \textbf{closed} string $(0\leq \sigma\leq 2\pi)$. In terms of modes these are:
\begin{eqnarray}
\label{numchargemodes}
U \ &\equiv& \ U_{-} \ = \ \sum_{n}:c_{n}b_{-n}: \ = \ \frac{1}{2}(c_{0}b_{0}-b_{0}c_{0}) \ + \
\sum_{n=1}^{\infty}(c_{-n}b_{n}-b_{-n}c_{n})\ , \nonumber \\
U \ &\equiv& \ U_{+} \ = \ \sum_{n}:{\tilde{c}}_{n}{\tilde{b}}_{-n}: \ = \ \frac{1}{2}({\tilde{c}}_{0}{\tilde{b}}_{0}
-{\tilde{b}}_{0}{\tilde{c}}_{0}) \ + \ \sum_{n=1}^{\infty}({\tilde{c}}_{-n}{\tilde{b}}_{n}-{\tilde{b}}_{-n}{\tilde{c}}_{n})\ .
\end{eqnarray}
For the \textbf{open} string, $U$ is (unsurprisingly) the same as either $U_{+}$ or $U_{-}$, due
to the fact that $c_{n}\equiv{\tilde{c}}_{n}$ and $b_{n}\equiv{\tilde{b}}_{n}$ for it.

In the above we can see that $b$ and $c$ enter fairly symmetrically into our expressions,
despite the asymmetrical tensor structures: $c^{\pm}$ and $b_{\pm\pm}$. They enter symmetrically
because on a \underline{flat} world sheet the ghost Lagrangian treats $b$ and $c$ symmetrically. This is
not so on a \underline{curved} world sheet. In fact in all deeper aspects of the theory, $b$ and $c$ enter
quite differently.

We know that our space of states contains both physical ones and ghosts, and the BRST quantization
procedure attempts to find conditions to separate the two. We'll now look at some of the states
and state conditions.

There are in fact \textit{two} distinct ghost ground states\footnote{This is due to the fact that
$c_{0}$ and $b_{0}$ both commute with the Hamiltonian, so the ground state has a
degeneracy that arises from the fact that it must furnish a representation of both these
operators.}:
\begin{equation}
\label{ghostground}
|\uparrow\,\rangle \quad \& \quad |\downarrow\,\rangle\ ,
\end{equation}
such that
\begin{eqnarray}
\label{c0}
c_{0}|\uparrow\,\rangle \ &=& \ 0\\
\label{b0}
b_{0}|\downarrow\,\rangle \ &=& \ 0\\
\label{c01}
c_{0}|\downarrow\,\rangle \ &=& \ |\uparrow\,\rangle\\
\label{b01}
b_{0}|\uparrow\,\rangle \ &=& \ |\downarrow\,\rangle\ .
\end{eqnarray}
We then have
\begin{equation}
\label{bcrest}
b_{n}|\uparrow\,\rangle \ = \
c_{n}|\uparrow\,\rangle \ = \
b_{n}|\downarrow\,\rangle \ = \
c_{n}|\downarrow\,\rangle \ = \ 0
\quad \forall \quad n>0\ .
\end{equation}
These ground states have ghost numbers according to:
\begin{eqnarray}
\label{up}
U|\uparrow\,\rangle \ &=& \ \frac{1}{2}|\uparrow\,\rangle\\
\label{down}
U|\downarrow\,\rangle \ &=& \ -\frac{1}{2}|\downarrow\,\rangle\ ,
\end{eqnarray}
which can be easily seen from (\emph{e.g.}):
\begin{eqnarray}
U|\uparrow\,\rangle \ &=& \ \left(\frac{1}{2}(c_{0}b_{0}-b_{0}c_{0}) \ + \
\sum_{n=1}^{\infty}(c_{-n}b_{n}-b_{-n}c_{n})\right)|\uparrow\,\rangle \nonumber \\
 \ &=& \ \frac{1}{2}c_{0}b_{0}|\uparrow\,\rangle \nonumber \\
 \ &=& \ \frac{1}{2}c_{0}|\downarrow\,\rangle \nonumber \\
 \ &=& \ \frac{1}{2}|\uparrow\,\rangle \ . \nonumber
\end{eqnarray}
Therefore $|\uparrow\,\rangle$ has ghost number $+1/2$, and similarly $|\downarrow\,\rangle$
has ghost number $-1/2$. In fact we can see that the ghost number of any state is:
\begin{equation}
\label{anyghostno}
n^{c}-n^{b}\pm\frac{1}{2}\ ,
\end{equation}
where $n^{c}$ is the number of $c$ operators acting, $n^{b}$ is the number of $b$ operators acting,
and we have $+1/2$ if the operators are acting on a $|\uparrow\,\rangle$ state, and $-1/2$ if
the operators are acting on a $|\downarrow\,\rangle$ state. For example
\begin{equation*}
\label{anyghostnoexplicit}
U\left(b_{-1}|\downarrow\,\rangle\right) \ = \ -\frac{3}{2}b_{-1}|\downarrow\,\rangle  \ ,
\end{equation*}
and so has ghost number $-3/2$ \footnote{Note that $[U,c_{n}]=c_{n}$ and $[U,b_{n}]=-b_{n}$.}.

The basic property of the BRST operator $Q$ is that $Q^2=0$. Classically this is ensured by the
Jacobi identity:

Consider any physical system with symmetry operators $K_{i}$ that form a closed Lie algebra $G$,
\begin{equation}
\label{lie}
\left[K_{i},\, K_{j}\right] \ = \ {f_{ij}}^{k}K_{k}\ ,
\end{equation}
with ${f_{ij}}^{k}$ being the structure constants of $G$. BRST quantization involves the introduction
of `antighosts' $b_{i}$, which transform in the adjoint representation of $G$, and `ghosts' $c^{i}$,
which transform in the dual of the adjoint of $G$. They obey the canonical anticommutation relations
\begin{equation}
\label{classghostanticomm}
\left\{c^{i},\, b_{j}\right\} \ = \ {{\delta}^{i}}_{j}\ .
\end{equation}
The ghost number is then defined as
\begin{equation}
\label{classghostnumber}
U \ = \ \sum_{i}c^{i}b_{i}\ ,
\end{equation}
and the BRST operator as
\begin{equation}
\label{classbrst}
Q \ = \ c^{i}K_{i} \ - \ \frac{1}{2}{f_{ij}}^{k}c^{i}c^{j}b_{k}\ .
\end{equation}
$Q^{2}=0$ is then ensured by the Jacobi identity
\begin{equation}
\label{jacobi}
{f_{ij}}^{m}{f_{mk}}^{l} \ + \
{f_{jk}}^{m}{f_{mi}}^{l} \ + \
{f_{ki}}^{m}{f_{mj}}^{l} \ = \ 0\ .
\end{equation}
At the quantum level we can use our definitions of $Q$ to write
\begin{equation}
\label{Q2quantum}
Q^{2} \ = \ \frac{1}{2}\left\{Q,\, Q\right\} \ = \ \frac{1}{2}\sum_{-\infty}^{\infty}
\left([L_{m}^{\textrm{tot}},\, L_{n}^{\textrm{tot}}] \ - \ (m-n)L_{m+n}^{\textrm{tot}}\right)
c_{-m}c_{-n}\ .
\end{equation}
So, for $d=26$ and $a=1$ (where the total Virasoro algebra is anomaly-free), $Q^{2}=0$. Yet another
demonstration of these magic numbers\footnote{For more on this see \cite{GSW1}, Chapter 3.}.

We thus hope that the physical states (states that arise in the sector with no ghost
excitations) can be characterized as BRST cohomology classes of some definite ghost number. Since we
expect that physical states need not contain ghost excitations, it should be possible (after a
possible transformation $\psi \rightarrow \psi + Q\lambda$) to put a physical state $\psi$ into a
form in which the ghost wave function is proportional to one of the two ground states $|\uparrow\rangle$
and $|\downarrow\rangle$. Consequently the possible choices for the ghost number of a physical state are
$\pm 1/2$. The correct choice turns out to be that physical states have ghost number $-1/2$.

Let $|\chi\rangle$ be a state that is annihilated by the ghost and antighost annihilation operators:
\begin{equation}
\label{chiconditions}
c_{n}|\chi\rangle \ = \ b_{n}|\chi\rangle \ = \ 0 \quad \forall \quad n>0\ .
\end{equation}
Furthermore, suppose that $|\chi\rangle$ has ghost number $-1/2$ so that it is
annihilated by $b_{0}$. Now
\begin{equation}
\label{qchi}
Q|\chi\rangle \ = \ \left[\sum_{n}\left(c_{n}L_{-n}^{(\alpha)} \ + \
\frac{1}{2}:c_{n}L_{-n}^{(bc)}:\right) \ - \ c_{0}\right]|\chi\rangle\ .
\end{equation}
$c_{n}$ and $L_{m}^{\alpha}$ commute, so after some manipulation:
\begin{equation}
\label{qchi2}
Q|\chi\rangle \ = \ \left[c_{0}\left(L_{0}^{(\alpha)}-1\right) \ + \ \sum_{n>0}c_{-n}L_{n}^{(\alpha)}\right]
|\chi\rangle\ .
\end{equation}
Thus the single condition $Q|\chi\rangle = 0$ reproduces all the physical conditions from before.
If we had chosen $|\chi\rangle$ to have ghost number $+1/2$, it would be annihilated by $c_{0}$
and we would not get all the physical state conditions. For the \textbf{closed} string, we also
need a ${\tilde{Q}}|\chi\rangle=0$ condition, where $\tilde{Q}$ is the analogous BRST operator
with $\tilde{c}$'s and $\tilde{L}$'s.

As a statement, we can say that \textit{physical states in Bosonic string theory are BRST
cohomology classes of $Q$ (and $\tilde{Q}$) with ghost number $-1/2$}:

\begin{equation}
\label{physicalstate}
|\chi\rangle \ = \ |\phi{\rangle}_{\alpha}\otimes|\downarrow{\rangle}_{bc}\ .
\end{equation}

Following this we could clearly always change any physical state into another one by transforming
\begin{equation}
\label{cohomxfm}
|\chi\rangle \rightarrow |\chi'\rangle \ = \ |\chi\rangle + Q|\chi\rangle\ ,
\end{equation}
and as $Q^{2}=0$, this satisfies the physical state conditions if $|\chi\rangle$ does. However,
if $|\chi\rangle$ satisfies the physical state conditions \textbf{and} can be written as
$|\chi\rangle=Q|\lambda\rangle$ for some $|\lambda\rangle$, then this requires that:
\begin{equation}
\label{chispurious}
|\chi\rangle \ = \ \sum_{n>0}L_{-n}^{(\alpha)}|{\lambda}_{n}\rangle
\end{equation}
for some $|{\lambda}_{n}\rangle$. This in turn implies that $|\chi\rangle$ is a null state, since
\begin{eqnarray}
\langle\chi|\chi\rangle \ &=& \ \Big\langle\sum_{n>0}L_{n}{\lambda}_{n}|\chi\Big\rangle \nonumber \\
\label{chinorm}
\ &=& \ \sum_{n>0}\langle{\lambda}_{n}|L_{n}|\chi\rangle \ = \ 0\ .
\end{eqnarray}
In fact these states are precisely the physical spurious states that we discussed before. Such
states are `pure gauge' and are trivial as a cohomology class of $Q$.

\section{Strings in Background Fields}

To describe strings in background fields we must include all the massless states of the closed string (not just the
graviton) as part of the background. The relevant closed-string fields are the antisymmetric tensor $B_{\mu\nu}(X)$,
the dilaton $\Phi(X)$, as well as the graviton $g_{\mu\nu}(X)$. We can have:

\begin{eqnarray}
\label{s-background}
S \ = \ S_1 \ + \ S_2 \ + \ S_3 \ ,
\end{eqnarray}
where
\begin{eqnarray}
\label{s1}
S_1 \ &=& \ -\frac{T}{2}\int\!d^{2}\xi \sqrt{-\det{\gamma}}\ {\gamma}^{\alpha\beta}
{\partial}_{\alpha}X^{\mu}{\partial}_{\beta}X^{\nu}g_{\mu\nu}\ , \\
\label{s2}
S_2 \ &=& \ -\frac{T}{2}\int\!d^{2}\xi {\epsilon}^{\alpha\beta}
{\partial}_{\alpha}X^{\mu}{\partial}_{\beta}X^{\nu}B_{\mu\nu}\ , \\
\label{s3}
S_3 \ &=& \ \frac{T}{2}\int\!d^{2}\xi \sqrt{-\det{\gamma}}\ {\alpha}'\Phi R^{(2)}\ .
\end{eqnarray}

$S_1$ incorporates the effects of 26-dimensional gravity, $S_2$ makes use of the world-sheet antisymmetric
tensor density ${\epsilon}^{\alpha\beta}$ and gives a way to incorporate the effects of the antisymmetric tensor field
$B_{\mu\nu}$, while $S_3$ is the correct way to include the 26-dimensional dilaton field into the $\sigma$-model. $S_3$
must be of this form (as opposed to the form of the Einstein-Hilbert action in 2-dimensions, say, $\chi = \frac{1}{4\pi}
\int\!d^2\xi\sqrt{-\det{\gamma}}\ R^{(2)}$) to make the theory more generally renormalizable
\footnote{See also footnote 7.}.

It is worth noting that classically, $S_1$ and $S_2$ are (manifestly) Weyl-invariant, whereas $S_3$ is not.
The Weyl-invariance of the first two is in general violated by quantum effects. In fact, under a Weyl
transformation ($\gamma_{\alpha\beta}\rightarrow e^{\rho}\gamma_{\alpha\beta}$), $R^{(2)}\rightarrow R^{(2)}\!-\Box\rho$.

When we quantize this sort of a theory, our main issue is with Weyl-invariance. Depending on the form of
$g_{\mu\nu}$, Weyl-invariance breaks down because there is no way to regularize the theory while preserving
conformal invariance. However, Weyl-invariance implies global scale invariance, which in turn implies vanishing
of the beta function and thus ultraviolet finiteness. The two calculations -- finiteness and Weyl-invariance -- are
thus essentially equivalent. The demanding of Weyl-invariance on a curved world-sheet necessarily implies the
vanishing of the renormalization group $\beta$-function, and hence finiteness.

We shall choose the gauge ${\gamma}_{\alpha\beta}=e^{2\phi}{\eta}_{\alpha\beta}$, and then use dimensional
regularization in $(2+\varepsilon)$-dimensions to calculate the possible breakdown of Weyl-invariance. $S_{1}$
becomes:
\begin{eqnarray}
\label{s1tilde}
S_1 \rightarrow\ {\tilde{S}}_1  &=&  -\frac{T}{2}\int\!d^{2+\varepsilon}\xi\ e^{\varepsilon\phi}
{\partial}_{\alpha}X^{\mu}{\partial}^{\alpha}\!X^{\nu}g_{\mu\nu}\ ,
\end{eqnarray}
and we wish to investigate whether or not the $\phi$ dependence goes to zero as $\varepsilon\rightarrow 0$.

We choose local inertial coordinates (of the background) at a point, and include the higher order terms in the
metric:
\begin{eqnarray}
\label{gcorrected}
g_{\mu\nu}(X)\ = \ {\eta}_{\mu\nu}-\frac{\alpha'}{3}R_{\mu\lambda\nu\rho}(X_{0})X^{\lambda}X^{\rho}+
\mathcal{O}(X^3)\ .
\end{eqnarray}
Then
\begin{eqnarray}
\label{s1tildeagain}
{\tilde{S}}_1\ = \ -\frac{T}{2}\int\! d^{2+\varepsilon}\xi\left(1+\varepsilon\phi+\mathcal{O}({\phi}^2)\right)
\left({\partial}_{\alpha}X^{\mu}{\partial}^{\alpha}\! X^{\nu}\right)\\
\left({\eta}_{\mu\nu}-\frac{\alpha'}{3}R_{\mu\lambda\nu\rho}(X_0)X^{\lambda}X^{\rho}+\mathcal{O}(X^3)\right)\ .
\end{eqnarray}
The kinetic terms ${\partial}_{\alpha}X^{\mu}{\partial}^{\alpha}\! X^{\nu}$ are associated with the propagator
and give divergences at one-loop:
\begin{eqnarray}
\label{prop}
\langle X^{\mu}X^{\nu}\rangle \sim \frac{{\eta}^{\mu\nu}}{\varepsilon}\ .
\end{eqnarray}
There are also terms of the form:
\begin{eqnarray}
\label{alphaprime}
\frac{\alpha'}{3}R_{\mu\lambda\nu\rho}X^{\lambda}X^{\rho}{\partial}_{\alpha}X^{\mu}{\partial}^{\alpha}\! X^{\nu}
(1+\varepsilon\phi)\ ,
\end{eqnarray}
which when contracted over $\langle X^{\lambda}X^{\rho}\rangle$ give
\begin{eqnarray}
\label{contraction}
\frac{\alpha'}{3}R_{\mu\lambda\nu\rho}\frac{{\eta}^{\lambda\rho}}{\varepsilon}
{\partial}_{\alpha}X^{\mu}{\partial}^{\alpha}\! X^{\nu}(1+\varepsilon\phi)
=\frac{\alpha'}{3\varepsilon}R_{\mu\nu}{\partial}_{\alpha}X^{\mu}{\partial}^{\alpha}\! X^{\nu}
(1+\varepsilon\phi)\ .
\end{eqnarray}
Clearly one of the terms here (that with $\varepsilon/\varepsilon$) will be finite, but
it depends on the conformal factor $\phi$. If we wish to ensure Weyl-invariance we will have
to require that:
\begin{eqnarray}
\label{Ricci}
R_{\mu\nu}(X)&=& 0 \\
\Rightarrow R_{\mu\nu}-\frac{1}{2}g_{\mu\nu}R \equiv G_{\mu\nu} &=& 0\ .
\end{eqnarray}
We can see that Einstein's field equations of GR must hold for the quantum theory to be Weyl-invariant
(and hence make sense).

If we now consider $S_1$, $S_2$ and $S_3$ together, the conditions for Weyl-invariance to hold
in two dimensions in the lowest non-trivial approximation in ${\alpha}'$ turn out to be:
\begin{eqnarray}
\label{weylinv1}
0 &=& R_{\mu\nu}+\frac{1}{4}{H_{\mu}}^{\lambda\rho}H_{\nu\lambda\rho}-2D_{\mu}D_{\nu}\Phi+\mathcal{O}(\alpha') \\
\label{weylinv2}
0 &=& D_{\lambda}{H^{\lambda}}_{\mu\nu}-2\left(D_{\lambda}\Phi\right){H^{\lambda}}_{\mu\nu}+\mathcal{O}(\alpha') \\
\label{weylinv3}
0 &=& 4D_{\mu}\Phi D^{\mu}\Phi-4D_{\mu}D^{\mu}\Phi+R+\frac{1}{12}H_{\mu\nu\rho}H^{\mu\nu\rho}+\frac{d-26}{3\alpha'}
+\mathcal{O}(\alpha')\ ,
\end{eqnarray}
where
\begin{eqnarray}
\label{Hfield}
H_{\mu\nu\rho}\ = \ {\partial}_{\mu}B_{\nu\rho}+{\partial}_{\rho}B_{\mu\nu}+{\partial}_{\nu}B_{\rho\mu}\ ,
\end{eqnarray}
and $D_{\mu}$ is the space-time covariant derivative. It is worth noting that these conditions are
equivalent to the vanishing of the $\beta$-functions for the 3 actions, $S_1$, $S_2$ and $S_3$. (\ref{weylinv1}) is equivalent to
${\beta}_{g_{\mu\nu}}=0$. (\ref{weylinv2}) is equivalent to ${\beta}_{{B_{\mu\nu}}}=0$, and (\ref{weylinv3})
is equivalent to ${\beta}_{\Phi}=0$.

These constraints must all make sense, and in fact it can be seen that in 26-dimensions they are the
Euler-Lagrange equations coming from the 26-dimensional action:
\begin{eqnarray}
\label{sugra}
S_{26}\ = \ -\frac{1}{2{\kappa}^2}\int\!d^{26}X\sqrt{-\det{g}}\ e^{-2\Phi}
\left\{R-4D_{\mu}\Phi D^{\mu}\Phi+\frac{1}{12}H_{\mu\nu\rho}H^{\mu\nu\rho}\right\}\ ,
\end{eqnarray}
where $\kappa$ is the gravitational coupling constant with $2{\kappa}^2=(\alpha')^{12}$ and $G_N\sim g_{s}^2
{\kappa}^2$ is Newton's constant \footnote{Note that $g_{s}\equiv e^{{\Phi}}$ is the string coupling
constant.}.

\section{Interactions (Scattering) in String Theory}

When dealing with string interactions, the usual Feynman diagrams of quantum field theory are easily
generalized by thickening their lines. In the case of open strings this means that the lines become strips. For
closed strings they become cylinders. In order to deal with these things one usually continues to a
Euclidean metric, where the stringy Feynman diagrams can be classified by their topology. The order of
a particular diagram in the perturbation expansion is determined by the number of handles (as well as the number
of windows in the case of theories with open strings) and the number of external lines. The leading terms for a
given process correspond to the tree-level approximation. Once a tree-level approximation (corresponding
to a consistent classical field theory) is known, the full quantum theory is in principle determined
by unitarity.

The amplitude for the scattering of $n$ on-mass-shell closed-string states is described in the tree-level
approximation by a world-sheet that is topologically a sphere with the $n$ external particles attached at
$n$ specific points $z_{i}$ on the surface. In principle, one integrates over all geometries at this topology
and all values of the points $z_{i}$, up to conformal equivalence.

In the case of the open-string, the world-sheet has boundaries, and emitted open-string states are attached to
boundaries. The cyclic ordering among sets of particles attached to common boundaries is meaningful. The
tree-level approximation corresponds to the disk diagram, which has one boundary. In this case the cyclic ordering
of all the open-string states must be specified. Closed-string states can be attached to the interior of the
surface at the same time to describe a tree amplitude for a mixed process.

In order to describe scattering, we need two essential ingredients: propagators and vertices. For an ordinary
Bosonic scalar field, $\phi$, with mass $m$, which obeys the Klein-Gordon equation, $(\Box+m^2)\phi=0$, the standard
Feynman propagator is just the inverse of the Klein-Gordon operator. For free open strings, the closest analogue
of the Klein-Gordon equation is the mass-shell condition $(L_0-1)|\phi\rangle=0$, which can be regarded as an
infinite-component generalization of the Klein-Gordon equation. A plausible guess for a propagator would therefore
be \cite{GSW1}:
\begin{eqnarray}
\label{propagator}
\Delta \ = \ \frac{1}{L_0-1} \ = \ \int_{0}^{1}\!dz\ z^{L_0-2}\ .
\end{eqnarray}
We also need an interaction vertex (or `vertex operator' in the language of conformal field theory) for emission or absorption of an external state
by an internal line. For example, some Tachyon vertex operators are:
\begin{eqnarray}
\label{tachyonclosed}
V_0(\xi,k) \ = \ e^{ik\cdot X(\xi)}\ ,
\end{eqnarray}
for the closed string, or
\begin{eqnarray}
\label{tachyonopen}
V_0(y,k) \ = \ e^{ik\cdot X(y)}\ ,
\end{eqnarray}
for the open string. (\ref{tachyonclosed}) will have $k^2=-M^2=4/\alpha'$, while (\ref{tachyonopen}) will
have $k^2=-M^2=1/\alpha'$. For an open-string vector state we will have:
\begin{eqnarray}
\label{vectoropen}
V_{{\zeta}_{\mu}}(y,k) \ = \ {\zeta}_{\mu}\frac{dX^{\mu}}{dy}\ e^{ik\cdot X(y)}\ ,
\end{eqnarray}
while for a closed-string tensor state we will have:
\begin{eqnarray}
\label{tensorclosed}
V_{{\zeta}_{\mu\nu}}(\xi,k) \ = \ {\zeta}_{\mu\nu}{\partial}_{\alpha}X^{\mu}{\partial}^{\alpha}X^{\nu}
e^{ik\cdot X(\xi)}\ .
\end{eqnarray}

To describe scattering we insert a `vertex operator' at each point which will encode the momentum and
species of particle. At tree-level the scattering amplitude reduces to:
\begin{eqnarray}
\label{openscattering}
A_{n} \ \sim \ g_{s}^{n-2}\int\prod_{i=1}^{n}dy_{i}\left\langle
V_{{\Lambda}_{1}}(y_1,k_1)V_{{\Lambda}_{2}}(y_2,k_2)\ldots V_{{\Lambda}_{n}}(y_n,k_n)\right\rangle\ ,
\end{eqnarray}
for scattering of $n$ open string states where the ${\Lambda}_{i}$ denote the type of state (\emph{i.e.}
${\Lambda}_{i}=0$ for a Tachyon, ${\Lambda}_{i}={\zeta}_{\mu}$ for a vector, ${\Lambda}_{i}=
{\zeta}_{\mu\nu}$ for a Tensor etc). For scattering of $n$ closed-string states at tree-level the
expression is:
\begin{eqnarray}
\label{closedscattering}
A_{n} \ \sim \ g_{s}^{n-2}\int\prod_{i=1}^{n}d^2{\xi}_{i}\left\langle
V_{{\Lambda}_{1}}(\xi_1,k_1)V_{{\Lambda}_{2}}(\xi_2,k_2)\ldots V_{{\Lambda}_{n}}(\xi_n,k_n)\right\rangle\ .
\end{eqnarray}
Note here that in Euclidean space $\langle A\rangle = \int \mathcal{D}XA\ e^{-S}$, where $S$ is the
Euclidean action.

\subsection{Tree-Level Tachyon Scattering}
As an example of the above discussion, let us consider the tree-level scattering of 4 Tachyons (which, we
should note is easily generalized to the $n$-particle case). At tree-level, in closed-string theory, the
scattering of $n$ Tachyons is given by:
\begin{eqnarray}
\label{closedntachyons}
A_{n}(k_{1},\ldots,k_{n})\ = \ \frac{1}{{\textrm{Vol}}_{SL(2,\mathbb{C})}}g_{s}^{n-2}
\int\prod_{i=1}^{n}d^2{\xi}_{i}\left\langle
V_{0}(\xi_1,k_1)V_{0}(\xi_2,k_2)\ldots V_{0}(\xi_n,k_n)\right\rangle\ ,
\end{eqnarray}
which trivially reduces to:
\begin{eqnarray}
\label{closed4tachyons}
A_{4}\ = \ \frac{1}{{\textrm{Vol}}_{SL(2,\mathbb{C})}}g_{s}^{2}
\int\prod_{i=1}^{4}d^2{\xi}_{i}\left\langle
V_{0}(\xi_1,k_1)V_{0}(\xi_2,k_2)V_0(\xi_3,k_3)V_{0}(\xi_4,k_4)\right\rangle\ ,
\end{eqnarray}
for 4 particles. Now, recall that
\begin{eqnarray}
\label{vertexop}
V_0(\xi,k)\ = \ e^{ik_{\mu}X^{\mu}(\xi)}
\end{eqnarray}
and
\begin{eqnarray}
\label{action}
S\ = \ \frac{1}{4\pi\alpha'}\int\!d^2\xi\ {\partial}_{\alpha}X^\mu{\partial}^{\alpha}X_{\mu}\ ,
\end{eqnarray}
so
\begin{eqnarray}
\label{expectation}
\left\langle V_{0}(1)V_{0}(2)V_0(3)V_{0}(4)\right\rangle =
\int\!\mathcal{D}X\ e^{-\frac{1}{4\pi\alpha'}\int\!d^2\xi\ {\partial}_{\alpha}X^\mu{\partial}^{\alpha}X_{\mu}}
\ e^{i\sum_{j=1}^{4}k_{j}^{\mu}X_{\mu}(\xi_{j})}\ ,
\end{eqnarray}
and we are ignoring the ghost action for the moment. We can write:
\begin{eqnarray}
\label{expvertops}
\sum_{j=1}^{4}k_{j}^{\mu}X_{\mu}(\xi_{j})\ &=& \ \int\!d^2\xi
\sum_{j=1}^{4}k_{j}^{\mu}X_{\mu}(\xi){\delta}^2(\xi-\xi_j)\nonumber \\
\label{expvertops2}
&=& \ \int\!d^2\xi J^{\mu}(\xi)X_{\mu}(\xi)\ ,
\end{eqnarray}
where
\begin{eqnarray}
\label{J}
J^{\mu}(\xi)\ = \ \sum_{j=1}^{4}k_{j}^{\mu}\, {\delta}^2(\xi-\xi_j)\ .
\end{eqnarray}
\begin{eqnarray}
\label{again}
\therefore\ \left\langle V_{0}(1)V_{0}(2)V_0(3)V_{0}(4)\right\rangle \ = \
\int\!\mathcal{D}X\ e^{\frac{1}{4\pi\alpha'}\int\!d^2\xi X^{\mu}{\partial}^{2}X_{\mu}
+i\int\!d^2\xi J^{\mu}X_{\mu}}\ .
\end{eqnarray}

Now, a standard finite-dimensional result is:
\begin{eqnarray}
\label{Bosonicgenexp}
\int\! du_1\ldots du_n\ e^{-\frac{1}{2}u^{\textrm{T}}Au\ +\ b^{\textrm{T}}u} \ = \
\frac{(2\pi)^{n/2}}{\sqrt{\det{A}}}\ e^{\frac{1}{2}b^{\textrm{T}}A^{-1}b}\ .
\end{eqnarray}
So, by taking $u\rightarrow X$, $A\rightarrow -{\partial}^2/2\pi\alpha'$, $A^{-1}\rightarrow -2\pi\alpha'/
{\partial}^2$ and $b\rightarrow iJ$, we can generalize this to our infinite-dimensional case, and write:
\begin{eqnarray}
\label{dopathint}
\left\langle V_{0}(1)V_{0}(2)V_0(3)V_{0}(4)\right\rangle \ \sim \
\frac{1}{\sqrt{\det{(-{\partial}^2/2\pi\alpha')}}}\ e^{\pi\alpha'\int\! d^2\xi d^2\xi' J^{\mu}\frac{1}{{\partial}^2}\, J_{\mu}}\ .
\end{eqnarray}
Here, $1/{\partial}^2$ is the propagator $G$ satisfying ${\partial}^2G={\delta}^{2}(\xi_i-\xi_j)$, \emph{i.e.}
\begin{eqnarray}
\label{2dimprop}
G\ = \ \frac{1}{2\pi}\log{\left|\xi_i-\xi_j\right|}\ .
\end{eqnarray}
It is interesting to note that we can arrive at the same conclusion by simply re-writing the integrand of the exponential in
(\ref{again}) as
\begin{eqnarray}
\label{rewrite}
\frac{1}{4\pi\alpha'}(X^{\mu}+i2\pi\alpha'J^{\mu}\frac{1}{{\partial}^2}){\partial}^{2}
(X_{\mu}+i2\pi\alpha'\frac{1}{{\partial}^2}J_{\mu})\ +\ \pi\alpha'J^{\mu}\frac{1}{{\partial}^2}\, J_{\mu}\ .
\end{eqnarray}
The second term is independent of $X$ so we can take it outside the $X$ integral. The first term we identify as being
$\sim 1/\sqrt{\det{(-{\partial}^2)}}$ (once the $X$ integral has been done). We subsume the first term into the
normalization (as per usual), and so writing things out we have:
\begin{eqnarray}
\label{amp} A_{4}\ &=& \
g_{s}^{2}\,\mathcal{N}\!\int\prod_{i=1}^{4}d^2{\xi}_{i}\
\exp{\left\{\frac{\alpha'}{2}\int\! d^2\xi d^2\xi'\sum_{j,l}k_{j}\!\cdot\!
k_{l}\ {\delta}^2(\xi_j-\xi)\log{\left|\xi-\xi'\right|}
{\delta}^2(\xi'-\xi_l)\right\}}\ ,\nonumber \\
\end{eqnarray}
which we can then work out to be
\begin{eqnarray}
\label{fudge}
A_4\ &=& \ g_{s}^{2}\,\mathcal{N}\!\int\prod_{i=1}^{4}d^2{\xi}_{i}\
\exp{\left\{\frac{\alpha'}{2}\sum_{j\neq l}
k_{j}\!\cdot\! k_{l}\log{\left|\xi_j-\xi_l\right|}\right\}}\nonumber\\
\ &=& \ g_{s}^{2}\,\mathcal{N}\!\int\prod_{i=1}^{4}d^2{\xi}_{i}\
\exp{\left\{\log\left(\prod_{j\neq l}\left|\xi_j
-\xi_l\right|^{\frac{\alpha'k_{j}\cdot k_{l}}{2}}\right)\right\}}\nonumber\\
\ &=& \ g_{s}^{2}\,\mathcal{N}\!\int\prod_{i=1}^{4}d^2{\xi}_{i}\
\prod_{j\neq l}\left|\xi_j
-\xi_l\right|^{\frac{\alpha'k_{j}\cdot k_{l}}{2}}\ \nonumber \\
\ &=& g_{s}^{2}\,\mathcal{N}\!\int\prod_{i=1}^{4}d^2{\xi}_{i}\
\prod_{j < l}\left|\xi_j -\xi_l\right|^{\alpha'k_{j}\cdot k_{l}}\ .
\end{eqnarray}

If we now regard the ${\xi}_{i}$ as complex variables (\emph{i.e.} $\xi_i\rightarrow z_i$, with
$z_i=\tau_i-i\sigma_i$ and $\bar{z}_i=\tau_i+i\sigma_i$)\footnote{Notice that the Jacobian for this
transformation is simply $1/2i$, which can be subsumed into the normalization $\mathcal{N}$.},
then it turns out that the amplitude is invariant under $SL(2,\mathbb{C})$ transformations:
\begin{eqnarray}
\label{sl2c}
z_{i}\rightarrow \frac{az_{i}+b}{cz_i+d}\ ,
\end{eqnarray}
where $a,b,c,d$ are complex and satisfy $ad-bc=1$ \footnote{Note that this symmetry has already been hinted
at in (\ref{closedntachyons}), by dividing out by the volume of this group, and as is suggested here the
symmetry is not special to the 4-point amplitude}. For example with $a=d=1,b=c=0$
we have the identity transformation, under which (\ref{fudge}) is trivially invariant. With $a=d=0,b=-1,c=1$ on the
other hand we have $z_{i}\rightarrow -1/z_{i}$. In this case, (\ref{fudge}) becomes
\begin{eqnarray}
\label{inversion}
g_{s}^{2}\,\mathcal{N}\!\int\prod_{i=1}^{4}d^2{z}_{i}\
|\tilde{z}_1\tilde{z}_2\tilde{z}_3\tilde{z}_4|^4\ \prod_{j <
l}\left|\tilde{z}_j-\tilde{z}_l\right|^{\alpha'k_{j}\cdot k_{l}}\
,
\end{eqnarray}
where we have used momentum conservation
$\left(\sum_{i}k_i^{\mu}=0\right)$ and the on-shell condition for
closed-string tachyons $(\alpha' k_i^2=4)$. Noting that the
Jacobian for transforming the measure is
$\prod_{i=1}^{4}\left(\tilde{z}_{i}\bar{\tilde{z}}_i\right)^{-2}=\prod_{i=1}^{4}|\tilde{z}_i|^{-4}$
we can again see that the amplitude is invariant. Moreover as
there are 4 unknowns in these transformations (\ref{sl2c}) and one
constraint, it means we can fix 3 of the 4 $z_i$, and thus set
$z_1=\infty,z_2=0,z_3=z,z_4=1$ to get:
\begin{eqnarray}
\label{fixed} g_{s}^{2}\,\mathcal{N}\!\int d^2z\
|z_1|^{-4}\left|1-\frac{1}{z_1}\right|^{\alpha'k_1\cdot
k_2}&|z|^{\alpha'k_3\cdot k_4}&
\left|1-\frac{z}{z_1}\right|^{\alpha'k_1\cdot k_3}|1-z|^{\alpha'k_2\cdot k_3}\nonumber \\
\sim g_{s}^{2}\tilde{\mathcal{N}}\!\int d^2z &|z|^{\alpha'k_3\cdot
k_4}&|1-z|^{\alpha'k_2\cdot k_3}\ .
\end{eqnarray}
Now, using the identity
\begin{eqnarray}
\label{gammastuff}
\int d^2z|z|^{-A}|1-z|^{-B} \ = \ B(1-\frac{A}{2},1-\frac{B}{2},\frac{A+B}{2}-1)\ ,
\end{eqnarray}
where
\begin{eqnarray}
\label{gammastuff2}
B(a,b,c)=\pi\frac{\Gamma(a)\Gamma(b)\Gamma(c)}{\Gamma(a+b)\Gamma(b+c)\Gamma(c+a)}\ ,
\end{eqnarray}
and the relations
\begin{eqnarray}
\label{invariants}
k_{i}^2&=& \frac{4}{\alpha'}\ ,\nonumber\\
s &=& -(k_1+k_2)^2\ ,\nonumber\\
t &=& -(k_1+k_4)^2\ ,\nonumber\\
u &=& -(k_1+k_3)^2\ ,\nonumber\\
s+t+u &=& -\sum_{i=1}^{4}k_i^2 \nonumber\\
&=& -\frac{16}{\alpha'}\ ,
\end{eqnarray}
we finally have that
\begin{eqnarray}
\label{atlast}
A_4 \sim g_{s}^{2}\frac{\Gamma\left(-1-\frac{\alpha's}{4}\right)\Gamma\left(-1-\frac{\alpha't}{4}\right)
\Gamma\left(-1-\frac{\alpha'u}{4}\right)}{\Gamma\left(2+\frac{\alpha's}{4}\right)
\Gamma\left(2+\frac{\alpha't}{4}\right)\Gamma\left(2+\frac{\alpha'u}{4}\right)}\ ,
\end{eqnarray}
up to a momentum-independent factor. The analogous result for scattering of 4 open-string tachyons is
\begin{eqnarray}
\label{atlastopen}
A_4^{open}\sim g_s\frac{\Gamma(-1-\alpha's)\Gamma(-1-\alpha't)}{\Gamma(-2-\alpha's-\alpha't)}\ .
\end{eqnarray}
These formul{\ae} are closely related to the famous scattering amplitudes \cite{Veneziano} proposed by Veneziano in 1968 which 
were very significant in the initiation and development of string theory. It is interesting to note that (\ref{atlastopen}) can be 
written in terms of the Beta function $A_4^{open}\sim g_s B(-1-\alpha's,-1-\alpha't)$. More information about the Beta 
function can be found in Appendix \ref{mathappendix}.

\section{The Superstring}

Everything we have looked at so far is all very well, but the Bosonic string has a number of
shortcomings. In particular there is a distinct absence of Fermions, and since matter is mostly
made up of Fermions, this is clearly a big problem. There is also the presence of tachyons (particles
with negative mass-squared) which are physically undesirable. If we introduce Fermions, and at the same time
introduce world-sheet supersymmetry which relates the space-time coordinates $X^{\mu}(\sigma,\tau)$ to Fermionic
partners ${\psi}^{\mu}(\sigma,\tau)$ (2-component spinors), then an action principle with $\mathcal{N}\!=1$
supersymmetry\footnote{$\mathcal{N}=N$ supersymmetry means that there are $N$ supersymmetries.}
gives rise to a consistent string theory with critical dimension $d=10$ \footnote{In the following,
we will often use units in which $\alpha'=1/2$ (for $0\leq\sigma\leq \pi$). If one has $0\leq\sigma\leq 2\pi$, then
one would have $\alpha'=2$, which reflects the fact that $\sigma_{max}\sim l_s$. One can then easily re-instate factors
of $\alpha'$ by dimensional analysis.}. Unless otherwise specified, we will be considering the open
superstring in what follows.

We consider the action:
\begin{eqnarray}
\label{superstring}
S\ &=& \ A^{(X)}+A^{(\psi)}\nonumber \\
&=& -\frac{T}{2}\int\! d^2\xi\left\{{\partial}_{\alpha}X^{\mu}{\partial}^{\alpha}X_{\mu}
\ + \ {\bar{\psi}}^{\mu}{\gamma}^{\alpha}{\partial}_{\alpha}{\psi}_{\mu}\right\}\ ,
\end{eqnarray}
where the ${\psi}^{\mu}$ are 2-component spinors in 2-dimensions and space-time vectors in 10-dimensions, while
the ${\gamma}^{\alpha}$ are 2-dimensional Dirac matrices.

We shall use the following representation for the two-dimensional Dirac matrices:
\begin{eqnarray}
\label{gammas}
{\gamma}^{0}=\left(\begin{array}{cc}
0 & 1 \\
-1 & 0 \\
\end{array}\right)\quad ; \quad
{\gamma}^{1}=\left(\begin{array}{cc}
0 & 1 \\
1 & 0 \\
\end{array}\right)\quad ; \quad
{\gamma}_{3}\equiv {\gamma}^{0}{\gamma}^{1}=
\left(\begin{array}{cc}
1 & 0 \\
0 & -1 \\
\end{array}\right)\ ,
\end{eqnarray}
which obey the anticommutation relation
\begin{eqnarray}
\label{clifford}
\{{\gamma}^{\alpha},\, {\gamma}^{\beta}\}\ = \ 2{\eta}^{\alpha\beta}\ .
\end{eqnarray}
For these matrices it is often useful to note that we also have the 2-dimensional identity
\begin{eqnarray}
\label{id}
{\gamma}^{\alpha}{\gamma}^{\beta}{\gamma}_{\alpha}\ = \ 0\ .
\end{eqnarray}
In the representation of (\ref{gammas}) it also makes sense to demand that the components of the
world-sheet spinor ${\psi}^{\mu}$ should be real. Such a 2-component real spinor is known as a
Majorana spinor. For Majorana spinors we have that
\begin{eqnarray}
\label{barpsi}
\bar{\psi}\equiv {\psi}^{\dag}{\gamma}^{0}\ = \ {\psi}^{T}{\gamma}^{0}\ ,
\end{eqnarray}
since $\psi$ is real. So:
\begin{eqnarray}
\label{barid}
\bar{\chi}\psi\ = \ \bar{\psi}\chi \ ,
\end{eqnarray}
using antisymmetry of ${\gamma}^{0}$ and anticommutation of the Majorana spinors. For the
representation of (\ref{gammas}) we also have the easily verifiable result:
\begin{eqnarray}
\label{bargammapsi}
\bar{\chi}{\gamma}^{\alpha}\psi \ = \ -\bar{\psi}{\gamma}^{\alpha}\chi \ .
\end{eqnarray}

\subsection{Symmetries}

In addition to the global symmetries of the Bosonic action in conformal gauge, there is also
a new global symmetry which is world-sheet supersymmetry (SUSY). The action (\ref{superstring})
is invariant under the SUSY transformations:
\begin{eqnarray}
\label{susy}
\delta X^{\mu} \ &=& \ \bar{\epsilon}{\psi}^{\mu}\ ,\nonumber \\
\delta{\psi}^{\mu} \ &=& \ {\gamma}^{\alpha}\epsilon\, {\partial}_{\alpha}X^{\mu}\ .
\end{eqnarray}
where $\epsilon$ is a real constant Grassman spinor parameter $(\bar{\epsilon}={\epsilon}^{T}
{\gamma}^{0})$. If we implement these transformations, then after some algebra we find that
\begin{eqnarray}
\label{deltasusy}
\delta S\ &=& \ -\frac{1}{2\pi\alpha'}\int\!d^2\xi\left(-{\bar{\psi}}^{\mu}\epsilon\,{\partial}^2X_{\mu}\
+ \ {\bar{\psi}}^{\mu}\epsilon\,{\partial}^2X_{\mu}\right)\nonumber \\
&=& 0\ ,
\end{eqnarray}
as required.

\subsection{Equations of Motion}

We can further derive the equations of motion for the Fermionic fields by treating $\psi$ and
$\bar{\psi}$ as independent fields as usual. We find that they obey the massless Dirac equation
in 2-dimensions:
\begin{eqnarray}
\label{masslessdirac}
\partialslash\psi\ = \ 0\ ,
\end{eqnarray}
where $\partialslash = \gamma\cdot\partial$ as usual, and the dot-product is of course a sum over
world-sheet variables.

We can further show that the above SUSY transformations (\ref{susy}) satisfy the algebra:
\begin{eqnarray}
\label{susyalgebra}
\left[{\delta}_1,{\delta}_2\right]Y^{\mu}\ = \ a^{\alpha}{\partial}_{\alpha}Y^{\mu}\ ,
\end{eqnarray}
where $a^{\alpha}=-2{\bar{\epsilon}}_1{\gamma}^{\alpha}{\bar{\epsilon}}_2$ and $Y^{\mu}$ is any of the
world-sheet fields (\emph{i.e.} $Y^{\mu}=X^{\mu}$ or $Y^{\mu}={\psi}^{\mu}$). It is well-known that the
commutator of two supersymmetries is a translation, so (\ref{susyalgebra}) confirms this and highlights
the fact that SUSY is an extension of Poincar\'{e} symmetry. However, closure of the algebra
(\emph{i.e.} for $Y^{\mu}={\psi}^{\mu}$) requires the use of the equations of motion (\ref{masslessdirac}). This
is known as on-shell SUSY.

If we now allow the SUSY parameter $\epsilon$ to vary (\emph{i.e.} assume that it is no-longer a constant),
 then we can use the Noether procedure to obtain a conserved SUSY current:
 \begin{eqnarray}
 \label{deltassusy}
 \delta S\ = \ \frac{1}{\pi\alpha'}\int\!d^2\xi({\partial}_{\alpha}\bar{\epsilon})_aJ_a^{\alpha}\ ,
 \end{eqnarray}
 where the supercurrent $J_a^{\alpha}$ takes the form
 \begin{eqnarray}
 \label{susyj}
 J^{\alpha}\ = \ \frac{1}{2}{\gamma}^{\beta}{\gamma}^{\alpha}{\psi}^{\mu}{\partial}_{\beta}X_{\mu}\ .
 \end{eqnarray}
 It satisfies conservation (using the equations of motion):
 \begin{eqnarray}
 \label{conservation}
 {\partial}_{\alpha}J^{\alpha}\ = \ 0\ ,
 \end{eqnarray}
 and `tracelessness':
 \begin{eqnarray}
 \label{traceless}
 {\gamma}^{\alpha}J_{\alpha}\ = \ 0\ .
 \end{eqnarray}

 The total energy-momentum tensor is now:
 \begin{eqnarray}
 \label{susyemtensor}
 T_{\alpha\beta}\ = \ {\partial}_{\alpha}X^{\mu}{\partial}_{\beta}X_{\mu}
 -\frac{1}{2}{\bar{\psi}}^{\mu}{\gamma}_{(\alpha}{\partial}_{\beta)}{\psi}_{\mu}
 -\frac{1}{2}\left({\partial}_{\rho}X^{\mu}{\partial}^{\rho}X_{\mu}
-\frac{1}{2}{\bar{\psi}}^{\mu}{\gamma}_{\rho}{\partial}^{\rho}{\psi}_{\mu}\right)
{\gamma}_{\alpha\beta}\ ,
 \end{eqnarray}
 and ${\gamma}_{\alpha\beta}$ is the world-sheet metric. We know that this is the conserved current
 related to changes in the world-sheet space-time. It results from varying the action with respect to the
 metric, and is thus related to the world-sheet graviton. The supersymmetry current has a similar partner -- the
 gravitino ${\chi}_a^{\alpha}$. In conformal gauge we fix ${\gamma}_{\alpha\beta}={\eta}_{\alpha\beta}$ by using
 the reparametrization and Weyl invariance. In superconformal gauge, we also fix ${\chi}_a^{\alpha}={\gamma}^{\alpha}
 {\rho}_a$, where $\rho$ is a spinor field, by using the local SUSY and super-Weyl invariance.

 \subsection{Solving the Equations of Motion}

 Let us begin by defining
 \begin{eqnarray}
 \label{psilr}
 \psi\ \equiv \ \left(\begin{array}{c}
 {\psi}_{L}\\
 {\psi}_{R}\\
 \end{array}\right)\ ,
 \end{eqnarray}
 and let us consider the massless Dirac equation $\partialslash\psi =  0$. Now in our basis:
 \begin{eqnarray}
 \label{infull}
 \partialslash\ &=& \ {\gamma}\cdot\partial\nonumber \\
 &=&\ {\gamma}^{0}{\partial}_{0}+{\gamma}^{1}{\partial}_1\nonumber \\
 &=& \ \left(\begin{array}{cc} 0 & {\partial}_{\tau}\\ -{\partial}_{\tau} & 0\\ \end{array}\right)
 +\left(\begin{array}{cc} 0 & {\partial}_{\sigma}\\ {\partial}_{\sigma} & 0\\ \end{array}\right)\nonumber \\
 &=& \ -2\left(\begin{array}{cc} 0 & -{\partial}_{+}\\ {\partial}_{-} & 0\\ \end{array}\right)\ .
 \end{eqnarray}
 So the dirac equation is
 \begin{eqnarray}
 \label{diracfull}
 \left(\begin{array}{cc} 0 & -{\partial}_{+}\\ {\partial}_{-} & 0\\ \end{array}\right)
 \left(\begin{array}{c} {\psi}_{L}\\ {\psi}_{R}\end{array}\right)\ &=& \
 \left(\begin{array}{c} -{\partial}_{+}{\psi}_{R}\\ {\partial}_{-}{\psi}_{L}\\ \end{array}\right)\nonumber \\
 &=& \ \left(\begin{array}{c} 0 \\ 0 \\ \end{array}\right)\ ,
 \end{eqnarray}
\emph{i.e.}
\begin{eqnarray}
\label{leftright}
{\psi}_{L}\ &=& \ {\psi}_{L}({\xi}^{+})\ ;\nonumber \\
{\psi}_{R}\ &=& \ {\psi}_{R}({\xi}^{-})\ ,
\end{eqnarray}
as our original notation anticipated. Note that the left-moving and right-moving components of
$\psi$ can be obtained using the chirality projectors $P^{\pm}=(1\pm{\gamma}_{3})/2$, \emph{i.e.}
\begin{eqnarray}
\label{pplus}
P^{+}\ &=& \ \left(\begin{array}{cc} 1 & 0\\ 0 & 0\\ \end{array}\right)\ ;\\
\label{pminus}
P^{-}\ &=& \ \left(\begin{array}{cc} 0 & 0\\ 0 & 1\\ \end{array}\right)\ .
\end{eqnarray}
So
\begin{eqnarray}
\label{p+comps}
P^{+}{\psi}\ &=& \ \left(\begin{array}{c} {\psi}_{L}\\ 0\\ \end{array}\right)\ ;\\
P^{-}{\psi}\ &=& \ \left(\begin{array}{c} 0 \\ {\psi}_{R}\\ \end{array}\right)\ .
\end{eqnarray}

It is also worth seeing that we can write the equations of motion in a more `light-cone'
orientated way. Define:
\begin{eqnarray}
\label{gammaplusminus}
{\gamma}^{\pm}\ = \ \frac{1}{2}\left({\gamma}^{0}\pm{\gamma}^{1}\right)\ ,
\end{eqnarray}
whose algebra is defined by ${({\gamma}^{+})}^2={({\gamma}^{-})}^{2}=0$ and
\begin{eqnarray}
\label{anticomm}
\left\{{\gamma}^+,{\gamma}^-\right\}_{ab}\ = \ \mathbb{I}_{ab}\ .
\end{eqnarray}
Then the Dirac equation implies that
\begin{eqnarray}
\label{plusr}
{\gamma}^{+}{\partial}_{+}{\psi}_{R}\ &=& \ 0\ ;\\
\label{minusr}
{\gamma}^{-}{\partial}_{-}{\psi}_{L}\ &=& \ 0\ ,
\end{eqnarray}
and thus the same dependence on ${\xi}^{\pm}$ as in (\ref{leftright}).

We will also need to consider the boundary conditions that are imposed on the Fermions in our
action. Ignoring overall factors, we can write the Dirac action as:
\begin{eqnarray}
\label{diracact}
A^{(\psi)}\ &\sim & \ \int\!d\sigma d\tau\ \bar{\psi}\partialslash\psi \nonumber \\
&\sim & \int\!d\sigma d\tau\ \left({\psi}_{L}{\partial}_{-}{\psi}_{L}+{\psi}_{R}{\partial}_{+}{\psi}_{R}\right)\nonumber \\
&\sim & \int\!d\sigma d\tau\ \big({\psi}_{L}({\partial}_{\tau}-{\partial}_{\sigma}){\psi}_{L}
+{\psi}_{R}({\partial}_{\tau}+{\partial}_{\sigma}){\psi}_{R}\big)\ .
\end{eqnarray}
Thus on varying to obtain the equations of motion, we will get a surface term of the form
\begin{eqnarray}
\label{sterm}
\int\!d\tau\left\{\psi_{R}{\delta\psi}_R-\psi_{L}{\delta\psi}_L\right\}\Big|_{\sigma}+
\int\!d\sigma\left\{\psi_{R}{\delta\psi}_R+\psi_{L}{\delta\psi}_L\right\}\Big|_{\tau}\ ,
\end{eqnarray}
and in a similar way to Section 4.3, we can see that these can be satisfied by making the choice
$\psi_R=\pm\psi_L$ and $\delta\psi_R=\pm\delta\psi_L$ at each end. The overall relative sign is a
matter of convention, so without loss of generality we set $\psi_R(0,\tau)=\psi_L(0,\tau)$. The
relative sign at the other end is now meaningful and there are two cases to consider. Consider the
\textbf{open} superstring:

\begin{enumerate}
\item \textbf{Ramond} (R) boundary conditions are:
\begin{eqnarray}
\label{r}
\psi_{R}(\pi,\tau)=\psi_L(\pi,\tau)\ ,
\end{eqnarray}
so that the mode expansions for the Fermions in the Ramond sector are
\begin{eqnarray}
\label{rmode}
\psi_R^{\mu}(\sigma,\tau)&=&\sum_{n\in \ \mathbb{Z}}\psi_{n}^{\mu}e^{-in(\tau-\sigma)}\ ;\\
\psi_L^{\mu}(\sigma,\tau)&=&\sum_{n\in \ \mathbb{Z}}\psi_{n}^{\mu}e^{-in(\tau+\sigma)}\ ,
\end{eqnarray}
where the sums run over all integers $n$.
\item \textbf{Neveu-Schwarz} (NS) boundary conditions on the other hand are:
\begin{eqnarray}
\label{ns}
\psi_{R}(\pi,\tau)=-\psi_L(\pi,\tau)\ ,
\end{eqnarray}
so that in the Neveu-Schwarz sector the mode expansions are
\begin{eqnarray}
\label{nsmode}
\psi_R^{\mu}(\sigma,\tau)&=&\sum_{r\in \ \mathbb{Z}+\frac{1}{2}}\psi_{r}^{\mu}e^{-ir(\tau-\sigma)}\ ;\\
\psi_L^{\mu}(\sigma,\tau)&=&\sum_{r\in \ \mathbb{Z}+\frac{1}{2}}\psi_{r}^{\mu}e^{-ir(\tau+\sigma)}\ ,
\end{eqnarray}
where now the sums run over the half-integers. We will often use $m$ and $n$ to represent integers in
these sums and $r$ and $s$ to represent half-integers\footnote{Note that the Ramond b.c.'s and integer
modes are appropriate to the description of string states which are space-time Fermions, whereas the
Neveu-Schwarz b.c.'s and half-integer modes give rise to space-time Bosons. These Bosonic states are,
of course, different to those of the Bosonic string.}.
\end{enumerate}

For \textbf{closed} superstrings $(0\leq\sigma\leq 2\pi)$, the surface terms vanish when the boundary
conditions are periodicity or anti-periodicity for each component of $\psi$ separately. We can thus have:
\begin{eqnarray}
\label{rightclosedmodes}
\psi_R^{\mu}=\sum\psi_n^{\mu}e^{-in(\tau-\sigma)}\quad \textrm{or}\quad
\psi_R^{\mu}=\sum\psi_r^{\mu}e^{-ir(\tau-\sigma)}
\end{eqnarray}
and
\begin{eqnarray}
\label{rightclosedmodes}
\psi_L^{\mu}=\sum\tilde{\psi}_n^{\mu}e^{-in(\tau+\sigma)}\quad \textrm{or}\quad
\psi_L^{\mu}=\sum\tilde{\psi}_r^{\mu}e^{-ir(\tau+\sigma)}\ .
\end{eqnarray}
Then there are four distinct closed-string sectors corresponding to the different pairings of left-moving
and right-moving modes that can be referred to as NS-NS, NS-R, R-NS and R-R. The first and the last cases
describe Bosonic states and the other two describe Fermions.

As the energy-momentum tensor now has Fermionic modes too, the super-virasoro generators will have the form:
\begin{eqnarray}
\label{supervirasorogens}
L_{m}=L_m^{(\alpha)}+L_m^{(\psi)}\ ,
\end{eqnarray}
which is defined in the usual way, \emph{e.g.} for \textbf{open superstrings}:
\begin{eqnarray}
\label{superlm}
L_{m} =
\frac{1}{2\pi\alpha'}\int_{0}^{2\pi}d\sigma{e^{im\sigma}}T_{++}\Bigg.\Bigg|_{\tau
= 0}\ .
\end{eqnarray}
For the Fermionic generators of the algebra, however, we define:
\begin{eqnarray}
\label{superfermiF}
F_{m}&=&\frac{2}{\pi\alpha'}\int_0^{2\pi}d\sigma e^{im\sigma}J_+\Bigg.\Bigg|_{\tau=0}\ ;\\
\label{superfermiG}
G_{r}&=&\frac{2}{\pi\alpha'}\int_0^{2\pi}d\sigma e^{ir\sigma}J_+\Bigg.\Bigg|_{\tau=0}\ ,
\end{eqnarray}
with similar extra modes for the closed superstring defined in terms of $T_{--}$ and $J_-$. Overall we thus have:
\begin{eqnarray}
\label{supervirasoro}
L_m^{(\alpha)}&=&\frac{1}{2}\sum_{-\infty}^{\infty}:{\alpha}_{-n}\cdot{\alpha}_{m+n}:\ \ ,\nonumber \\
L_m^{(\psi_n)}&=&\frac{1}{2}\sum_{-\infty}^{\infty}(n+\frac{1}{2}m):{\psi}_{-n}\cdot{\psi}_{m+n}:\ \ ,\nonumber \\
L_m^{(\psi_r)}&=&\frac{1}{2}\sum_{-\infty}^{\infty}(r+\frac{1}{2}m):{\psi}_{-r}\cdot{\psi}_{m+r}:\ \ ,\nonumber \\
F_m&=&\frac{1}{2}\sum_{-\infty}^{\infty}:{\alpha}_{-n}\cdot{\psi}_{m+n}:\ \ ,\nonumber \\
G_r&=&\frac{1}{2}\sum_{-\infty}^{\infty}:{\alpha}_{-n}\cdot{\psi}_{r+n}:\ \ ,
\end{eqnarray}
where $L_m^{(\psi_n)}$ and $F_m$ arise in the Ramond sector and $L_m^{(\psi_r)}$ and $G_r$ arise in the Neveu-Schwarz
sector. In the Ramond sector, the super-Virasoro algebra is then:
\begin{eqnarray}
\label{supervraramond}
\left[L_m,L_n\right]&=&(m-n)L_{m+n}+\frac{d}{8}m^3\delta_{m+n}\ ;\nonumber \\
\left[L_m,F_n\right]&=&(\frac{m}{2}-n)F_{m+n}\ ;\nonumber \\
\left\{F_m,F_n\right\}&=&2L_{m+n}+\frac{d}{2}m^2\delta_{m+n}\ .
\end{eqnarray}
In the Neveu-Schwarz sector on the other hand, the super-Virasoro algebra is:
\begin{eqnarray}
\label{supervirans}
\left[L_m,L_n\right]&=&(m-n)L_{m+n}+\frac{d}{8}(m^3-m)\delta_{m+n}\ ;\nonumber \\
\left[L_m,G_r\right]&=&(\frac{m}{2}-r)G_{m+r}\ ;\nonumber \\
\left\{G_r,G_s\right\}&=&2L_{r+s}+\frac{d}{2}(r^2-\frac{1}{4})\ .
\end{eqnarray}
See Appendix \ref{VirasoroAppendix} for methods which may be used to determine these commutators.

For these, we need as before the momentum conjugate to $\psi$:
\begin{eqnarray}
\label{psimom}
\frac{\delta S}{\delta(\partial_\tau\psi)}\sim \frac{T}{2}\,\psi\ ,\nonumber
\end{eqnarray}
which implies poisson commutation brackets of
\begin{eqnarray}
\label{superpoisson}
\left\{\psi_a^{\mu}(\sigma,\tau),\psi_b^{\nu}(\sigma',\tau)\right\}_{P.B.}=
2\pi{\delta}_{ab}\delta(\sigma-\sigma')\eta^{\mu\nu}\ ,
\end{eqnarray}
and thus on quantization we have:
\begin{eqnarray}
\label{supercomm}
\left\{\psi_a^{\mu}(\sigma,\tau),\psi_b^{\nu}(\sigma',\tau)\right\}=
-2\pi i{\delta}_{ab}\delta(\sigma-\sigma')\eta^{\mu\nu}\ ,
\end{eqnarray}
with
\begin{eqnarray}
\label{supermodes}
\left\{\psi_m^{\mu},\psi_n^{\nu}\right\}&=&-i\delta_{m+n}\eta^{\mu\nu}\ ;\nonumber \\
\left\{\psi_r^{\mu},\psi_s^{\nu}\right\}&=&-i\delta_{r+s}\eta^{\mu\nu}\ ,
\end{eqnarray}
with all other commutators vanishing and similar relations for $\tilde{\psi}_{n,r}$. We also have the reality
conditions $\psi_{-n,-r}=\psi_{n,r}^{\dag}$ and $\tilde{\psi}_{-n,-r}=\tilde{\psi}_{n,r}^{\dag}$. Thus
the $\psi_{n,r}$ are creation operators for $n,r<0$ and annihilation operators for $n,r>0$. For $n=0$
(in the Ramond sector), we have $\left\{\psi_0^{\mu},\psi_0^{\nu}\right\}=-i\eta^{\mu\nu}$. But this is
just the Dirac algebra in $d$-dimensions (up to normalization), so the zero modes $\psi_0^{\mu}$ are just
Dirac gamma matrices up to normalization: $\psi_0^{\mu}=\gamma^{\mu}/(\sqrt{2}e^{i3\pi/4})$.

\subsection{Spectrum of States in Type I (Open) Superstring Theory}

Given the above discussion of the super-Virasoro algebra, the physical state conditions are now:
\begin{eqnarray}
\label{superstateconsramond}
L_m|\phi\rangle_R=F_m|\phi\rangle_R&=&0 \quad m>0\\
\label{superstateconsns}
L_m|\phi\rangle_{NS}=G_r|\phi\rangle_{NS}&=&0 \quad m,r>0\ ,
\end{eqnarray}
where (\ref{superstateconsramond}) refers to the Ramond sector, and (\ref{superstateconsns})
refers to the Neveu-Schwarz sector. The zero mode constraints are modified as before due to the
normal ordering issue, and they take the form:
\begin{eqnarray}
\label{superzeromodesr}
\left(L_0-a_R\right)|\phi\rangle_R=F_0|\phi\rangle_R&=&0\\
\left(L_0-a_{NS}\right)|\phi\rangle_{NS}&=&0\ .
\end{eqnarray}
However, the super-Virasoro algebra in the Ramond sector (\ref{supervraramond}) implies that
$F_0^2=L_0$, so we must have that $a_R$ is in fact zero. These super-Virasoro conditions are
necessary and sufficient to show that all the negative norm states decouple and that the
physical spectrum is positive definite. That is to say (\emph{c.f.} light-cone gauge quantization):
\begin{eqnarray}
\label{superlightlike}
\alpha_n^{\mu}&\rightarrow&\alpha_n^{i}\quad(i=1,\ldots,d-2)\quad n\neq 0\nonumber \\
\psi_{n,r}^{\mu}&\rightarrow&\psi_{n,r}^{i}\quad(i=1,\ldots,d-2)\quad n\neq 0\ .
\end{eqnarray}
We are now ready to write down the spectrum of states of the superstring.

\begin{itemize}
\item Open (Type I) Superstring in the \textbf{NS Sector}.
\begin{enumerate}
\item Ground State (even number of $\psi$'s):\newline
$N^{(\alpha)}=N^{(\psi)}=0$ \footnote{$N^{(\alpha)}$ is defined in the same way as for the
Bosonic string, and similarly $N^{(\psi)}=\sum_{r=1/2}^{\infty}r\,\psi_{-r}\cdot\psi_{r}$},
and we have the state $|0;p\rangle$ with
\begin{eqnarray}
\label{type1ground}
\left(L_0-a_{NS}\right)|0;p\rangle&=&0\nonumber \\
\Rightarrow\left(N^{(\alpha)}+N^{(\psi)}+\alpha'p^2-a_{NS}\right)|0;p\rangle&=&0\nonumber \\
\Rightarrow M^2=-\frac{a_{NS}}{\alpha'}\ .
\end{eqnarray}
\item First Excited State (odd number of $\psi$'s):\newline
$N^{(\alpha)}=0$, $N^{(\psi)}=1/2$ and we have the state(s):
\begin{eqnarray}
\label{type1firstexcited}
\underbrace{\zeta_i^{(1)}\psi_{-1/2}^i}_{\textrm{8 states in $d\!=\!10$}}\!\!\!\!\!|0;p\rangle\nonumber \\
M^2=-\frac{1}{\alpha'}(a_{NS}-\frac{1}{2})\ .
\end{eqnarray}
We see that we have 8 states at the first excited level.
\item Second Excited State (even number of $\psi$'s):\newline
$N^{(\alpha)}+N^{(\psi)}=1$, and
\begin{eqnarray}
\label{type1secondexcited}
\underbrace{\zeta_i^{(2)}\alpha_{-1}^{i}}_{\textrm{8 states}}|0;p\rangle+
\underbrace{\zeta_{ij}^{(2)}\psi_{-1/2}^i\psi_{-1/2}^j}_{\textrm{$\binom{8}{2}=28$ states}}|0;p\rangle
\nonumber \\
M^2=-\frac{1}{\alpha'}(a_{NS}-1)\ .
\end{eqnarray}
Here we have a total of $8+28=36$ states. (Compare with an antisymmetric tensor in $d=10$,
which has $(9\times8)/2=36$ independent components.)
\item Third Excited State (odd number of $\psi$'s):\newline
$N^{(\alpha)}+N^{(\psi)}=3/2$, so
\begin{eqnarray}
\label{type1thirdexcited}
\underbrace{\zeta_i^{(2)}\zeta_j^{(1)}\alpha_{-1}^{i}\psi_{-1/2}^j}_{\textrm{$8\times8=64$ states}}|0;p\rangle+
\underbrace{\zeta_{ijk}^{(3)}\psi_{-1/2}^i\psi_{-1/2}^j\psi_{-1/2}^k}_{\textrm{$\binom{8}{3}=56$ states}}|0;p\rangle+
\underbrace{\zeta_i^{(1)}\psi_{-3/2}^{i}}_{\textrm{8 states}}|0;p\rangle\ ,
\end{eqnarray}
giving a total of $64+56+8=128$ states at this level.
\end{enumerate}

All these states represent space-time Bosons as we are in the Neveu-Schwarz sector. We would also
like to know what $a_{NS}$ is. Using a similar argument to the one we used for the Bosonic string, we can show that
\begin{eqnarray}
\label{ans}
a_{NS}=\frac{d-2}{16}\ ,
\end{eqnarray}
and again requiring that the first excited (vector) state is massless so that it does not acquire a
longitudinal polarization when Lorentz boosted tells us that
\begin{eqnarray}
\label{deeequals10}
a_{NS}=\frac{1}{2}\Rightarrow d=10\ .
\end{eqnarray}
Note that in order to normal-order the Fermionic mode operators it is necessary to
deal with divergent sums over half-integers which we may do as follows:
\begin{eqnarray}
\label{rsum}
\sum_{1/2}^{\infty}r&=&\frac{1}{2}\left(1+3+5+\ldots\right)\nonumber \\
&=& \frac{1}{2}\left(\sum_{n=1}^{\infty}n-\sum_{n=1}^{\infty}2n\right)\nonumber \\
&=& -\frac{1}{2}\sum_1^{\infty}n\nonumber \\
&=& \frac{1}{24}\ ,
\end{eqnarray}
where in the last step we have used zeta function regularisation.
%\pagebreak

\item Open (Type I) Superstring in the \textbf{Ramond Sector}
\begin{enumerate}
\item Ground State:\newline
$N^{(\alpha)}=N^{(\psi)}=0$, with $N^{(\psi)}=\sum_{m=1}^{\infty}m\,\psi_{-m}\cdot\psi_{m}$, and:
\begin{eqnarray}
\label{ramondground}
L_{0}|\phi\rangle=0\Rightarrow\alpha'p^2=0\Rightarrow M^2=0
\end{eqnarray}
(recall that $a_{R}=0$). Now, the Ramond sector describes space-time Fermions, so the ground state must
be a spinor wavefunction in $d\,(=\!\!\!10)$ dimensions $|\phi\rangle=u_{a}|a,0;p\rangle$, where $u_a$ is a
spinor in 10 dimensions and thus has 16 physical components (if it is a Majorana spinor obeying
the Dirac equation)\footnote{See Section \ref{masslessstatesappsection} for more.}.
\item First Excited State:\newline
$N^{(\alpha)}+N^{(\psi)}=1$:
\begin{eqnarray}
\label{firstexcitedramond}
\underbrace{\zeta_i^{(1)}\alpha_{-1}^{i}u_a}_{\textrm{$8\times16=64$ states}}\!\!\!\!|a,0;p\rangle+
\underbrace{\zeta_i^{(2)}\psi_{-1}^iu_a}_{\textrm{$8\times16=64$ states}}\!\!\!\!|a,0;p\rangle\ ,
\end{eqnarray}
which gives a total of 256 (Fermionic) states.
\end{enumerate}
\end{itemize}

%Notice that the number of states at the ground level of the Ramond sector is the same as the number of
%states at the first excited level of the Neveu-Schwarz sector, and that the number of states at the
%first excited level of the Ramond sector is the same as the number of states at the third excited level
%of the Neveu-Schwarz sector. This is a hint at supersymmetry!

\subsection{The GSO Projection}

This is all very well, but it turns out that even in $d=10$ and with $a_{NS}=1/2$ and $a_{R}=0$, the RNS
model described above is still an inconsistent quantum field theory. In addition it has some unattractive
features that we would like to get rid-of. To start with, the theory has a Tachyon which we would like to
eliminate. Secondly, although there is no actual conflict with the spin-statistics theorem, it seems
unnatural to have world-sheet Fermions $\psi_a^{\mu}$ that are also space-time vectors. In order to fix
these `problems' and make a consistent theory, we must truncate the spectrum in a way first proposed by
Gliozzi, Scherk and Olive.

Essentially, the GSO conditions boil down to a chirality condition in the Ramond sector and an
elimination of all states with an even number of $\psi$'s in the Neveu-Schwarz sector. Mathematically
this is accomplished by defining:
\begin{eqnarray}
\label{gsons}
P^{NS}&=&\frac{1}{2}\left(1+(-1)^F\right)\ ;\\
\label{F}
F&=&\sum_{r=1/2}^{\infty}\psi_r^{\dag}\cdot \psi_r\ ,
\end{eqnarray}
where $(-1)^F\psi_r=-\psi_r(-1)^F$ and $(-1)^F|0;p\rangle=-|0;p\rangle$. $P^{NS}$ is thus a projection
operator that kills states with an even number of $\psi$'s. In the Ramond sector we have:
\begin{eqnarray}
\label{gsor}
P^R&=&\frac{1}{2}\left(1\pm(-1)^G\gamma_{11}\right)\ ;\\
\label{G}
G&=&\sum_{n>0}\psi_n^\dag\cdot\psi_n\ ,
\end{eqnarray}
where the sign in (\ref{gsor}) can be chosen arbitrarily.

Thus the GSO projection kills the ground state in the NS sector (\ref{type1ground}) and the second
excited state in the NS sector (\ref{type1secondexcited}). What we are left with are the 8 states of the
original first excited level (\ref{type1firstexcited}) as our new ground states and the 128 states of the
original third excited level as our new first excited states. In the Ramond sector the projection acts
as a chirality projection, demanding that our spinors must be further constrained and actually be Weyl spinors.
This halves the number of degrees of freedom in the $u_a$\footnote{Note for future reference that $u_a$ lives in the $\mathbf{8_s}$ or
$\mathbf{8_c}$ spinor representations of $SO(8)$ depending on it's chirality.} and thus gives us 8 ground states and 128 states at the
first excited level. We see that the number of states in the Ramond and Neveu-Schwarz sectors now
match level by level. As the NS sector describes spacetime Bosons and the
R sector describes spacetime Fermions, we see that the number of Bosonic degrees of freedom must match the
number of Fermionic ones in this theory. Thus we can see that the Type I superstring has space-time
supersymmetry. It is in fact $\mathcal{N}\!=\!1$ supersymmetry.

The massless states of this theory are (as we have already seen) a vector in $d=10$, $\zeta_i|\,i\rangle$ and
a spinor in $d=10$, $u_a|\,a\rangle$, which both have 8 physical degrees of freedom. If we now look at the
low energy effective theory by discarding massive states (we usually say that we integrate out these states), then
these massless states become $A_{\mu}$ and $\psi$ -- the vector potential and spinor field of a super Yang-Mills
multiplet in $d=10$. This low energy theory is in fact 10-dimensional $\mathcal{N}\!=\!1$ super Yang-Mills. Its
action is
\begin{eqnarray}
\label{effaction}
S=\int\!d^{10}X\left(-\frac{1}{4}F^2+\frac{1}{2}\bar{\psi}\,\slash\!\!\!\!D\psi\right)\ ,
\end{eqnarray}
where $F_{\mu\nu}=\partial_{\mu}A_{\nu}-\partial_{\nu}A_{\mu}+g_{YM}\left[A_{\mu},A_{\nu}\right]$ and
$D$ is the Yang-Mills covariant derivative $(D_{\mu}\psi)^{A}=\partial_{\mu}\psi^{A}+g_{YM}{f^{A}}_{BC}
A^B_{\mu}\psi^{C}$. It is invariant under the supersymmetry transformations:
\begin{eqnarray}
\label{ymsusy}
\delta A_{\mu}=\frac{1}{2}\bar{\epsilon}\gamma_{\mu}\psi\ ,\nonumber \\
\delta\psi=-\frac{1}{4}F_{\mu\nu}\gamma^{\mu\nu}\epsilon\ ,
\end{eqnarray}
where $\epsilon$ is a Majorana-Weyl spinor in $d=10$ and $\gamma^{\mu\nu}=(\gamma^{\mu}\gamma^{\nu}
-\gamma^{\nu}\gamma^{\mu})/2$.

\subsection{Massless States in Type II (Closed) Superstring Theory}

As in the case of the Bosonic string we have a doubling-up of the states, so we can write the closed-string
states in terms of direct products of open-string states:
$|\,\phi\rangle\otimes|\,\tilde{\phi}\rangle$.\footnote{Note that in this section we are going to consider
open-string states that have already been GSO projected. If we did not do this we would need to think
about GSO projecting our final sets of closed-string states.}
Thus the \textbf{massless} states in the \textbf{Bosonic} sector are:

\begin{enumerate}
\item NS-NS sector
\begin{eqnarray}
\label{nsns}
\zeta_{ij}\,|\,i\rangle\otimes|\,\tilde{j}\rangle\ ,
\end{eqnarray}
where the roman letters $(i,j,k\ldots)$ indicate that this is a direct product of vector states. In fact these states
($|\,i\rangle$) -- which are just those of the Type I superstring at it's massless level in the NS sector --
are in the $\mathbf{8_v}$ representation of $SO(8)$ (\emph{i.e.} the fundamental vector representation). As we've
discussed before when looking at the states of the Bosonic string, this is because the physical constraints
mean that we can effectively ignore our light-cone co-ordinates -- our symmetry group is reduced from $SO(1,9)$
to $SO(8)$. Thus $\zeta_{ij}$ has $8\times8=64$ components and so we have 64 states. As before, $\zeta_{ij}$
decomposes into $G_{(ij)}$ with 35 degrees of freedom -- this is the graviton -- the $B$-field, $B_{[ij]}$ with
28 degrees of freedom, and the Dilaton $\Phi={\zeta_i}^i$ with 1 degree of freedom. This is just the
 group theoretic decomposition: $\mathbf{8_v}\otimes\mathbf{8_v}=\mathbf{35_v}+\mathbf{28}+\mathbf{1}$.

\item R-R sector
\begin{eqnarray}
\label{rr}
\zeta_{a\tilde{a}}\,|\,a\rangle\otimes|\,\tilde{a}\rangle\ ,
\end{eqnarray}
where this time the letters $(a,b,c\ldots)$ indicate spinor states. In $SO(8)$ there are two inequivalent
fundamental spinor representations, and $a$ labels $\mathbf{8_c}$ representations, while $\tilde{a}$ labels
$\mathbf{8_s}$ representations. As such we have two inequivalent possibilities and thus two different Type II
theories. In Type IIA the states $|\,a\rangle$ and $|\,\tilde{a}\rangle$ have opposite spacetime chiralities,
and as such the theory as a whole is \textit{non-chiral}:
\begin{eqnarray}
\label{oppchiral}
\frac{1}{2}\left(1+{\hat{\gamma}}_9\right)|\,a\rangle=\frac{1}{2}\left(1-{\hat{\gamma}}_9\right)
|\,\tilde{a}\rangle=0\ ,
\end{eqnarray}
where $\hat{\gamma}_9=\prod_{i=1}^8\hat{\gamma}_i$ and the $\hat{\gamma}_i$ are Dirac matrices of $SO(8)$.
Here the representations decompose as $\mathbf{8_s}\otimes\mathbf{8_c}=\mathbf{8_v}+\mathbf{56_v}$, so that\footnote{
Here, things like $\hat{\gamma}^{ij}$ stand for antisymmetrised products of gamma matrices. \emph{i.e.} up to normalisation
we have $\hat{\gamma}_{i_1i_2\ldots i_k}\sim \hat{\gamma}_{[i_1}\hat{\gamma}_{i_2}\ldots\hat{\gamma}_{i_k]}$.}
\begin{eqnarray}
\label{scdecomp}
\zeta_{a\tilde{a}}=\underbrace{{\hat{\gamma}^i}_{b\tilde{b}}\hat{\zeta}_{\tilde{b}b}}_{\textrm{8 states}}\,{\hat{\gamma}^i}_{a\tilde{a}}\ +
\underbrace{{\hat{\gamma}^{ijk}}_{b\tilde{b}}\hat{\zeta}_{\tilde{b}b}}_{\textrm{$\binom{8}{3}=56$ states}}\!\!\!\!\!
{\hat{\gamma}^{ijk}}_{a\tilde{a}}\ ,
\end{eqnarray}
with 64 states in total. ${\hat{\gamma}^i}_{b\tilde{b}}\hat{\zeta}_{\tilde{b}b}$ is a 1-form\footnote{$p$-forms are
just totally antisymmetric $p$-tensors.} and is more commonly
referred to as $C^{(1)}$ in the literature. Similarly ${\hat{\gamma}^{ijk}}_{b\tilde{b}}\hat{\zeta}_{\tilde{b}b}$ is
the 3-form $C^{(3)}$.

In Type IIB, however, $|\,a\rangle$ and $|\,\tilde{a}\rangle$ have the same spacetime chirality, and thus the
IIB theory \textit{is chiral}. We have:
\begin{eqnarray}
\label{samechiral}
\frac{1}{2}\left(1+{\hat{\gamma}}_9\right)|\,a\rangle=\frac{1}{2}\left(1+{\hat{\gamma}}_9\right)
|\,\tilde{a}\rangle=0\ .
\end{eqnarray}
Here we have $\mathbf{8_s}\otimes\mathbf{8_s}=\mathbf{1}+\mathbf{28}+\mathbf{35_v}$, so that
\begin{eqnarray}
\label{ocdecomp}
\zeta_{a\tilde{a}}=\underbrace{\delta_{b\tilde{b}}\hat{\zeta}_{\tilde{b}b}}_{\textrm{1 state}}\,
\delta_{a\tilde{a}}\ +\underbrace{{\hat{\gamma}^{ij}}_{b\tilde{b}}\hat{\zeta}_{\tilde{b}b}}
_{\textrm{$\binom{8}{2}=28$ states}}\!\!\!\!\!\!{\hat{\gamma}}_{a\tilde{a}}^{ij}\ +
\underbrace{{\hat{\gamma}}_{b\tilde{b}}^{ijkl}\hat{\zeta}_{\tilde{b}b}}_{\frac{1}{2}\binom{8}{4}\,=\,35\textrm{ states}}
\!\!\!\!\!{\hat{\gamma}_{a\tilde{a}}^{ijkl}}\ ,
\end{eqnarray}
where this time we have a 0-form (scalar) $C^{(0)}$, a 2-form $C^{(2)}$ and a so-called `self-dual'\footnote{This is
the reason that it has half the expected number of components.} 4-form $C^{(4)}$. These form-fields of IIA and IIB are
intimately related to the existence of \textit{stable} $D$-branes of different dimensionalities.
\end{enumerate}

The \textbf{Fermionic} sectors are the NS-R or R-NS sectors, and the \textbf{massless} states are thus spinor-vectors:
\begin{eqnarray}
\label{spinvec}
\zeta_{i\tilde{a}}\,|\,i\rangle\otimes|\,\tilde{a}\rangle\, ;\\
\label{spinvec2}
\zeta_{ai}\,|\,a\rangle\otimes|\,\tilde{i}\rangle\ .
\end{eqnarray}
For (\ref{spinvec}) we have $\mathbf{8_v}\otimes\mathbf{8_s}=\mathbf{8_c}+\mathbf{56_s}$ and thus
64 massless Fermionic states. For (\ref{spinvec2}) on the other hand we have
$\mathbf{8_c}\otimes\mathbf{8_v}=\mathbf{8_s}+\mathbf{56_c}$ and again 64 Fermionic states. In IIA we
actually have $(\mathbf{8_v}\otimes\mathbf{8_c})+(\mathbf{8_s}\otimes\mathbf{8_v})=\mathbf{8_c}+\mathbf{56_c}
+\mathbf{8_s}+\mathbf{56_s}$ and therefore a total of 128 Fermionic states. Both chiralities are present
equally and so the Fermionic sector is also non-chiral. In IIB we have $(\mathbf{8_v}\otimes\mathbf{8_s})+
(\mathbf{8_s}\otimes\mathbf{8_v})=\mathbf{8_s}+\mathbf{56_s}+\mathbf{8_s}+\mathbf{56_s}$. Here only one
chirality is present and so again the Fermionic sector is chiral with 128 states. Thus, if we compare the
Bosonic and Fermionic sectors for both IIA and IIB, we find that we have 128 Bosons and 128 Fermions and thus
spacetime supersymmetry as before. It is worth noting that the particles in the $\mathbf{56}$ representations
are spin-$3/2$ Gravitinos.\footnote{In a compact form, the massless sectors are: IIA:
$(\mathbf{8_v}+\mathbf{8_s})\otimes(\mathbf{8_v}+\mathbf{8_c})$; IIB: $(\mathbf{8_v}+\mathbf{8_s})\otimes
(\mathbf{8_v}+\mathbf{8_s})$.}

The Type I theory has $\mathcal{N}\!=\!1$ supersymmetry, but our Type II multiplets here are direct products
of the Type I multiplets and so the Type II theories actually have $\mathcal{N}\!=\!2$ supersymmetry.

%\newpage

%%%%%%%%%%%%%%%%%%%%%%%%%%%%%%%%%%%%%%%%%%%%%%%%%%%%%%%%%%%%%%%%%%%%%%%%%%%%%%%%%%%%%%%%%%

\newpage

\startappendix
\section{Advanced Topics}
\subsection{The Hagedorn Temperature}
Let us consider the high-temperature counting of states for the Bosonic string, following the general line of reasoning taken in 
\cite{GSW1}, Chapter 2. Recall that the masses of open string
states are given by $\alpha'M^2=n-1$, where $n$ is the eigenvalue of the number operator. As Einstein's mass-energy relation is 
$E=M$ (in natural units), we can conveniently describe the energies of open string states by referring to the level, $n$. 
If we denote the number of states at level-$n$ by $d_n$, then this is conveniently described by the coefficient of $\omega^n$ in
$\textrm{Tr}\,\omega^N$, where $N=\sum_{m=1}^{\infty}\alpha_{-m}\cdot\alpha_m$ is the number operator. We may think of this as the partition function of the 
theory: $\textrm{Tr}$ represents the sum over states, $\omega\sim e^{-\frac{E}{k_{B}T}}$ and the power of $N$ allows the encoding of the different numbers of states 
at each energy level. Taking the high-temperature limit, $T\rightarrow\infty$, will therefore correspond to taking $\omega\rightarrow 1$.

We wish to consider only
transverse (physical) states, so the oscillators are $\alpha_m^i$ with $i=1,\ldots,24$ and we may define our generating function $F(\omega)$ as:
\begin{eqnarray}
\label{genfun}
F(\omega)=\sum_{n=0}^{\infty}d_n\omega^n=\textrm{Tr}\,\omega^N\ .
\end{eqnarray}
Now:
\begin{eqnarray}
\label{stuff}
\textrm{Tr}\,\omega^N&=&\textrm{Tr}\,\omega^{\sum_{m=1}^{\infty}\alpha_{-m}\cdot\alpha_m}=
\prod_{m=1}^{\infty}\textrm{Tr}\,\omega^{\alpha_{-m}\cdot\alpha_m}\nonumber \\
&=&\prod_{m=1}^{\infty}\prod_{i=1}^{24}\textrm{Tr}\,\omega^{\alpha_{-m}^i\alpha_m^i}=
\left(\prod_{m=1}^{\infty}\textrm{Tr}\,\omega^{\alpha_{-m}^1\alpha_m^1}\right)^{24}\ .\nonumber
\end{eqnarray}
But
\begin{eqnarray}
\label{geomprog}
\textrm{Tr}\,\omega^{\alpha_{-m}^1\alpha_m^1}=\sum_{n=0}^{\infty}\omega^{mn}=
\frac{1}{1-\omega^m}\ ,
\end{eqnarray}
since we are considering Bose-Einstein statistics and this is a geometric progression.
\begin{eqnarray}
\label{eulerform}
\therefore\ \textrm{Tr}\,\omega^N=\left[\prod_{m=1}^{\infty}(1-\omega^m)\right]^{-24}=F(\omega)\ .
\end{eqnarray}

It is interesting to note that from this form we can see that $F(\omega)=\phi(\omega)^{-24}$, where
\begin{eqnarray}
\phi(t)=\prod_{k=1}^{\infty}(1-t^k)
\end{eqnarray}
is the Euler function. It is well known that the coefficient $p_n$ in the formal power series 
expansion of $1/\phi(t)=\sum_{n=0}^{\infty}p_nt^n$ gives the number of partitions of $n$; that is, the 
number of ways of writing an integer $n$ as a sum of component integers:
\begin{center}
Partitions of $n$ for $1\leq n\leq 4$
\begin{displaymath}
\begin{array}{|c|c|}
\hline n & \textrm{Partitions} \\
\hline \textrm{1} & \textrm{1}\\
\hline \textrm{2} & \textrm{2, 1+1}\\
\hline \textrm{3} & \textrm{3, 2+1, 1+1+1}\\
\hline \textrm{4} & \textrm{4, 3+1, 2+2, 2+1+1, 1+1+1+1}\\
\hline
\end{array}
\end{displaymath}
\end{center}
What we are trying to do is exactly this, since the number of states at a given level $n$ is 
simply the number of ways of writing creation operators for that level
\begin{center}
Arrangements of $\alpha_{-n}$ for $1\leq n\leq 4$
\begin{displaymath}
\begin{array}{|c|c|}
\hline n & \textrm{Partitions} \\
\hline 1 & \alpha_{-1}\\
\hline 2 & \alpha_{-2},\quad \alpha_{-1}\alpha_{-1}\\
\hline 3 & \alpha_{-3},\quad \alpha_{-2}\alpha_{-1},\quad \alpha_{-1}\alpha_{-1}\alpha_{-1}\\
\hline 4 & \alpha_{-4},\quad \alpha_{-3}\alpha_{-1},\quad \alpha_{-2}\alpha_{-1}\alpha_{-1},\quad \alpha_{-1}\alpha_{-1}\alpha_{-1}\alpha_{-1}\\
\hline
\end{array}
\end{displaymath}
\end{center}
The power of 24 in (\ref{eulerform}) simply encodes the fact that $i$ runs from 1 to 24 in $\alpha_{-n}^i$ and thus we actually have 24 creation 
operators for each $n$. So in actual fact we could have started by considering the different arrangements of creation operators at each level, 
realised that they are just partitions of $n$ and therefore written down the generating function $F(\omega)=\left(\frac{1}{\phi(\omega)}\right)^{24}$.

In order to estimate the asymptotic density of states, we need the behaviour of $F(\omega)$ as
$\omega\rightarrow1$. Crudely:
\begin{eqnarray}
\label{crude}
\phi(\omega)&=&\prod_{m=1}^{\infty}(1-\omega^m)=\exp\left({\sum_{m=1}^{\infty}\log(1-\omega^m)}\right)\nonumber \\
&=&\exp\left(-\sum_{m,n}\frac{\omega^{mn}}{n}\right)=\exp\left(-\sum_{n=1}^{\infty}\frac{\omega^n}{n(1-\omega^n)}\right)\ .
\end{eqnarray}
\newpage
As $\omega$ is close to 1, we write $\omega=1-\epsilon$ (with $\epsilon\ll1$). Thus
$\omega^n=(1-\epsilon)^n\sim1-n\epsilon$, and $1-\omega^n\sim n\epsilon$.
\begin{eqnarray}
\label{fomega}
\Rightarrow -\sum_{n=1}^{\infty}\frac{\omega^n}{n(1-\omega^n)}&\sim&-\sum_n\frac{1-n\epsilon}{n^2\epsilon}\
\sim-\sum_n\frac{1}{n^2\epsilon}=-\frac{1}{1-\omega}\sum_{n=1}^{\infty}\frac{1}{n^2}\nonumber \\
\Rightarrow \phi(\omega)&\sim& \exp\left({-\frac{\pi^2}{6(1-\omega)}}\right)\nonumber \\
\Rightarrow F(\omega)&\sim& \exp\left({\frac{4\pi^2}{1-\omega}}\right)\ ,
\end{eqnarray}
where in getting from the first line to the second we have used the fact that $\sum_{n=1}^{\infty}\frac{1}{n^2}=\zeta(2)=\frac{\pi^2}{6}$.

However, we can get a \emph{better} approximation by considering the Dedekind eta function, which is defined by:
\begin{eqnarray}
\label{dedekind}
\eta(\tau):=e^{i\pi\tau/12}\prod_{n=1}^{\infty}\left(1-e^{2\pi in\tau}\right)\ .
\end{eqnarray}
If we let $\omega=e^{2\pi i\tau}$ (\emph{i.e.} $\tau=\frac{\log\omega}{2\pi i}$), then we have the equivalent form in terms of $\omega$
\begin{eqnarray}
\label{dedekind2}
\eta(\tau)=\omega^{1/24}\prod_{n=1}^{\infty}\left(1-\omega^n\right)\ .
\end{eqnarray}
Now the Dedekind eta function is what's known as a modular form and as such has the following transformational property:
\begin{eqnarray}
\label{modtransform}
\eta(-1/\tau)=(-i\tau)^{1/2}\eta(\tau)\ .
\end{eqnarray}
It will therefore also be useful for us to also have $\eta(-1/\tau)$ in terms of $\tau$:
\begin{eqnarray}
\label{dedekind3}
\eta(-1/\tau) = e^{-i\pi/12\tau}\prod_{n=1}^{\infty}\left(1-e^{-2\pi in/\tau}\right)\ ,
\end{eqnarray}
and also in terms of $q^2$, where $q^2=e^{-2\pi i/\tau}$:
\begin{eqnarray}
\label{dedekind4}
\eta(-1/\tau) = (q^2)^{1/24}\prod_{n=1}^{\infty}\left(1-(q^2)^n\right)\ .
\end{eqnarray}

We can thus re-write $\phi(\omega)$ as:
\begin{eqnarray}
\phi(\omega) &=&\omega^{-1/24}\eta(\tau)\nn\\
&=& \omega^{-1/24}(-i\tau)^{-1/2}\eta(-1/\tau)\nn\\
&=& \left(\frac{-\log\omega}{2\pi}\right)^{-1/2}\omega^{-1/24}(q^2)^{1/24}\phi(q^2)\ .
\end{eqnarray}
Now when we make our high-temperature approximation, $\omega\rightarrow 1$ so we must have $\tau\rightarrow 0$, which in turn 
means that $q^2\rightarrow 0$. In terms of $\omega$, $q^2=\exp\left(\frac{4\pi^2}{\log\omega}\right)$, and noting that as 
$\omega\rightarrow 1$, $\log\omega\rightarrow \omega-1$, so $(q^2)^{1/24}\rightarrow \exp\left(-\frac{\pi^2}{6(1-\omega)}\right)$. Clearly
$\phi(q^2)\rightarrow 1$ and $\omega^{-1/24}\rightarrow 1$, so in this limit
\begin{eqnarray}
\label{dedasymptote1}
\phi(\omega)\sim(2\pi)^{1/2}(1-\omega)^{-1/2}\exp\left(\frac{-\pi^2}{6(1-\omega)}\right)\ ,
\end{eqnarray}
and
\begin{eqnarray}
\label{dedasymptote}
F(\omega)\sim(2\pi)^{-12}(1-\omega)^{12}\exp\left(\frac{4\pi^2}{1-\omega}\right)\ .
\end{eqnarray}

Now we want the large-$n$ behaviour for $d_n$. It is easy to see that we can project out $d_n$ from
$F(\omega)=\sum d_m\omega^m$ by a contour integral around a small circle encompassing the origin of
the complex-$\omega$ plane:
\begin{eqnarray}
\label{contour}
d_n=\frac{1}{2\pi i}\oint\!d\omega\frac{F(\omega)}{\omega^{n+1}}\ .
\end{eqnarray}
Using our asymptotic form for $F(\omega)$, we can estimate this for large $n$ by a saddle-point
evaluation. Writing $\omega^{-(n+1)}$ as an exponential and re-casting $\omega-1$ as $\log\omega$ gives the asymptotic integral
\begin{eqnarray}
\label{saddleint}
d_n\sim \frac{1}{2\pi i}\oint\!d\omega\left(\frac{-\log\omega}{2\pi}\right)^{12}\exp\left[-(n+1)\left(\log\omega+\frac{4\pi^2}{(n+1)\log\omega}\right)\right]\ .
\end{eqnarray}

In terms of the quantities defined under the saddle-point evaluation in Appendix \ref{mathappendix} we have
\begin{eqnarray}
\label{quantities}
s &\rightarrow& n+1\\
f(z) &\rightarrow& f(\omega) = -\left(\log\omega+\frac{4\pi^2}{(n+1)\log\omega}\right)\\
g(z) &\rightarrow& g(\omega) = \left(\frac{-\log\omega}{2\pi}\right)^{12}\ ,
\end{eqnarray}
and it is readily found that $f'(\omega)=0$ gives $(\log\omega)^2=4\pi^2/(n+1)$. The negative 
square root then gives the saddle-point $\log\omega_0=-2\pi/(\sqrt{n+1})$, with
\begin{eqnarray}
\label{easyfind}
f''(\omega_0)&=&f''(\omega)\Big|_{\omega=e^{-2\pi/\sqrt{n+1}}}=\left(\frac{\sqrt{n+1}}{\pi}\right)e^{4\pi/\sqrt{n+1}}\\
g(\omega_0)&=&g(\omega)\Big|_{\omega=e^{-2\pi/\sqrt{n+1}}}=\left(\frac{1}{\sqrt{n+1}}\right)^{12}\ .
\end{eqnarray}
Furthermore as $f''(\omega_0)$ is real, $\arg f''(\omega_0)=0$, so $\alpha=\left(\pi-\arg f''(\omega_0)\right)/2=\pi/2$ and 
$e^{i\pi/2}=i$. Therefore, performing the saddle-point evaluation
\begin{eqnarray}
\label{largen1}
d_n\sim \frac{1}{\sqrt{2}}(n+1)^{-27/4}e^{4\pi\sqrt{n}}e^{-2\pi/\sqrt{n+1}}\ ,
\end{eqnarray}
and one thus finds that as $n\rightarrow\infty$ (where $n+1\sim n$ and $e^{-2\pi/\sqrt{n+1}}\rightarrow 1$):
\begin{eqnarray}
\label{largen}
d_n\sim n^{-27/4}e^{4\pi\sqrt{n}}\ .
\end{eqnarray}

Now, for our open strings, $M^2=(n-1)/\alpha'\sim n/\alpha'$ for large $n$. This implies that
$(E_n)^2\sim n/\alpha'\Rightarrow n\sim \alpha'E_n^2\Rightarrow dn\sim 2\alpha'E_ndE_n$, and we can
calculate the partition function:
\begin{eqnarray}
\label{partitionfunc}
Z=\sum_nd_ne^{-E_n/(k_BT)}&\rightarrow& \int\!dn\,d_ne^{-E_n/(k_BT)}\nonumber \\
&\sim&2\alpha'\int\!\,dE_n\,d_nE_ne^{-\beta E_n}\ ,
\end{eqnarray}
where $\beta=1/(k_BT)$. Thus:
\begin{eqnarray}
\label{zfinally}
Z&\sim&2\alpha'\int\!dE\,n^{-27/4}e^{4\pi\sqrt{n}}Ee^{-\beta E}\nonumber \\
&\sim&2\alpha'(\alpha')^{-27/4}\int\!dE\,E^{-27/2}Ee^{4\pi\sqrt{\alpha'}E-\beta E}\nonumber \\
&\sim&\int \!dE\,E^{-25/2}e^{(4\pi\sqrt{\alpha'}-\beta)E}\ .
\end{eqnarray}
The Hagedorn temperature in theoretical physics is the temperature above which the partition function diverges, and here it is 
clear that the integrand diverges as $4\pi\sqrt{\alpha'}\rightarrow\beta$. The Hagedorn temperature for the Bosonic string is thus:
\begin{eqnarray}
\label{hagedorntemp}
T_H=\frac{1}{4\pi k_B\sqrt{\alpha'}}\ .
\end{eqnarray}
This temperature may denote a limiting temperature of the theory or there may be a phase transition there. See also \cite{Zwiebach} for an
alternative approach to calculating the Hagedorn temperature.

\newpage

\subsection{Compactification and T-Duality}

Imagine we are in closed Bosonic string theory for a moment. This lives in 26 dimensions, and yet
we know from experience that we see only 4 dimensions. It would seem at first sight that string theory
is way off the mark. However, it is not inconsistent if the extra 22 spatial dimensions are too small
to be noticed in everyday life (compactified)\footnote{Note that another possibility is that our
4-dimensional world is some sort of a hypersurface in a larger 26-dimensional universe.}. So, as a
toy model, let us imagine that we compactify one of our space directions, $X^{25}$ say, on a circle of
radius $R$. This is about the simplest thing we can imagine doing.

From Chapter 4, we have:
\begin{eqnarray}
\label{total}
X^{\mu}(\sigma,\tau)&=&X_L^{\mu}(\sigma+\tau)+X_R^{\mu}(\sigma-\tau)\ ;\\
\label{left}
X_{L}^{\mu} &=& \frac{1}{2}x^{\mu} +
\frac{1}{2}{\alpha}'p_{L}^{\mu}{\xi}^{+} +
i\sqrt{\frac{\alpha'}{2}}\sum_{n \neq
0}{\frac{{\tilde{\alpha}}_{n}^{\mu}}{n}e^{-in{\xi}^{+}}}\ ,\\
X_{R}^{\mu} &=& \frac{1}{2}x^{\mu} +
\frac{1}{2}{\alpha}'p_{R}^{\mu}{\xi}^{-} +
i\sqrt{\frac{\alpha'}{2}}\sum_{n \neq
0}{\frac{{\alpha}_{n}^{\mu}}{n}e^{-in{\xi}^{-}}}\ ,
\end{eqnarray}
where $0\leq\sigma\leq 2\pi$. And because of our compactification we require:
\begin{eqnarray}
\label{compactcondition}
X^{25}(\sigma,\tau)&=&X^{25}(\sigma+2\pi,\tau)+2\pi Rn^{25}\ ;\\
\label{noncompactcondition}
X^{\nu}(\sigma,\tau)&=&X^{\nu}(\sigma+2\pi,\tau)\quad \nu=0,1,\ldots,24\ ,
\end{eqnarray}
where $n^{25}$ is the winding number of the (compactified) 25 direction. So, as:
\begin{eqnarray}
\label{whatisx25}
X^{25}(\sigma+2\pi,\tau)=x^{25}&+&\frac{\alpha'}{2}p_L^{25}\xi^+
+\frac{\alpha'}{2}p_R^{25}\xi^-+\frac{\alpha'}{2}2\pi(p_L^{25}-p_R^{25})\nonumber\\
\qquad\qquad&+&i\sqrt{\frac{\alpha'}{2}}\sum_{n\neq 0}\frac{1}{n}
\left[\alpha_n^{25}e^{-in\xi^+}+{\tilde{\alpha}}_n^{25}e^{-in\xi^-}\right]
\nonumber\\
=X^{25}(\sigma,\tau)&+&\pi\alpha'(p_L^{25}-p_R^{25})\ ,
\end{eqnarray}
\begin{eqnarray}
\label{implies}
\Rightarrow p_L^{25}-p_R^{25}=-\frac{2Rn^{25}}{\alpha'}\ .
\end{eqnarray}
Furthermore, from the invariance of the wavefunction $e^{ip\cdot X}$ under
$X^{25}\rightarrow X^{25}+2\pi Rn^{25}$, we must have that $p^{25}=m^{25}/R$.
Identifying $p^{\mu}$ with $(p_L^{\mu}+p_R^{\mu})/2$ for $\mu=0,1,\ldots,25$,
we can therefore see that
\begin{eqnarray}
\label{pleftright}
p_L^{25}&=&\frac{m^{25}}{R}-\frac{Rn^{25}}{\alpha'}\ ;\nonumber \\
p_R^{25}&=&\frac{m^{25}}{R}+\frac{Rn^{25}}{\alpha'}\ .
\end{eqnarray}

Recall now that we have a Virasoro algebra with $L_n$ and $\tilde{L}_n$ as usual,
where $\alpha_0^{\mu}=\sqrt{\alpha'/2}\,p_L^{\mu}$,
$\tilde{\alpha}_0^{\mu}=\sqrt{\alpha'/2}\,p_R^{\mu}$ and $L_0=N+{\alpha}_0^2/2$ and
$\tilde{L}_0=\tilde{N}+\tilde{\alpha}_0^2$. From the $L_0-\tilde{L}_0$ physical
state condition, we can easily see that
\begin{eqnarray}
\label{physstatecompact}
\left(N+\tilde{N}+\frac{\alpha'}{4}\left(p_L^{\nu}p_{L\,\nu}-p_R^{\nu}p_{R\,\nu}\right)+
\frac{\alpha'}{4}\left({(p_L^{25})}^2-{(p_R^{25})}^2\right)\right)|\phi\rangle=0\ ,
\end{eqnarray}
where $\nu\neq 25$. As $p_L^{\nu}=p_R^{\nu}$, we can thus deduce that physical states must
satisfy the `level-matching' condition
\begin{eqnarray}
\label{levelmatch}
\left(N-\tilde{N}-m^{25}n^{25}\right)|\phi\rangle=0\ .
\end{eqnarray}

If we also consider the $L_0$ and $\tilde{L}_0$ conditions separately, we quickly find:
\begin{eqnarray}
\label{lolotilde}
\left(\frac{N+\tilde{N}}{2}+\frac{{(m^{25})}^2\alpha'}{4R^2}+\frac{{(n^{25})}^2R^2}{4\alpha'}+
\frac{\alpha'}{4}p_{A}^{\nu}p_{A\,\nu}-1\right)|\phi\rangle\ ,
\end{eqnarray}
where $A$ is $L$ or $R$ and is \textit{not} summed over and again $\nu\neq 25$.
And since $p_L^{\nu}=p_R^{\nu}=q^{\nu}$, it is really the same equation and tells us
that the 25-dimensional mass spectrum is:
\begin{eqnarray}
\label{uncompactmassspec}
M^2=-q^2=\frac{2(N+\tilde{N})}{\alpha'}-\frac{4}{\alpha'}+\frac{{(m^{25})}^2}{R^2}
+\frac{{(n^{25})}^2R^2}{\alpha'^2}\ .
\end{eqnarray}

This mass spectrum is clearly invariant under $R\leftrightarrow \alpha'/R$ and
$m^{25}\leftrightarrow n^{25}$. In fact as $\alpha_0$ and $\tilde{\alpha}_0$ are
related to $p_L$ and $p_R$, we have that
\begin{eqnarray}
\label{alphas}
\alpha_0^{25}&=&\left(\frac{m^{25}}{R}-\frac{Rn^{25}}{\alpha'}\right)\sqrt{\frac{\alpha'}{2}}\ ;\nonumber \\
\tilde{\alpha}_0^{25}&=&\left(\frac{m^{25}}{R}+\frac{Rn^{25}}{\alpha'}\right)\sqrt{\frac{\alpha'}{2}}\ ,
\end{eqnarray}
and thus we must also send $\alpha_0^{25}\rightarrow-\alpha_0^{25}$. This collection of
exchanges is known as T-dualising, and the theory is invariant under it. It is a symmetry of
the interacting theory if the coupling constant is also re-scaled $(e^{\Phi}\rightarrow
\sqrt{\alpha'}e^{\Phi}/R)$.

\newpage

\section{Central Extension}\label{VirasoroAppendix}

In this appendix we use methods similar to those in \cite{Lust} to derive the full quantum Virasoro algebra for the Bosonic string. 
See also \cite{GSW1,Polchinski1,Zwiebach} for alternatives. These techniques can 
similarly be applied to calculate the quantum Virasoro algebra of the superstring, although a far more elegant and 
less unwieldy calculation uses the operator product expansion of CFT.

\subsection{Bosonic Algebra}\label{Bosonicalgebra}

Classically the Virasoro generators are given by
\begin{eqnarray}
L_m=\frac{1}{2}\sum_{n=-\infty}^{\infty}\alpha_{m-n}\cdot\alpha_n\ ,
\end{eqnarray}
but quantum mechanically we must use the normal ordered expression
\begin{eqnarray}
L_m=\frac{1}{2}\sum_{n=-\infty}^{\infty}:\alpha_{m-n}\cdot\alpha_n:
\end{eqnarray}
As can be seen in the above expression, there is some potential ambiguity in the normal ordering 
depending on how the value of $m$ compares with the value of $n$. For example, if $m>n>0$ there 
is seemingly little to say which ordering of operators we should choose. Similarly for $m<n<0$.
To avoid this we can define our normal ordering as follows:
\begin{eqnarray}\label{howtonormalorder}
:\alpha_{p}^{\mu}\alpha_q^{\nu}: \ =\ \left\{
\begin{array}{ll} \alpha_{p}^{\mu}\alpha_q^{\nu} & q\geq 0 \\  \alpha_{q}^{\mu}\alpha_p^{\nu} & q<0\ ,
\end{array} \right. 
\end{eqnarray}

In order to work out the full quantum Virasoro algebra we may begin as in Section \ref{virasalgebra} by showing 
the useful relation:
\begin{eqnarray}
\left[L_m,\alpha_n^{\mu}\right]=-n\alpha_{m+n}^{\mu}\ ,
\end{eqnarray}
which also holds in the quantum theory when $L_{m}$ is normal ordered.
Using this our aim will be to compute the commutator while explicitly ordering the $\alpha$'s, much in the 
same spirit as the computation of the number operator earlier in the book and in Section \ref{normalorderingsection}. Thus:
\begin{eqnarray}
\label{startoff}
\left[L_m,L_n\right] &=&\left[\frac{1}{2}\sum_{p=-\infty}^{\infty}:\alpha_{m-p}\cdot\alpha_p:,L_n\right]\nonumber\\
&=&\frac{1}{2}\sum_{p=-\infty}^{-1}\left[\alpha_{p}\cdot\alpha_{m-p},L_n\right]+\frac{1}{2}\sum_{p=0}^{\infty}\left[\alpha_{m-p}\cdot\alpha_{p},L_n\right]\nonumber\\
&=&\frac{1}{2}\sum_{p=-\infty}^{-1}\Big((m-p)\alpha_p\cdot\alpha_{m+n-p}+p\,\alpha_{n+p}\cdot\alpha_{m-p}\Big)\nonumber\\
&{}&+\frac{1}{2}\sum_{p=0}^{\infty}\Big(p\,\alpha_{m-p}\cdot\alpha_{n+p}+(m-p)\alpha_{m+n-p}\cdot\alpha_{p}\Big)\ ,
\end{eqnarray}
where in the second line we have explicitly normal-ordered $L_m$ and in the third line we have expanded the commutators and used the 
$[L_m,\alpha_n]$ commutator to rewrite the terms. Now change variables in the second and third terms of (\ref{startoff}) to $q=n+p$ and then re-label 
$q$ as $p$. We therefore arrive at:
\begin{eqnarray}
\label{startoff2}
\left[L_m,L_n\right]&=&\frac{1}{2}\Bigg(\sum_{p=-\infty}^{-1}(m-p)\alpha_p\cdot\alpha_{m+n-p}+\sum_{p=-\infty}^{n-1}(p-n)\,\alpha_{p}\cdot\alpha_{n+m-p}\Bigg)\nonumber\\
&{}&+\frac{1}{2}\Bigg(\sum_{p=n}^{\infty}(p-n)\,\alpha_{n+m-p}\cdot\alpha_{p}+\sum_{p=0}^{\infty}(m-p)\alpha_{m+n-p}\cdot\alpha_{p}\Bigg)\ .
\end{eqnarray}
Consider the case $n\leq 0$. In this case, it is clear that terms 1, 2 and 4 of the above expression are already normal-ordered. Part of the third term, however, is not. Specifically 
the $\sum_{p=n}^{-1}$. Thus
\begin{eqnarray}
\label{startoff3}
\sum_{p=n}^{\infty}(p-n)\,\alpha_{n+m-p}\cdot\alpha_{p} &=& \sum_{p=n}^{-1}(p-n)\,\alpha_{n+m-p}\cdot\alpha_{p}\ +\  
\sum_{p=0}^{\infty}(p-n)\,\alpha_{n+m-p}\cdot\alpha_{p}\nonumber\\
&=& \sum_{p=n}^{-1}(p-n)\,\Big(\alpha_{p}\cdot\alpha_{n+m-p}+\eta_{\mu\nu}[\alpha^{\mu}_{n+m-p},\alpha^{\nu}_{p}]\Big)\nonumber\\
 &{}&+\  \sum_{p=0}^{\infty}(p-n)\,\alpha_{n+m-p}\cdot\alpha_{p}\nonumber\\
 &=& \sum_{p=n}^{-1}(p-n)\,\alpha_{p}\cdot\alpha_{n+m-p}+\sum_{p=n}^{-1}(p-n)(n+m-p)\delta_{n+m}\eta_{\mu\nu}\eta^{\mu\nu}\nonumber\\
 &{}&+\  \sum_{p=0}^{\infty}(p-n)\,\alpha_{n+m-p}\cdot\alpha_{p}\ .\nonumber\\
\end{eqnarray}
Now, adding the third term of (\ref{startoff3}) to the fourth term of (\ref{startoff2}) straightforwardly gives
\begin{eqnarray}
\label{startpositivesum}
\sum_{p=0}^{\infty}(m-n)\,\alpha_{n+m-p}\cdot\alpha_{p}\ .
\end{eqnarray}
Similarly, adding the first term of (\ref{startoff3}) to the second term of (\ref{startoff2}) straightforwardly gives
\begin{eqnarray}
\label{startneg1}
\sum_{p=-\infty}^{-1}(p-n)\,\alpha_{p}\cdot\alpha_{n+m-p}\ ,
\end{eqnarray}
which can be combined with the first term of (\ref{startoff2}) to give
\begin{eqnarray}
\label{startneg2}
\sum_{p=-\infty}^{-1}(m-n)\,\alpha_{p}\cdot\alpha_{n+m-p}\ .
\end{eqnarray}
(\ref{startneg2}) and (\ref{startpositivesum}) then combine together (along with the overall factor of $1/2$ from (\ref{startoff2})) to give
\begin{eqnarray}
\label{finalnormo}
\frac{1}{2}\sum_{p=-\infty}^{\infty}(m-n)\, :\alpha_{n+m-p}\cdot\alpha_{p}:\ =\ L_{m+n}\ .
\end{eqnarray}
Lastly, note that $\eta_{\mu\nu}\eta^{\mu\nu}=d$ (the dimension of spacetime) and that the $\delta_{n+m}$ in the second term of (\ref{startoff3}) imposes 
$n=-m$. This means that as we have assumed $n<0$, we must have $m>0$, and this term can be dealt with as follows:
\begin{eqnarray}
\sum_{p=n}^{-1}(p-n)(n+m-p)\delta_{n+m}\eta_{\mu\nu}\eta^{\mu\nu} &=& \sum_{p=-m}^{-1}d(p+m)(-p)\delta_{n+m}\nonumber\\
&=& \sum_{-q=-m}^{-1}d(-q+m)(q)\delta_{n+m}\nonumber\\
&=& \sum_{q=m}^{1}d(mq-q^2)\delta_{n+m}\nonumber\\
&=& \sum_{q=1}^{m}d(mq-q^2)\delta_{n+m}\nonumber\\
&=& dm\delta_{n+m}\sum_{q=1}^{m}q-d\delta_{n+m}\sum_{q=1}^{m}q^2\nonumber\\
&=& d\delta_{n+m}\Big(m\frac{1}{2}m(m+1)-\frac{1}{6}m(m+1)(2m+1)\Big)\nonumber\\
&=& \frac{d}{6}m\delta_{n+m}\Big(3m^2+3m-2m^2-3m-1\Big)\nonumber\\
&=& \frac{d}{6}(m^3-m)\delta_{n+m}\ .
\end{eqnarray}
Put this together with the factor of $1/2$ from (\ref{startoff2}) and equation (\ref{finalnormo}) we have:
\begin{eqnarray}
\left[L_m,L_n\right]\ =\ L_{m+n} + \frac{d}{12}(m^3-m)\delta_{m+n}\ ,
\end{eqnarray}
as required.

Strictly speaking, we only know for sure that this applies when $n<0$, since that is what we started off assuming. When $n> 0$, we may go 
back to (\ref{startoff2}) and note that in this case, terms 1, 3 and 4 are already normal-ordered. Similarly to the case $n\leq 0$, we may then split the 
second term up into a $\sum_{-\infty}^{-1}$ and a $\sum_{0}^{n-1}$. Explicitly normal-ordering the second of these two terms produces two further terms, one of 
which combines with the $\sum_{-\infty}^{-1}$ and then with terms 1, 3 and 4 to produce the standard $L_{m+n}$ term in the algebra. The other term yields
\bea
\frac{1}{2}\sum_{p=0}^{n-1}p(p-n)d\delta_{m+n}\ ,\nn
\eea
which can be straightforwardly shown to give the same anomaly as before.

\subsection{Ghost Algebra}

Recall that quantum mechanically we have the following non-zero (anti)-commutation relations:
\bea
\{c_m,b_n\}=\delta_{m+n}\ ,
\eea
and the normal ordered expression for $L_{m}^{(bc)}$
\bea
L_m^{(bc)}=\sum_{p=-\infty}^{\infty} :b_{m+p}c_{-p}:\ ,
\eea
where similarly to $:\alpha_i\alpha_j :$
\begin{eqnarray}
:b_ic_j: \ =\ \left\{
\begin{array}{ll} b_ic_j & j\geq 0 \\  -c_jb_i & j<0\ ,
\end{array} \right. 
\end{eqnarray}
although there is now a minus sign in the normal ordering as the ghosts are Fermionic. Using these, one can show the 
useful results that
\bea
\left[L_{m}^{(bc)},b_n\right] &=& (m-n)b_{m+n}\ ,\\
\left[L_{m}^{(bc)},c_n\right] &=& -(2m+n)c_{m+n}\ .
\eea

Proceeding in the same fashion as in Section \ref{Bosonicalgebra}:
\bea
\left[L_m^{(bc)},L_n^{(bc)}\right] &=& \sum_{p=-\infty}^{0}(m-p)(2n-p)b_{m+p}c_{n-p} -\sum_{p=-\infty}^0(m-p)(n-m-p)b_{m+n+p}c_{-p}\nn\\
&{}& +\sum_{p=1}^{\infty}(m-p)(n-m-p)c_{-p}b_{m+n+p}-\sum_{p=1}^{\infty}(m-p)(2n-p)c_{n-p}b_{m+p}\nn\ ,
\eea
which upon re-labelling $p-n\rightarrow p$ in terms 1 and 4 becomes
\bea
\label{ghostsnormal}
\left[L_m^{(bc)},L_n^{(bc)}\right] &=& \sum_{p=-\infty}^{-n}(m-n-p)(n-p)b_{m+n+p}c_{-p} -\sum_{p=-\infty}^0(m-p)(n-m-p)b_{m+n+p}c_{-p}\nn\\
&{}& +\sum_{p=1}^{\infty}(m-p)(n-m-p)c_{-p}b_{m+n+p}-\sum_{p=-n+1}^{\infty}(m-n-p)(n-p)c_{-p}b_{m+n+p}\nn\ .
\eea
Now if $n<0$, terms 2, 3 and 4 are all normal ordered already. Term 1 can be split up into two parts, one $\sum_{-\infty}^{0}$ and one $\sum_{1}^{-n}$, and 
when the second of these is explicitly normal ordered, one part contributes to the standard commutator along with the $\sum_{-\infty}^{0}$ and terms 2, 3 and 4 of 
(\ref{ghostsnormal}) thus:
\bea
&{}&\sum_{p=-\infty}^0\left((m-n-p)(n-p)+(m+p-n)(m-p)\right)b_{m+n+p}c_{-p}\nn\\
&{}&\qquad-\sum_{p=1}^{\infty}\left((m+p-n)(m-p)+(m-n-p)(n-p)\right)c_{-p}b_{m+n+p}\nn\\
&{}&\qquad\qquad\qquad\qquad\qquad\qquad\qquad\qquad\qquad =(m-n)\sum_{p=-\infty}^{\infty}(m+n-p): b_{m+n+p}c_{-p}:\nn\\
&{}&\qquad\qquad\qquad\qquad\qquad\qquad\qquad\qquad\qquad =(m-n)L_{m+n}^{(bc)}\ .
\eea
The other term generated by explicit normal ordering yields
\bea
\sum_{p=1}^{-n}(m-n-p)(n-p)\delta_{m+n} = \frac{1}{6}(m-13m^3)\delta_{m+n}\ .
\eea
A similar calculation for $n>0$ gives the same, thus verifying that
\bea
\left[L_m^{(bc)},L_n^{(bc)}\right] = (m-n)L_{m+n}^{(bc)}+\frac{1}{6}(m-13m^3)\delta_{m+n}\ .
\eea

\newpage

\section{Orientifolds}\label{orientifoldappendix}

This appendix contains an introduction to world-sheet parity and
orientifolds. Topics covered include some explanation of the
basics, $D3$-brane spectra, the $D5$-$D9$ setup in Type I and the
$D3$-$D7$ setup in Type IIB. Some of this follows \cite{Polchinski1,Polchinski2} 
and \cite{Uranga}, and a (slightly) greater familiarity with $D$-branes is assumed than that 
touched upon in the body of these notes. Readers are also referred in particular to 
\cite{Gimon,Dabholkar,Giveon}.

Unless otherwise specified closed strings are treated as having
$0\leq\sigma\leq 2\pi$ and open strings as having
$0\leq\sigma\leq\pi$. So, summarising the results from Chapter 4, for \textbf{closed} strings we can write
the solutions to the 2-d wave-equation as a mode-expansion:
\begin{eqnarray}
\label{closed1} X^{\mu} = x^{\mu} + {\alpha}'p^{\mu}\tau +
i\sqrt{\frac{\alpha'}{2}}\sum_{n \neq
0}{\frac{1}{n}\left({\alpha}_{n}^{\mu}e^{-in(\tau-\sigma)} +
{\tilde{\alpha}}_{n}^{\mu}e^{-in(\tau+\sigma)}\right)}.
\end{eqnarray}
For \textbf{open} strings, on the other hand, we have the
independent choices of Neuman \mbox{$({\acute{X}}^{\mu}=0)$}, or
Dirichlet $({\dot{X}}^{\mu} = 0)$. The possibilities are then:
\begin{itemize}
\item With NN boundary conditions
\begin{eqnarray}
\label{neumann1} X^{\mu} = x^{\mu} + 2{\alpha}'p^{\mu}\tau +
i\sqrt{2{\alpha}'}\sum_{n \neq
0}{\frac{{\alpha}_{n}^{\mu}}{n}e^{-in\tau}\cos{n\sigma}}.
\end{eqnarray}
\item With DD boundary conditions
\begin{eqnarray}
\label{dirichlet1} X^{\mu} = a^{\mu} + \frac{1}{\pi}(b^{\mu} -
a^{\mu})\sigma + \sqrt{2{\alpha}'}\sum_{n \neq
0}{\frac{{\alpha}_{n}^{\mu}}{n}e^{-in\tau}\sin{n\sigma}}.
\end{eqnarray}
\item With ND boundary conditions
\begin{eqnarray}
\label{ND1} X^{\mu} = b^{\mu} + i\sqrt{2{\alpha}'}\sum_{r \in
\,\mathbb{Z} +
\frac{1}{2}}{\frac{{\alpha}_{r}^{\mu}}{r}e^{-ir\tau}\cos{r\sigma}}.
\end{eqnarray}
\item With DN boundary conditions
\begin{eqnarray}
\label{DN1} X^{\mu} = a^{\mu} + \sqrt{2{\alpha}'}\sum_{r \in
\,\mathbb{Z} +
\frac{1}{2}}{\frac{{\alpha}_{r}^{\mu}}{r}e^{-ir\tau}\sin{r\sigma}}.
\end{eqnarray}
\end{itemize}
This is the Bosonic sector, common to Bosonic strings and
superstrings.

For the superstring, we also have the $\psi^{\mu}$ which are
spacetime Bosons and worldsheet Fermions\footnote{What we're
really saying here is that they're worldsheet spinors and
spacetime vectors which also anticommute. From that point of view
they're worldsheet Fermions and something strange in spacetime. In
fact one of the ideas of the GSO projection is to deal with their
strange spacetime properties.}. From the 2-d point of view they
are thus 2-component spinors and as such we can split them into a
left-moving part $(\psi_L)$ and a right-moving part $(\psi_R)$.
The massless Dirac equation, which the $\psi^{\mu}$ obey,
necessarily dictates that the $\psi_{L,R}$ are left- and
right-moving respectively.

For the \textbf{open} superstring, vanishing of the surface term
from varying the action can be satisfied by making the choice
$\psi_R=\pm\psi_L$ and $\delta\psi_R=\pm\delta\psi_L$ at each end.
The overall relative sign is a matter of convention, so without
loss of generality we set $\psi_R(0,\tau)=\psi_L(0,\tau)$. The
relative sign at the other end is now meaningful and there are two
cases to consider:

\begin{enumerate}
\item \textbf{Ramond} (R) boundary conditions are:
\begin{eqnarray}
\label{r} \psi_{R}(\pi,\tau)=\psi_L(\pi,\tau)\ ,
\end{eqnarray}
so that the mode expansions in the Ramond sector are
\begin{eqnarray}
\label{rmode1}
\psi_R^{\mu}(\sigma,\tau)&=&\sum_{n\in \ \mathbb{Z}}\psi_{n}^{\mu}e^{-in(\tau-\sigma)}\ ;\\
\psi_L^{\mu}(\sigma,\tau)&=&\sum_{n\in \
\mathbb{Z}}\psi_{n}^{\mu}e^{-in(\tau+\sigma)}\ ,
\end{eqnarray}
where the sums run over all integers $n$. \item
\textbf{Neveu-Schwarz} (NS) boundary conditions on the other hand
are:
\begin{eqnarray}
\label{ns} \psi_{R}(\pi,\tau)=-\psi_L(\pi,\tau)\ ,
\end{eqnarray}
so that in the Neveu-Schwarz sector the mode expansions are
\begin{eqnarray}
\label{nsmode1}
\psi_R^{\mu}(\sigma,\tau)&=&\sum_{r\in \ \mathbb{Z}+\frac{1}{2}}\psi_{r}^{\mu}e^{-ir(\tau-\sigma)}\ ;\\
\psi_L^{\mu}(\sigma,\tau)&=&\sum_{r\in \
\mathbb{Z}+\frac{1}{2}}\psi_{r}^{\mu}e^{-ir(\tau+\sigma)}\ ,
\end{eqnarray}
where now the sums run over the half-integers. We will often use
$m$ and $n$ to represent integers in these sums and $r$ and $s$ to
represent half-integers\footnote{Note that the Ramond b.c.'s and
integer modes are appropriate to the description of string states
which are space-time Fermions, whereas the Neveu-Schwarz b.c.'s and
half-integer modes give rise to space-time Bosons. These Bosonic
states are, of course, different to those of the Bosonic string.}.
\end{enumerate}
It is very important to note that the above solutions only apply
when you have NN or DD b.c.'s in the $X$-sector. If you have ND or
DN b.c.'s there then it turns out that the Ramond sector now has
half-integer mode expansions (and so describes spacetime Bosons),
while the NS sector now has integer mode expansions and thus
describes spacetime Fermions. These sorts of changes of property
of the integer/half-integer nature of the mode expansions are
already clearly present in the forms of the solutions for the
$X$-fields in changing NN or DD b.c.'s to ND or DN b.c.'s -- see
equations (\ref{neumann1})-(\ref{DN1}).

For \textbf{closed} superstrings, the surface terms vanish when
the boundary conditions are periodicity or anti-periodicity for
each component of $\psi$ separately. We can thus have:
\begin{eqnarray}
\label{rightclosedmodes1}
\psi_R^{\mu}=\sum\psi_n^{\mu}e^{-in(\tau-\sigma)}\quad
\textrm{or}\quad \psi_R^{\mu}=\sum\psi_r^{\mu}e^{-ir(\tau-\sigma)}
\end{eqnarray}
and
\begin{eqnarray}
\label{rightclosedmodes1}
\psi_L^{\mu}=\sum\tilde{\psi}_n^{\mu}e^{-in(\tau+\sigma)}\quad
\textrm{or}\quad
\psi_L^{\mu}=\sum\tilde{\psi}_r^{\mu}e^{-ir(\tau+\sigma)}\ .
\end{eqnarray}
Then there are four distinct closed-string sectors corresponding
to the different pairings of left-moving and right-moving modes
that can be referred to as NS-NS, NS-R, R-NS and R-R. The first
and the last cases describe Bosonic states and the other two
describe Fermions. Again these closed-string solutions are given with
the assumption of NN or DD b.c.'s in the $X$-sector.

We'll be mostly interested in the open-string states and it may be useful to recall that the $\alpha_{-n/r}$ and
$\psi_{-n/r}$ act as raising operators on a ground state and thus
create the states of the string.

\subsection{Worldsheet Parity}

The action of the world-sheet parity operator is:
\begin{eqnarray}
\label{parity} \Omega:\sigma\rightarrow l-\sigma\,\, ; \,\,
\tau\rightarrow \tau\ ,
\end{eqnarray}
where $l$ is the periodicity in question, \emph{i.e.} $2\pi$ for closed
strings and $\pi$ for open strings. If we wish for the $X$'s and
$\psi$'s to be invariant under this operation, it is
clear that we must have
\begin{eqnarray}
\label{closedparity}
\Omega:\alpha_n^{\mu}\leftrightarrow\tilde{\alpha}_n^{\mu}\ ,
\end{eqnarray}
for closed strings. Open strings will then transform as
\begin{itemize}
\item NN
\begin{eqnarray}
\label{NNomega} \Omega:\alpha_n^{\mu}\rightarrow
(-1)^n\alpha_n^{\mu}.
\end{eqnarray}
\item DD
\begin{eqnarray}
\label{NNomega} \Omega:\alpha_n^{\mu}\rightarrow
(-1)^n\alpha_n^{\mu}\,\, ; \,\, a^{\mu}\leftrightarrow b^{\mu}.
\end{eqnarray}
The exchange of $a^{\mu}$ and $b^{\mu}$ here is something like the
exchange of the Chan-Paton factors as $a^{\mu}$ and $b^{\mu}$ are
the locations of the ends of the DD string. \item ND and DN
\begin{eqnarray}
\label{NDomega} \textrm{ND}\qquad\Omega:\alpha_r^{\mu}&\rightarrow
&
e^{i\pi r}\alpha_r^{\mu}\nonumber \\
\label{DNomega} \textrm{DN}\qquad\Omega:\alpha_r^{\mu}&\rightarrow
& e^{-i\pi r}\alpha_r^{\mu}
\end{eqnarray}
and the ND and DN strings are interchanged. \item R \& NS for open strings
\begin{eqnarray}
\label{RNSomega}
\Omega:\psi_n^{\mu}&\rightarrow &(-1)^n\psi_n^{\mu}\\
\Omega:\psi_r^{\mu}&\rightarrow &e^{\pm i\pi r}\psi_r^{\mu}\ ,
\end{eqnarray}
\end{itemize}
where the sign in the second relation is correlated with whether
you are a left- or right-mover. Left-movers and right-movers are
interchanged in the R and NS sectors too.

\subsubsection{Normal-Tangential Sign Difference}
There is actually
a minus sign difference between NN strings and DD
strings\footnote{See \textit{e.g.} page 197 of \cite{Polchinski2}.}. This
is seemingly a great deal easier to understand from the viewpoint of CFT. We know that
in CFT, states such as $\alpha_{-n/r}^{\mu}|0;p\rangle$ and correspond to vertex operators such
as $\int\partial_{\alpha}X^{\mu}e^{ik\cdot X}$. In this expression, $\alpha$ is either $\tau$ if $\mu$
is a direction with NN b.c.'s, or $\sigma$ if $\mu$ is a direction
with DD b.c.'s. The basic idea is that $\tau$ derivatives should
pick up a sign under worldsheet parity while $\sigma$ derivatives
should not. The diagrams of the string worldsheet below give an 
idea of why that should be.

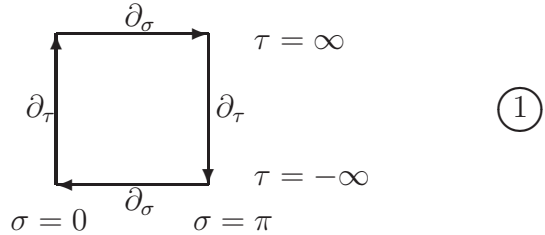
\begin{center}
\setlength{\unitlength}{2cm}
\begin{picture}(3,3)
\thicklines \put(1,1){\vector(0,1){1}} \put(1,2){\vector(1,0){1}}
\put(2,2){\vector(0,-1){1}} \put(2,1){\vector(-1,0){1}}
\put(0.7,0.7){$\sigma =0$} \put(1.9,0.7){$\sigma =\pi$}
\put(2.3,1.0){$\tau =-\infty$} \put(2.3,1.9){$\tau =\infty$}
\put(1.45,0.83){$\partial_{\sigma}$}
\put(1.45,2.05){$\partial_{\sigma}$}
\put(0.8,1.45){$\partial_{\tau}$}
\put(2.05,1.45){$\partial_{\tau}$} \put(4,1.45){$1$}
\put(4.05,1.5){\circle{0.3}}
\end{picture}

\setlength{\unitlength}{2cm}
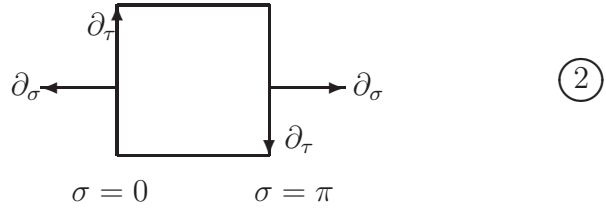
\begin{picture}(3,3)
\thicklines \put(1,1){\vector(0,1){1}} \put(1,2){\line(1,0){1}}
\put(2,2){\vector(0,-1){1}} \put(2,1){\line(-1,0){1}}
\put(1,1.45){\vector(-1,0){0.5}} \put(2,1.45){\vector(1,0){0.5}}
\put(0.7,0.7){$\sigma =0$} \put(1.9,0.7){$\sigma =\pi$}
%\put(2.3,1.0){$\tau =-\infty$}
%\put(2.3,1.9){$\tau =\infty$}
\put(0.3,1.4){$\partial_{\sigma}$}
\put(2.55,1.4){$\partial_{\sigma}$}
\put(0.8,1.8){$\partial_{\tau}$} \put(2.1,1.05){$\partial_{\tau}$}
\put(4,1.45){$2$} \put(4.05,1.5){\circle{0.3}}
\end{picture}

\setlength{\unitlength}{2cm}
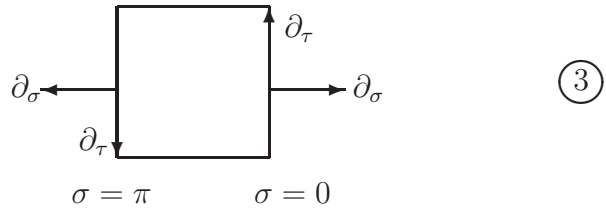
\begin{picture}(3,3)
\thicklines \put(1,2){\vector(0,-1){1}} \put(1,2){\line(1,0){1}}
\put(2,1){\vector(0,1){1}} \put(2,1){\line(-1,0){1}}
\put(1,1.45){\vector(-1,0){0.5}} \put(2,1.45){\vector(1,0){0.5}}
\put(0.7,0.7){$\sigma =\pi$} \put(1.9,0.7){$\sigma =0$}
%\put(2.3,1.0){$\tau =-\infty$}
%\put(2.3,1.9){$\tau =\infty$}
\put(0.3,1.4){$\partial_{\sigma}$}
\put(2.55,1.4){$\partial_{\sigma}$}
\put(0.75,1.05){$\partial_{\tau}$}
\put(2.1,1.8){$\partial_{\tau}$} \put(4,1.45){$3$}
\put(4.05,1.5){\circle{0.3}}
\end{picture}
\end{center}

Diagram 2 is a re-drawing of diagram 1 appropriate to focussing on
the spatial boundary of the worldsheet and the normal/tangential
derivatives there. Then, diagram 3 is diagram 2 after worldsheet
parity. Note that in going from diagrams 1 to 2 a right-handed (say) corkscrew following the
arrows pointing into the page has been maintained. In 1 this applies to following the
arrows round the edge of the worldsheet while in 2 it applies to
the `local' arrows at the spatial boundaries. Note also that we
would find the same thing (\textit{i.e.} a relative sign between
normal and tangential derivatives) if we were to focus on the
timelike boundary and what happens there under worldsheet parity.

If you imagine a string world-sheet with $\sigma=0$ and
$\sigma=\pi$ as its spatial boundaries, then it is fairly clear
that $\partial_{\tau}$ is the derivative tangential to these
boundaries while $\partial_{\sigma}$ is the derivative
perpendicular to them. You can think of it a bit like
$\partial_{\tau}$ being the time-evolution operator on the
world-sheet, which clearly points in the $\tau$ direction and thus
is parallel to the spatial boundaries. Similarly
$\partial_{\sigma}$ evolves you in space on the world-sheet and is
thus perpendicular to the spatial boundaries. We should also be
careful to pick a consistent orientation so that if
$\partial_{\tau}$ is future-pointing on the $\sigma=0$ boundary it
should be past-pointing on the $\sigma=\pi$ boundary. One should
be able to `follow the derivatives' round the edge of the
world-sheet consistently. If you imagine a world-sheet with
$\sigma$ going from left to right with a range from 0 to $\pi$,
and with $\tau$ going `up' from $-\infty$ to $\infty$, then
picking $\partial_{\tau}$ pointing towards the future (up) at
$\sigma=0$ means that it should point towards the past at
$\sigma=\pi$. $\partial_{\sigma}$ should point from left to right
at the time-like boundary of $\tau=\infty$ and from right to left
at the time-like boundary of $\tau=-\infty$. We can then follow
the arrows round the boundary of the world-sheet in a clockwise
fashion.

All this is a very long-winded way of saying that while
the tangential derivative at the spatial boundaries must point in
opposite directions on each, the normal derivative will point in
the same direction (\emph{i.e.} either both away from the world-sheet or
both into the world-sheet). Now, under parity we basically
interchange $\sigma=0$ and $\sigma=\pi$. This means that where
$\partial_{\tau}$ was pointing up it will now be pointing down and
vice-versa. $\partial_{\tau}\rightarrow -\partial_{\tau}$. The
perpendicular derivative will, however, remain unchanged.
$\partial_{\sigma}\rightarrow \partial_{\sigma}$. This indicates 
that NN and DD b.c.'s should behave differently -- by a
sign -- under world-sheet parity. See \cite{Polchinski2}, pages 196-197 for
more.

\subsubsection{States}
As per \cite{Polchinski1}, pages 189-192, consider
the open Bosonic string for a moment with only NN b.c.'s (DD or
NN+DD b.c.'s will work in a very similar way except we should be a
little careful about the different sectors and such like). As
mentioned before, the states are created by the $\alpha_{-n}^i$
\footnote{Here $i$ runs from 1 to 25 (1 to 8 for the superstring)
which gives us the physical states via light-cone gauge quantization.}. This essentially means that the states are
characterized by the number operator
$N=\sum_{n=1}^{\infty}\alpha_{-n}^i\alpha_n^i$. When this operator
acts on a state, it returns a number times that state. So, for the
ground state this number is 0. For the first excited state,
created by $\alpha_{-1}$ it is 1. For the second excited state,
created by $\alpha_{-1}\alpha_{-1}$ and $\alpha_{-2}$, it is 2.
It's not hard to see that a state created by a `maximal' operator
of $\alpha_{-n}$ (and thus other stuff like
$\alpha_{-n+1}\alpha_{-1}$ \emph{etc.}) has an eigenvalue $n$ of $N$.

It is clear from this
and the above discussion that the action of $\Omega$ on a general
state is then
\begin{eqnarray}
\label{stateomega} \Omega:|n;k\rangle\rightarrow
(-1)^{-n}|n;k\rangle\ .
\end{eqnarray}
So, the eigenvalue of $\Omega$ is $(-1)^{-n}$, which we'll
call $\omega_n$. We can re-write this using the mass-shell
relation $M^2=-(1-n)/\alpha'$ \footnote{This is for the Bosonic
sector and 26-dimensions of course. With the superstring and 10
dimensions we have to consider the $\alpha$'s and $\psi$'s
together and with NN b.c.'s we would get $a_{NS}=1/2$ and $a_R=0$.
These $a$'s give the vacuum energy in the different sectors and it
will be very important to take care to work them out in different
cases to work out the massless states.} as
$\omega_n=(-1)^{1+\alpha'M^2}$. Notice that the action of $\Omega$
here naturally squares to 1 as it should.

It's also worth noting at this point that the form of the
groundstate will be different when we consider the (open)
superstring. For the open superstring it is 
\begin{eqnarray}
\label{susyground}
|0\rangle_{\alpha}\otimes|0\rangle_{\psi}(n;k;ij)\ .
\end{eqnarray}
This indicates the fact that the ground-state factorises between
the sectors where states are created by $\alpha$'s (the mode
operators for the Bosonic string) and $\psi$'s (the extra mode
operators added to make the superstring). The whole state is then
a function of the `level' at which the state is created
(\textit{i.e.} number of $\alpha$ and $\psi$ mode operators \emph{etc.}
which also depends on their integer/half-integer properties), the
momentum of the state, the Chan-Paton degrees of freedom (see
below) \textit{etc.}

This is all very well, but we know that the open string endpoints
have Chan-Paton degrees of freedom too, so we should really denote
the states as $|n;k;ij\rangle$. In fact, following Polchinski (\!\!\cite{Polchinski1}
page 186) one can define a `basis' as
\begin{eqnarray}
\label{basis} |n;k;a\rangle = \sum_{i,j=1}^{N_c}|n;k;ij\rangle
\lambda_{ji}^a,
\end{eqnarray}
where $i$ and $j$ clearly run from 1 to $N_c$\footnote{The $ij\, ji$ ordering is used for consistency in writing
things as a trace structure.}. For example, in the adjoint there are as many generators as there are
degrees of freedom in the matrices, so one can see how (\ref{basis}) makes sense in this case. Thus, for the adjoint of
$U(N_c)$, $a$ goes from 1 to $N_c^2$.

It is clear that since the Chan-Paton factors label the ends of the
string, they should be exchanged under world-sheet parity. The
action of $\Omega$ on a state is therefore more like:
\begin{eqnarray}
\label{morelike}
\Omega:|n;k;ij\rangle &\rightarrow &\omega_n|n;k;ij\rangle^T\nonumber \\
&=& \omega_n|n;k;ji\rangle\ ,
\end{eqnarray}
so for (\ref{basis}) we have:
\begin{eqnarray}
\label{basisxfm}
\Omega:|n;k;a\rangle &\rightarrow & \omega_n\sum_{i,j}|n;k;ij\rangle^T(\lambda_{ji}^a)^T\nonumber \\
&=& \omega_n\sum_{i,j}|n;k;ji\rangle{(\lambda^a)^T}_{ij}\nonumber \\
&=& \omega_n\sum_{i,j}|n;k;ij\rangle{{(\lambda^a)}^T}_{ji}\nonumber \\
&=& \omega_ns^a\sum_{i,j}|n;k;ij\rangle\lambda_{ji}^a\nonumber \\
&=& \omega_ns^a|n;k;a\rangle\ ,
\end{eqnarray}
with $s^a=\pm 1$ according to whether $\lambda=\pm\lambda^T$.

One thing that we can now do is to impose that our theory should
be invariant under $\Omega$. We'll be interested in the massless
states of the string which clearly have $\omega_n=-1$, and thus
this would require $\lambda^a=-(\lambda^a)^T$ (\emph{i.e.} $s^a=-1$). In
summary this requires $\lambda$ antisymmetric and thus gauge group
$SO(N_c)$. Why are we talking about gauge groups? Well, we've been
thinking about the NN strings and thus a stack of
branes.

\subsubsection{Parity}

This is all very well, but it is clear from (\ref{basis}) that we
actually have more symmetry than that discussed above. We can also
perform a rotation, under which (in some shorthand notation)
$\sum|\ldots\rangle\lambda\rightarrow\sum\gamma|\ldots\rangle^T\gamma^{-1}\gamma\lambda^T\gamma^{-1}$,
which will leave the theory invariant because of the trace
structure of the summation. This sort of transformation follows
the same logic as (\ref{basisxfm}) except that we now have
$\gamma$'s participating.
 We specify this new rotation,
combined with $\Omega$ as $\Omega_{\gamma}$. The effect on the
states is:
\begin{eqnarray}
\label{rotation}
\Omega_{\gamma}:|n;k;ij\rangle\rightarrow\omega_n\gamma_{jj'}|n;k;j'i'\rangle\gamma^{-1}_{i'i}\
.
\end{eqnarray}
For good reasons (see \cite{Polchinski1} pages 191-192), the form of
$\gamma$ is determined by requiring $\Omega_{\gamma}^2=1$. Let's
look at this:
\begin{eqnarray}
\label{omegammasq} \Omega_{\gamma}^2:|n;k;ij\rangle &\rightarrow &
\omega_n^2\gamma_{ii''}\left[\gamma_{j''j'}|n;k;j'i'\rangle\gamma^{-1}_{i'i''}\right]^T\gamma^{-1}_{j''j}
\nonumber \\
&=&\omega_n^2\gamma_{ii''}\left(\gamma^{-1}_{i'i''}\right)^T|n;k;i'j'\rangle\left(\gamma_{j''j'}\right)^T
\gamma^{-1}_{j''j}\nonumber \\
&=&\omega_n^2\gamma_{ii''}\left(\gamma^T\right)^{-1}_{i''i'}|n;k;i'j'\rangle\gamma^T_{j'j''}\gamma^{-1}_{j''j}
\nonumber \\
&=&\omega_n^2\left(\gamma(\gamma^T)^{-1}\right)_{ii'}|n;k;i'j'\rangle\left(\gamma^T\gamma^{-1}\right)_{j'j}\
.
\end{eqnarray}
$\omega_n^2$ is 1 as $n$ is an integer here, so for the whole
thing to be invariant we need $\gamma^T=\pm\gamma$. \emph{i.e.} $\gamma$
should be symmetric or antisymmetric. Note that here we have been
looking at the matrices that generate the transformation,
$\gamma$, \textit{not} the Chan-Paton wavefunction $\lambda$.

If $\gamma$ is symmetric, then it is always possible to
choose a basis where $\gamma=I$, the $N_c\!\!\times\!\!N_c$
identity matrix. In this case, the action of $\Omega_{\gamma}$ is
essentially just the same as that described previously for
$\Omega$ and the $\lambda$ should be \textit{antisymmetric},
giving $SO(N_c)$ on the branes. A \textbf{symmetric} projection
($\gamma=\gamma^T$) gives an \textbf{antisymmetric} gauge group
($\lambda=-\lambda^T$).

If, on the other hand, $\gamma$ is antisymmetric, then you can choose a basis in which
\begin{eqnarray}
\label{antisymmg}
\gamma=M=i\left(\begin{array}{cc} 0 & I_k \\
-I_k & 0 \end{array}\right)\ ,
\end{eqnarray}
where $I_k$ is the $k\!\times\! k$ identity matrix and we must
have $N_c=2k$. The reason for this is that in order to do the
transformation (\ref{rotation}) at all, we must be able to invert
$\gamma$. Thus we need $\det(\gamma)\neq 0$, and for $\gamma$
antisymmetric its determinant is only non-zero if $\gamma$ is a
$2k\!\times\! 2k$ matrix -- it must be even-dimensional.

Now what about $\lambda$ in this case? It is easily verified that
$M^{-1}=M$, so under the rotation, $\lambda\rightarrow M\lambda^T
M$. Again we're looking at what happens to the massless states
under $\Omega_{\gamma}$, and we want to require that they are
invariant under it. Since $\omega_n=-1$ for these states as
before, and we want to require $M|\ldots\rangle^TM=|\ldots\rangle$
by comparison with the third line of (\ref{basisxfm}), so we must
have $M\lambda^T M=-\lambda$ for invariance under
$\Omega_{\gamma}$. Considering a generic $\lambda$, it is
straightforward to see that
\begin{eqnarray}
\label{matrixes}
\lambda &\rightarrow & M\lambda^T M\nonumber \\
\nonumber \\
\left(\begin{array}{cc} A_k & B_k \\
C_k & D_k \end{array}\right)&\rightarrow&
\left(\begin{array}{cc} D_k^T & -B_k^T \\
-C_k^T & A_k^T \end{array}\right)\nonumber \\
\nonumber \\
&=&-\left(\begin{array}{cc} A_k & B_k \\
C_k & D_k \end{array}\right)\ .
\end{eqnarray}
We can thus represent $\lambda$ by
\begin{eqnarray}
\label{lambdamatrix}
\lambda=\left(\begin{array}{cc} A_k & B_k \\
C_k & -A_k^T \end{array}\right)\ ,
\end{eqnarray}
with $B_k^T=B_k$ and $C_k^T=C_k$. $A_k$ has $k^2$ independent
components, while $B_k$ and $C_k$ each have $k(k+1)/2$ as they are
symmetric. We thus have a total of $[2\times
k(k+1)/2]+k^2=2k(2k+1)/2=N_c(N_c+1)/2$ degrees of freedom in
$\lambda$. This is the dimension of the adjoint of what is
referred to as $Sp(N_c/2)$ or $U\!Sp(N_c)$ in the literature, and
in fact this is the gauge group we get. Here, an
\textbf{antisymmetric} projection ($\gamma=-\gamma^T$) gives a
\textbf{symmetric} gauge group.

\subsection{Orientifolding}

So far, all we've done is considered the world-sheet parity
operation. Orientifolding requires this plus some orbifold action.
States invariant under this combined transformation are the ones
that are kept. For a simple $\mathbb{Z}_2$ orbifold, the action is
then:
\begin{eqnarray}
\label{z2} X^{I}(\sigma,\tau)\rightarrow -X^{I}(l-\sigma,\tau)\ ,
\end{eqnarray}
for $I=p+1\ldots 9$. This defines an $Op$-plane extending in the
$X^1\ldots X^p$ directions and time.

In the presence of an $Op$-plane, the transverse space
$\mathbb{R}^{9-p}$ is replaced by $\mathbb{R}^{9-p}/\mathbb{Z}_2$.
However, it is often convenient to continue to think of the
geometry as being $\mathbb{R}^{9-p}$, but to add a $\mathbb{Z}_2$
image for each object lying outside the fixed plane. The
$Op$-plane carries RR charge under the same $(p+1)$-form potential
and breaks the same half of the SUSY as a parallel $Dp$-brane. Its
charge is, however, different to that of a $Dp$ (see \emph{e.g.} \cite{Giveon}):
\begin{eqnarray}
\label{rrcharge} Q_{Op}=\pm 2\cdot 2^{p-5}Q_{Dp}\ ,
\end{eqnarray}
where the first 2 is really for the images of the $Dp$'s 
(this will be a little clearer in a moment). It turns out that the
sign is a choice that is correlated with the choice of an
antisymmetric or a symmetric projection on the $Dp$'s. The $+$ sign
gives an antisymmetric projection and thus $Sp$ gauge groups,
while the $-$ sign gives a symmetric projection and thus $SO$
gauge groups. You can read more about this in \cite{Uranga}.

\subsubsection{$O9$ in Type I}
Let's try to illustrate what is meant with a simple example.
Consider the Type I string with a space-filling $O9$-plane. As it
fills all of space, there is no transverse space to do the
orbifold action with, so we really just have the parity operation
discussed above. In the simplest case we have NN boundary
conditions in all 9 directions, so we really have some space
filling $D9$-branes too. In fact let's put $2^{9-5}=16$ of them so
that their RR charge has the same magnitude as the $O9$-plane. The
$2^{p-5}$ tells you the number of physical branes, while the extra
2 in (\ref{rrcharge}) counts their mirror images. Think of the branes as sharing the same world-volume
directions as the orientifold plane, but very slightly separated
from it in the transverse directions. This means you will have the
branes and their images, and strings stretching between branes and
their mirror images are ultimately important in realising the
gauge theory on the brane. Then, when you bring the branes
together on the orientifold plane, strings become light and you
get the gauge theory as in the usual way with coincident branes,
although the gauge group is now modified because of the
orientifolding. Obviously in the case where we have $D9$'s and an
$O9$, there are no transverse directions in which to actually separate the
branes, but it is still a useful picture to have in mind. It is also important,
that while there are $2^{p-5}$ physical branes, since you can have
strings stretching between branes and their mirror images, the
Chan-Paton indices can run up to $2\cdot 2^{p-5}$, so the $N_c$
discussed earlier in the appendix will be $N_c=2\cdot 2^{p-5}$.

To recap, we have an $O9$-plane and 16 $D9$-branes all parallel to
each other. As discussed before, the world-sheet parity operation
allows us to either do a symmetric or an antisymmetric projection
on the $D9$s, but it doesn't fix what this should be. This choice
is correlated with the choice of sign in (\ref{rrcharge}). For
tadpole cancellation (and because there are no other
directions for the RR charge to `leak' out into and decay at
infinity), we should choose the negative sign so that the overall
RR charge is zero. With a negative sign, we must perform a
symmetric projection on the $D9$s (see \cite{Uranga}), and thus we
end up with gauge group $SO(32)$. This is why Type I string theory has 
gauge group $SO(32)$.

\subsubsection{$D5$-$D9$ in Type I}

Note that much of this subsection uses results explained later on
and so may be clearer after reading the rest of the appendix. To make
progress, we in any case just need the final results in the discussion after
eq. (\ref{mess}).

The setup here is that we have Type I string theory with an
$O9$-plane, 16 $D9$-branes and a $D5$-brane. The 5-branes occupy
$X^0\ldots X^5$, say. We have 3 types of strings: 9-9 strings, 5-5
strings and 5-9 (or 9-5) strings. The 9-9 strings have NN b.cs in
all directions. The 5-5 strings have NN b.cs in $0\ldots 5$ and DD
b.c.'s in $6\ldots 9$. The 5-9 (or 9-5) strings on the other hand
have NN b.cs in $0\ldots 5$, but DN (or ND) b.cs in $6\ldots 9$.

Lets think about 9-9 strings to start off with. They are NN in all
directions so for the $X$s the arguments will be much the same as
before with the purely Bosonic string -- we'll get an antisymmetric
projection for these. Now for the NS sector. This has half-integer
modes in this case, so states are created by $\psi_{-r}$. Under
parity these are sent to $e^{\pm i\pi r}\psi_{-r}$ with the
sign depending on whether they are left-movers or a right-movers.
These creation operators are world-sheet Fermions but space-time
vectors. However, $\psi_{-r}|0;k\rangle$ is a Boson of both the
world-sheet and the space-time, so $|0;k\rangle$ is a world-sheet
Fermion. Following \cite{Polchinski2},
page 197, $|0;k\rangle$
should transform as $e^{\pm i\pi/2}|0;k\rangle$ under parity. Lets
take the $-$ sign for definiteness. As $r$ is half-integer valued,
$e^{\pm i\pi r}=\pm e^{i\pi r}$, so states in this sector will go
to $\pm e^{i\pi r}e^{-1/2}=\pm 1$. This is basically the same as
the $X$-sector where states changed by a sign under parity and the
operation squared to one. The same clearly happens here, although one will
end-up having to be a bit careful about some aspects. For example, the $r$-dependence, which
will affect what happens with the massless states given that
$a_{NS}=1/2$ in this sector. The point is that we can do either a
symmetric or an antisymmetric projection as before -- constrained
only by tadpole cancellation \emph{etc.}

The 5-5 sector works out in much the same fashion since NN and DD
strings have the same mode expansions in terms of their
integer/half-integer properties. One needs to be careful about the
spacetime indices of course and whether or not they belong to the
5-brane or the 9-brane. Again a symmetric or antisymmetric
projection can be done.

The 5-9 sector is more complicated however. For the 0-5 directions
we have NN b.c.'s and the situation is much like the 5-5 or 9-9
sectors described above. However, for
the 6-9 directions we have DN b.c.'s. This means half-integer
$\alpha_{-r}$, and thus integer $\psi_{-n}$ in the NS sector and
half-integer $\psi_{-r}$ in the R sector. It is clear that the NS
sector here behaves like the usual R sector for the Type I string
without any branes (where we also had integer $\psi_{-n}$).
Similarly, the R sector behaves like the usual NS sector when
there are no branes. One sometimes says that the 5-9 strings are
`twisted' and this gives rise to the swap in the roles of the two
sectors.

We want some information on $\omega^2$ and a trick can help us
here. The point is that if we have a vertex operator (or
equivalently a string state) then: $\Omega:V\rightarrow \Omega
V\Omega^{-1}=\omega V$. So, $\Omega:V^2\rightarrow \Omega V^2
\Omega^{-1}=\Omega V\Omega^{-1}\Omega V\Omega^{-1}=\omega V\cdot
\omega V=\omega^2V^2$. So the eigenvalue for the action of
$\Omega$ on $V^2$ will give us $\omega^2$. \textit{i.e.} $\Omega$
on $V^2$ gives us information about $\Omega^2$ on $V$.

Let's think about the NS sector. Now, one way to think about
Fermions in 2-dimensions is via Bosonization. It turns out that in
2-d, a system of $2n$ free Fermions is the same as a system of $n$
free Bosons. The basic idea is that we construct linear
combinations of our $\psi$'s \'{a} la eq.(\ref{cliff}) and then
write $\psi_a^{\pm}=e^{\pm iH_a/2}$. In our case we have
directions $6\ldots 9$ (with no time directions, so no need to go
to the light-cone) and thus we can have $e^{\pm iH_{3/4}/2}$ which
are the Bosonization of $\psi_3^+=(\psi_{6}\pm
i\psi_{7})/\sqrt{2}$ and $\psi_4^+=(\psi_{8}\pm
i\psi_{9})/\sqrt{2}$. To make our states we'll need only the
$\psi^{+}$ which act as raising operators, so in the end we only
need to consider $e^{iH_3/2}$ and $e^{iH_4/2}$. If we now think
about the GSO projection and project so that we get states created
by an even number of $\psi$'s here, then we find that we can have
either the vacuum $|0\rangle $ or $\psi_3^+\psi_4^+|0\rangle$ as
our states in this sector (see the following subsections for much more information). So, for our vertex operator $V$ we should consider
$\psi_3^{+}\psi_4^{+}\equiv e^{i(H_3+H_4)/2}$. This means, though,
that $V^2=e^{i(H_3+H_4)}$, which is in-turn the vertex operator
for the state
\begin{eqnarray}
\label{polchinski}
\frac{1}{2}(\psi_6+i\psi_7)_{-1/2}(\psi_8+i\psi_9)_{-1/2}|0\rangle
,
\end{eqnarray}
\textit{i.e.} a state created by Fermions with half-integer
modedness.

One might wonder why our original definition of $V$ had
a factor of $1/2$ in the exponential while our `new' vertex
operator $V'=V^2$ does not. From the point of
view that $V'$ is the square of $V$ this is trivial. However, more information can be found in \cite{Lust}. We should also
note that such half-integer modes do not exist in the 5-9 sector,
or rather they do (in the Ramond subsector) but they do not give
anything massless as $a_R=a_{NS}=0$ for the 5-9 strings (see
later in this appendix for details). However, the 5-5 strings and 9-9 strings
have the same $a_R$ and $a_{NS}$ which in-turn have the same
values as for the Type I string without any branes at all. In
particular, there \textit{are} massless states created by
$\psi_{-1/2}$. So eq.(\ref{polchinski}) is really talking
about states in the 5-5 or 9-9 sector.

In any case, we can finally deduce the eigenvalue of
$V^2$. $|0\rangle$ gives us +1, while each $\psi^+_{-1/2}$
contributes $\pm i$. The `$i$' comes from the fact that they are
$1/2$ integer-moded and thus we get $e^{-i\pi/2}$ under parity,
and the relative sign because we should recall that there is a
relative sign between NN and DD boundary conditions. 9-9 strings
are NN in the directions we are considering and thus give us $-i$
while 5-5 strings are DD in these directions and thus give us $i$.
Overall we would have $(\pm i)^2=-1$. Thus $\omega^2=-1$ in this
sector and parity will take:
\begin{eqnarray}
\label{mess} \Omega_{\gamma_5\gamma_9}^2:|\phi;iJ\rangle
&\rightarrow &
\Omega_{\gamma_5\gamma_9}:\omega_{59}{\gamma_{9}}_{JJ'}|\phi;J'i'\rangle{\gamma_5^{-1}}_{i'i}
\nonumber \\ &\rightarrow &
\omega_{59}^2{\gamma_5}_{ii''}\left[{\gamma_{9}}_{J''J'}|\phi;J'i'\rangle{\gamma_5^{-1}}_{i'i''}\right]^T{\gamma_9^{-1}}_{J''J}
\nonumber \\ &=&
\omega_{59}^2{\gamma_5}_{ii''}\left({\gamma_5^{-1}}_{i'i''}\right)^T|\phi;i'J'\rangle\left({\gamma_{9}}_{J''J'}\right)^T{\gamma_9^{-1}}_{J''J}
\nonumber \\ &=&
\omega_{59}^2{\gamma_5}_{ii''}\left(\gamma_5^T\right)^{-1}_{i''i'}|\phi;i'J'\rangle{\gamma_9^{T}}_{J'J''}{\gamma_9^{-1}}_{J''J}
\nonumber \\ &=&
\omega_{59}^2\left(\gamma_5{\gamma_5^T}^{-1}\right)_{ii'}|\phi;i'J'\rangle\left(\gamma_9^T\gamma_9^{-1}\right)_{J'J}\
,
\end{eqnarray}
where lowercase indices refer to the 5-brane and capitals to the
9-brane. We can now invoke tadpole cancellation to give us
$\gamma_9^T=\gamma_9$. This gives us the identity for the factor
on the right. We know from above that $\omega_{59}^2=-1$, so that
leaves us with $\gamma_5{\gamma_5^T}^{-1}=-I_5$, where $I_5$ is
the identity appropriate to the 5-brane indices. Thus
$\gamma_5^T=-\gamma_5$. We can see that a symmetric projection on
the 9-branes must give us an antisymmetric projection on the
5-branes. It is also clear that the converse must be true.

\subsection{Massless States}\label{masslessstatesappsection}
\subsubsection{$D3$s}

In order to work out what's going on in the $D3$-$D7$ case it'll
be necessary to consider what the massless states of open strings
in this configuration are. Let's start with a simple example that
we know very well -- $N_c$ $D3$s in flat space. It's pretty obvious
that strings stretching in the $0\ldots 3$ direction will have NN
boundary conditions -- this is after all the definition of a
$D3$-brane. Those in the $4\ldots 9$ directions will of course
have DD boundary conditions. Actually this doesn't make too much
odds to us. It affects the precise form of the mode-expansions
that you write down, but the main point is really (as usual) the
integer/half-integer modedness. As discussed before we will thus
have integer-modes in the Bosonic sector\footnote{\emph{i.e.} the sector described by 
$\alpha$'s.} and the Ramond
sector and half-integer-modes in the Neveu-Schwarz sector.
In order to determine what the massless states are it will be
important to determine what the vacuum energy is. The masses of
higher-excited states are then typically propertional to the
number operator minus this zero-point energy. And how does it
arise? It's essentially the anomaly that arises when normal-ordering
the number operator as per Section 5.2 and Section 8.4. A short review of this is provided in the following subsection.

\subsubsection{Normal Ordering}\label{normalorderingsection}
With NN or DD b.c.'s, the NS sector has:
\begin{eqnarray}
\label{normalordering}
N_{NS}&=&\tilde{N}^{(\alpha)}+\tilde{N}^{(\psi)}\nonumber \\
&=& \frac{1}{2}\sum_{n\in
\mathbb{Z}}\alpha_{-n}^{i}\alpha_{n}^i+\frac{1}{2}\sum_{r\in
\mathbb{Z}+\frac{1}{2}}r\psi_{-r}^i\psi_r^i \nonumber \\ &=&
\frac{1}{2}\sum_{1}^{\infty}\alpha_{-n}^i\alpha_n^i+\frac{1}{2}\sum_{-\infty}^{-1}\alpha_{-n}^i\alpha_n^i+
\frac{1}{2}\sum_{1/2}^{\infty}r\psi_{-r}^i\psi_r^i+\frac{1}{2}\sum_{-\infty}^{-1/2}r\psi_{-r}^i\psi_r^i\nonumber \\
&=&
\frac{1}{2}\sum_{1}^{\infty}\alpha_{-n}^i\alpha_n^i+\frac{1}{2}\sum_{1}^{\infty}\alpha_{n}^i\alpha_{-n}^i+
\frac{1}{2}\sum_{1/2}^{\infty}r\psi_{-r}^i\psi_r^i+\frac{1}{2}\sum_{1/2}^{\infty}-r\psi_{r}^i\psi_{-r}^i\nonumber \\
&=&
\frac{1}{2}\sum_{1}^{\infty}\alpha_{-n}^i\alpha_n^i+\frac{1}{2}\sum_{1}^{\infty}\left(\alpha_{-n}^i\alpha_{n}^i+
\left[\alpha_n^i,\alpha_{-n}^i\right]\right)\nonumber \\
&\,&+\,\,\frac{1}{2}\sum_{1/2}^{\infty}r\psi_{-r}^i\psi_r^i+\frac{1}{2}\sum_{1/2}^{\infty}\left(r\psi_{-r}^i\psi_{r}^i-
r\left\{\psi_{r}^i,\psi_{-r}^i\right\}\right)\nonumber \\
&=&
\sum_{1}^{\infty}\alpha_{-n}^i\alpha_n^i+\sum_{1/2}^{\infty}r\psi_{-r}^i\psi_r^i+\frac{1}{2}\sum_{1}^{\infty}n\eta^{ii}
-\frac{1}{2}\sum_{1/2}^{\infty}r\eta^{ii}\nonumber \\
&=& N^{(\alpha)}+N^{(\psi)}+\frac{d-2}{2}\left(\sum_1^{\infty}n-\sum_{1/2}^{\infty}r\right)\nonumber \\
&=& N^{(\alpha)}+N^{(\psi)}-a_{NS}\ ,
\end{eqnarray}
where we have used the commutation relations
\mbox{$\left[\alpha_n^{\mu},\alpha_m^{\nu}\right]=n\delta_{n+m}\eta^{\mu\nu}$}
and
\mbox{$\left\{\psi_r^{\mu},\psi_s^{\nu}\right\}=\delta_{r+s}\eta^{\mu\nu}$},
and we are in light-cone gauge so $i=1\ldots d-2$. In the NS
sector there are no zero-modes, and in the Bosonic sector we have
omitted them as they give us the $M^2$ operator. These sums are of
course ostensibly divergent, but we can take care of them in the
usual way by using zeta function regularization. The first sum is
$\zeta(-1)=-1/12$. On the other hand, referring to (\ref{rsum}) we have
\begin{eqnarray}
\label{rsumagain}
\sum_{1/2}^{\infty}r=\frac{1}{24}\ .
\end{eqnarray}

Considering (\ref{normalordering}), it is clear that
for the Bosonic string (where there are no $\psi$'s and $d=26$) you
get $-a=24/2\times(-1/12)=-1\Rightarrow a=1$. For the NS sector in 10
dimensions we will therefore get $-a_{NS}=8/2\times
(-1/12-1/24)=-1/2\Rightarrow a_{NS}=1/2$. Finally, in the Ramond sector we
can play the same game as above, except that this time the $\psi$'s
will be integer-moded. The minus sign between the two sums will
stay as they're still world-sheet Fermions, but the second sum
will now be the same as the first one. They'll cancel and we'll
get $a_R=0$. All of this discussion applies equally well
for space-filling $D9$-branes (NN b.c.'s everywhere) or our current set-up of
$N_c$ $D3$s.

\subsubsection{Bosonic States}
Considering our 3-3 strings then and the NS sector. As
$\alpha_{-n}$ have integer $n$ and $a_{NS}=1/2$, it is clear that
the $\alpha$'s can't give anything massless. The massless
states are thus created by $\psi_{-1/2}$. With 3-brane indices,
$\psi_{-1/2}^{\mu}|0;k;ij\rangle$ is a massless vector in 4-d with
2 physical degrees of freedom. In fact there are $N_c^2-1$ of
these because of the Chan-Paton indices -- it is a massless
vector in the adjoint of $SU(N_c)$. With indices of $I=4\ldots 9$,
$\psi_{-1/2}^{I}|0;k;ij\rangle$ are massless scalars in 4-d. There
are 6 of them. Again in the adjoint.

\subsubsection{Fermionic States}
Now for the Ramond sector. Let's just think about what happens in
10-dimensions with the Type I string to warm up. In this case,
$a_R=0$ so the massless state is the ground state $|0\rangle$. Now
if we think about our physical state conditions, they include
$\psi_{n}|0\rangle=0$ for $n>0$. But the algebra of the $\psi_{n}$
is
$\left\{\psi_n^{\mu},\psi_m^{\nu}\right\}=\delta_{r+s}\eta^{\mu\nu}$
which in-particular implies that $\psi_{n}\psi_0=-\psi_0\psi_n$
which in turn implies that $\psi_n(\psi_0|0\rangle)=0$. Clearly
$\psi_0|0\rangle$ is just as good a ground-state as $|0\rangle$,
and since there are quite a few different $\psi_0$ (because of
their spacetime indices which we have ommitted at the moment)
there is more than one different possible ground state. That is
to say, the ground state is degenerate. Now for these $\psi_0$
(the so-called zero-modes) the algebra that they obey is
\mbox{$\left\{\psi_0^{\mu},\psi_0^{\nu}\right\}=\eta^{\mu\nu}$}
which is precisely a Clifford algebra and seeing as the definition
of the ground state has this symmetry corresponding to the action
of $\psi_0$, it must lie in a representation of this Clifford
algebra.

The representations of Clifford algebras are well-known.
Essentially, to get the physical states we are interested in, what
we do is the following. Knock off 1 time and 1 space direction
from the $\mu=0\ldots 9$. This is effectively a light-cone
quantisation condition or equivalently the statement that massless
particles are in representations of the Little group $SO(d-2)$
associated with the Lorentz group $SO(d-1,1)$. If we don't have
any timelike directions then it's not necessary to do this (at
least for even space-like dimensions). \textit{i.e.} if we have
$SO(2k)$ then we just use $2k$ as the $d-2$ mentioned above. Then
take the remaining $d-2$ $\psi$'s and form $(d-2)/2$ (or just $k$
in the case of $SO(2k)$) linear combinations of them according to:
\begin{eqnarray}
\label{cliff} \psi_a^{\pm}=\frac{1}{\sqrt{2}}(\psi_{2a}\pm
i\psi_{2a+1})\ ,
\end{eqnarray}
with $a=1\ldots (d-2)/2$. It is easily shown that the $\psi^{\pm}$
act as Fermionic ladder operators and thus we introduce a `ground
state' $|0\rangle$ such that $\psi_a^{-}|0\rangle=0$ for our
`harmonic oscillator' to build a representation of the Clifford
algebra which we do by applying the $\psi_a^{+}$ in all possible
inequivalent ways. In the present case of $SO(9,1)\rightarrow
SO(8)$ we will get 4 $\psi_a^{+}$ and thus states with up-to 4
$\psi^{+}$s acting:
\begin{eqnarray}
\label{cliffstates} &|0\rangle&\nonumber \\
\psi_{a}^+\!\!\!\!&|0\rangle&\nonumber \\
\psi_{a_1}^+\psi_{a_2}^+\!\!\!\!&|0\rangle&\nonumber \\
\psi_{a_1}^+\psi_{a_2}^+\psi_{a_3}^+\!\!\!\!&|0\rangle&\nonumber \\
\psi_{a_1}^+\psi_{a_2}^+\psi_{a_3}^+\psi_{a_4}^+\!\!\!\!&|0\rangle&\
.
\end{eqnarray}
There are $\binom{(d-2)/2}{r}$ states when $r$ of the $\psi_{a}^+$
are applied and thus $2^{(d-2)/2}$ states in total. In our case
where $(d-2)/2=4$ we will get 16 possible states distributed as
$1,4,6,4,1$ (from top to bottom in (\ref{cliffstates})).

In 10-dimensions we have therefore found that the Ramond sector
ground state has 16 degrees of freedom. This splits into two
different spinor representations of the Little group $SO(8)$,
namely the $\mathbf{8_s}$ and $\mathbf{8_c}$. \textit{i.e.} we
have the $\mathbf{16}$ of $SO(9,1)$ which decomposes as
$\mathbf{16}=\mathbf{8_s}+\mathbf{8_c}$. $\mathbf{8_s}$
corresponds to the states created by an even number of $\psi^+$s
and $\mathbf{8_c}$ to the states with an odd number of $\psi^+$s.
However, we also have to GSO project our states, and in the Ramond
sector this removes half the states at each level and forces us to
choose $\mathbf{8_c}$ rather than $\mathbf{8_s}$\footnote{See pages 179-180 of
\cite{Uranga}}. This is good as we know that we want 8 Fermionic
degrees of freedom to saturate the 8 Bosonic ones that we already
have.

But what about our $D3$-branes? Essentially the situation is the
same as in 10-dimensions for the Type I string as we're only
considering the open string modes of IIB. $a_R$ is again zero so
we have the same problem with zero-modes, and \textit{all} of the
$\psi_0$ will still take one ground state into another regardless
of whether they have 3-brane valued spacetime indices (what we
called $\mu$ for the Bosonic states in the previous subsection) or
indices of the 6-dimensional transverse space (what we called $I$
in the previous subsection). Thus we again get 8 Fermionic states
after the GSO projection.

Actually we can do better than this and look at how the Lorentz
group decomposes when we have $D3$-branes. Here we can use results from
Appendix B of \cite{Polchinski2}. In terms of the Lorentz group, what we
have is:
\begin{eqnarray}
\label{decomp1} SO(9,1)\rightarrow SO(3,1)\times SO(6)\ ,
\end{eqnarray}
and for the representations of the Lorentz group we would have
\begin{eqnarray}
\label{decomp4}
\mathbf{16}=\mathbf{8_s}+\mathbf{8_c}\rightarrow\mathbf{(2,4)}+\mathbf{(2',4')}\
.
\end{eqnarray}
After GSO projection this would give us $\mathbf{(2,4)}$ or
$\mathbf{(2',4')}$ under the broken group. Thus we clearly get 4
spinors in the $\mathbf{2}$ (or $\mathbf{2'}$) representation of
the $SO(2)$ Little group of $SO(3,1)$.

From the 4-dimensional perspective a generic spinor has
$2^{d/2}=2^{4/2}=4$ complex components (see Appendix \ref{spinorappendix}). However, we know that our
spinors should be Weyl spinors which means that this number should
be halved, and obeying the Dirac equation makes the number of
physical degrees of freedom of such a spinor half of this again.
Thus it should have 1 complex or 2 real degrees of freedom. 8
Fermionic degrees of freedom can thus make 4 Weyl spinors in
4-dimensions -- perfect for the $\mathcal{N}\!=\!4$ SYM multiplet.

The $\mathcal{N}\!=\!4$ SYM multiplet consists of 1 vector
$A_{\mu}$ (2 real d.o.f.), 6 real scalars $\Phi^{I}$ (6 real
d.o.f.) and 4 Weyl spinors $\lambda_{\alpha}$ (8 real
d.o.f.)\footnote{$(1,4,6,4,1)$ in a helicity-basis of helicities $(-1,-\frac{1}{2},0,\frac{1}{2},1)$.}.
We know that a stack of branes in flat (10-d) space breaks half of
the supersymmetry. In 10-d we can have a maximum of
$\mathcal{N}\!=\!2$ with 32 supercharges which will thus be broken
to 16 supercharges. For the 4-d theory on the $D3$s, 16
supercharges means $\mathcal{N}\!=\!4$. We can see that the
multiplet is the one that we should get.
\begin{center}
$D3$-brane Massless State Counting
\begin{displaymath}
\begin{array}{|l|c|c|}
\hline \textrm{Directions} & $0-3$ & $4-9$ \\
\hline \textrm{3-3 b.cs} & \textrm{NN} & \textrm{DD} \\
\hline \textrm{NS} & \alpha_{-n}/\psi_{-r} &
\alpha_{-n}/\psi_{-r} \\
 a_{NS}=1/2 & \psi_{-1/2}^{\mu}|0\rangle &
\psi_{-1/2}^{I}|0\rangle \\
 \textrm{\#} & 2 & 6 \\
\hline \textrm{R} & \alpha_{-n}/\psi_{-r} &
\alpha_{-n}/\psi_{-r} \\
 a_R=0 & \psi_0^{\mu}|0\rangle & \psi_0^{I}|0\rangle \\
\cline{2-3} \textrm{\#}  & \multicolumn{2}{c|}{\textrm{8 total}}\\
\hline
\end{array}
\end{displaymath}
\end{center}

\subsection{$D3$-$D7$s}

The setup that we are looking at here is as follows: We are in Type IIB with an
$O7$-plane (in $0\ldots 7$ say), 4 (physical) $D7$-branes whose
world-volumes are coincident and aligned with the $O7$. We pick
the tension of the $O7$ to be negative so that the RR charge is
cancelled between the two. We also have $N_c$ physical
$D3$-branes, located in $0\ldots 3$ say.

It may also be useful to recall that the theory we \emph{should} get on
the $D3$-branes is $\mathcal{N}=2$ Supersymmetric gauge theory
with gauge group $U\!Sp(N_c)$, 1 vector multiplet $(1,2,2,2,1)$ in
the adjoint (for $U\!Sp(N_c)$ the adjoint is the symmetric
representation with dimension $N_c(N_c+1)/2$), 1
($\mathcal{N}\!=\!2$) hypermultiplet $(0,2,4,2,0)$ in the
antisymmetric (the antisymmetric has dimension $N_c(N_c-1)/2$) and
4 hypermultiplets in the fundamental. In some ways, these
fundamental hypermultiplets are perhaps best thought of as 8
fundamental $\mathcal{N}\!=\!1$ chiral multiplets $(0,1,2,1,0)$
which makes the $SO(8)$ symmetry a bit more manifest. Actually the
most natural split is probably into 2 sets of 4 chiral multiplets.

We now have quite a few different sectors to consider. In terms of
open strings, there are 3-3 strings, 7-7 strings and 3-7 or 7-3
strings. It is fairly clear (and easily worked out), that 3-3
string are NN in $0\ldots 3$ and DD in $4\ldots 9$. 7-7 strings
are NN in $0\ldots 7$ and DD in 8 and 9. 3-7 strings on the other
hand are NN in $0\ldots 3$, DN in $4\ldots 7$ and DD in 8 and
9\footnote{7-3 strings on the other hand will be NN in $0\ldots
3$, ND in $4\ldots 7$ and DD in 8 and 9.}. We're not too interested in what the theory on the $D7$s is so no need to worry
about 7-7 strings at the moment. The 3-3 strings will have vacuum
energies as above and so the multiplet will (in principle) be the
$\mathcal{N}\!=\!4$ one. We know that this can't be the case of
course as the $Dp$-$D(p+4)$ system (which is what we have) preserves
only 1/4 of the SUSY. An $O(p+4)$ parallel to the $D(p+4)$s
doesn't break any more, so this system preserves 8 supercharges.
In 4-d on the $D3$'s this is $\mathcal{N}\!=\!2$. The
resolution is that some of this multiplet will be projected out by
the orientifold action and let's have a look at this.

\subsubsection{3-3 Strings}
For the 3-3 strings in the NN directions, it is clear from above
that we should get one vector (2 degrees of freedom). The gauge
group is in principle $SU(N_c)$, but tadpole cancellation for the
$D7$-$O7$ setup tells us that we should get an $SO$ gauge-group on
the $D7$s and thus by complementarity an $Sp$ gauge group on the
$D3$s. The massless states are $\psi_{-1/2}^{\mu}|0;p\rangle$ with
$\mu=0\ldots 4$ and $N_c(N_c+1)/2$ degrees of freedom in the
Chan-Paton factors. These are the $\pm 1$ helicity degrees of
freedom in the vector multiplet.

Now for the DD directions.
Firstly $4\ldots 7$. Here we have $\psi^{M}_{-1/2}|0;p\rangle$
where $M=4\ldots 7$. From the discussion earlier there should be a sign difference in the action of
$\Omega_{\gamma}$ between the NN and DD sectors. In principle the
state we have just written down is identical to the one for the
gauge Bosons except that it has transverse indices and we are not
thinking about light-cone quantisation. This means that these
will give 4 scalars on the $D3$s. However, because of the sign
difference we have mentioned, it is clear that when thinking about
the action of $\Omega_{\gamma}$ we would have a sign
difference from before and would have concluded that we should
have an antisymmetric projection on the Chan-Paton wavefunction
here. That is to say, the Chan-Paton indices would be restricted
to run from $1\ldots N_c(N_c-1)/2$. Having looked at the gauge
Bosons already, we know that the gauge group should be
$U\!Sp(N_c)$, so these 4 transverse scalars will be in the
antisymmetric representation of $U\!Sp(N_c)$, which has dimension
$N_c(N_c-1)/2$. These are the 4 scalars of the hypermultiplet in
the antisymmetric.

Finally lets think about the DD directions of 8
and 9. Again we have massless states of the form
$\psi_{-1/2}^{m}|0;p\rangle$, but with $m=8,9$. Because they are
DD there should again be a sign difference between
them and the NN directions. However, the 8 and 9 directions also
pick up an extra minus sign because they are outside the
orientifold plane. So their transformation under $\Omega_{\gamma}$
will be the same as for the NN states and they will be in the
adjoint (symmetric) of $U\!Sp(N_c)$. The index $m$ is again
transverse to the $D3$-branes so these are the 2 scalars of the
vector multiplet.

For the Fermions, the
construction proceeds totally analogously with the $N_c$ $D3$ case
and then the orientifolding breaks the representations into the
symmetric and antisymmetric of the $U\!Sp(N_c)$ subgroup. See \cite{gava} for more.
\begin{center}
$D3$-$D7$ Massless 3-3 Strings
\begin{displaymath}
\begin{array}{|l|c|c|c|}
\hline \textrm{Directions} & 0-3 & 4-7 & 8-9\\
\hline \textrm{b.cs} & \textrm{NN} & \textrm{DD} & \textrm{DD}\\
\hline \textrm{NS} & \alpha_{-n}/\psi_{-r} & \alpha_{-n}/\psi_{-r}
& \alpha_{-n}/\psi_{-r}\\
 a_{NS}=1/2 & \psi_{-1/2}^{\mu}|0\rangle &
\psi_{-1/2}^{M}|0\rangle & \psi_{-1/2}^{m}|0\rangle\\
 \textrm{\#} & 2 & 4 & 2\\
\textrm{$U\!Sp(N_c)$ rep} & N_c(N_c+1)/2 & N_c(N_c-1)/2 &
N_c(N_c+1)/2\\\hline \textrm{R} & \alpha_{-n}/\psi_{-r} &
\alpha_{-n}/\psi_{-r}
& \alpha_{-n}/\psi_{-r}\\
 a_R=0 & \psi_0^{\mu}|0\rangle & \psi_0^{M}|0\rangle &
\psi_0^{m}|0\rangle\\
 \# & 2 & 4 & 2\\
\textrm{$U\!Sp(N_c)$ rep} & N_c(N_c+1)/2 &
N_c(N_c-1)/2 & N_c(N_c+1)/2\\
\hline
\end{array}
\end{displaymath}
\end{center}

Note that here it is easy to see how the spinor degrees of freedom arrange themselves 
under the decomposition of the Lorentz group $SO(9,1)\rightarrow SO(3,1)\times SO(4)\times SO(2)$. 
A spinor of $SO(3,1)$ has $2^{(3-1)/2}=2$ d.o.f., a spinor of $SO(4)$ has $2^{4/2}=4$ d.o.f. and a spinor 
of $SO(2)$ has $2^{2/2}=2$ d.o.f.

\subsection{3-7 Strings}
For the 3-7 strings we have to be more careful about the vacuum
energy. For these we have: $\alpha$ integer-moded for
$0\ldots 3$ and 8,9 but half-integer for $4\ldots 7$. In the NS
sector $\psi$ is half-integer-moded when $\alpha$ is integer-moded
and vice-versa. Vice-versa again for the Ramond sector. It's not too
hard to see that in the NS sector the normal ordering constant
will be:
\begin{eqnarray}
\label{norderND}
-a_{NS}&=&\frac{4}{2}\left(\sum_1^{\infty}n-\sum_{1/2}^{\infty}r\right)+\frac{4}{2}\left(\sum_{1/2}^{\infty}r-\sum_1^{\infty}n\right)
\nonumber \\ &=& 0\ .
\end{eqnarray}
The first term comes from the NN and DD conditions on the $0\ldots 4$
and 8,9 directions. You might wonder why we have a 4 there instead
of a 6. Well, as usual we have to do light-cone quantization for
the physical states so we must eliminate 2 degrees of freedom
(\textit{c.f.} the $d\!-\!2$ in the previous subsection). In the
Ramond sector on the other hand we'll have:
\begin{eqnarray}
\label{norderR}
-a_{R}&=&\frac{4}{2}\left(\sum_1^{\infty}n-\sum_{1}^{\infty}n\right)+\frac{4}{2}\left(\sum_{1/2}^{\infty}r-\sum_{1/2}^{\infty}r\right)
\nonumber \\ &=& 0\ .
\end{eqnarray}

We should also be careful to note that 3-7 strings are mapped into
7-3 strings under the orientifold projection\footnote{This
is fairly clear given that the ends of the strings are
interchanged and a 3-7 starts on a $D3$ and ends on a $D7$ while a
7-3 starts on a $D7$ and ends on a $D3$. It is also pretty
clear from the orientifold actions earlier in the appendix (see
eq.(\ref{DNomega}) in particular).} and so it is enough to keep
the degrees of freedom in the 3-7 sector and not perform any
projection. This is analogous to the 5-9 sector of strings in a
Type I setup with $D5$ and $D9$ branes as described in
\cite{Uranga} (see page 319 particularly).

Note that the
Chan-Paton indices should run over the $\mathbf{N_c}$ of
$U\!Sp(N_c)$ and the $\mathbf{8}$ of the $SO(8)$ generated by the
orientifold action on the 7-branes. (\textit{i.e.} what really
happens is that 3-7 and 7-3 strings give the same since they
are mapped into one another under the projection. We can thus
consider only one class of them (3-7 strings say) and \textit{no}
orientifold and double our counting at the end. If we
\textit{actually} weren't orientifolding then we would only really
have 4 $D7$-branes and so our strings would transform under the
vector of some $SU(4)$. We would get $4\times\textrm{(\# d.o.f of
a 3-7 string)}\times 2$. However, we actually \textit{are}
orientifolding of course which tells us to consider our strings as
transforming under the vector of $SO(8)$ and this does the same
doubling job for us.

The degrees of freedom are:

\begin{center}
$D3$-$D7$ 3-7 + 7-3 Strings
\begin{displaymath}
\begin{array}{|l|c|c|c|}
\hline \textrm{Directions} & 0-3 & 4-7 & 8-9\\
\hline \textrm{b.cs} & \textrm{NN} & \textrm{DN} & \textrm{DD}\\
\hline \textrm{NS} & \alpha_{-n}/\psi_{-r} & \alpha_{-r}/\psi_{-n}
& \alpha_{-n}/\psi_{-r}\\
a_{NS}=0 & - & \psi_0^{M}|0\rangle & -\\
\# & - & 2 & -\\
\textrm{$SO(8)$ rep} & - & \mathbf{8} & -\\
\hline \textrm{R} & \alpha_{-n}/\psi_{-n} & \alpha_{-r}/\psi_{-r}
& \alpha_{-n}/\psi_{-n}\\
a_R=0 & \psi_0^\mu|0\rangle & - & \psi_0^{m}|0\rangle\\
\cline{2-4} \# & \multicolumn{3}{c|}{\textrm{2 total}}\\
\textrm{$SO(8)$ rep} & \multicolumn{3}{c|}{\mathbf{8}}\\
\hline
\end{array}
\end{displaymath}
\end{center}

The states in the NS sector do not have any knowledge of 3-brane
indices and so are scalars of the 4-d theory. In the Ramond sector they
are the Fermionic superpartners. An $\mathcal{N}\!=\!1$ chiral
multiplet has states $(0,1,2,1,0)$ in a helicity decomposition and
we can see that we have constructed 8 of these (because of their
being in the vector of $SO(8)$) from the 3-7 and 7-3 strings. Here
the $SO(8)$ symmetry is manifest, but as we can make 2
($\mathcal{N}\!=\!2$) hypermultiplets from an
($\mathcal{N}\!=\!1$) chiral multiplet we see that we have found
our 4 hypermultiplets in the fundamental of $U\!Sp(N_c)$.

As a last remark it may seem more natural to \textit{actually}
consider no orientifold projection when thinking about these 3-7
strings and then double what you get to account for the 7-3
strings. In this case there would really only be 4 $D7$-branes and
thus the relevant Chan-Paton index would run from $1\ldots 4$.
Accounting for both 3-7 and 7-3 strings would give 2 lots of 4
chiral multiplets. See \cite{BPZ} for an alternative take on these spectra.

\subsection{$D5$-$D9$s in Type I Revisited}

Let's revisit the $D5$-$D9$ setup in Type I. Firstly, let's pre-empt ourselves slightly and state the
multiplets of 6-dimensional $\mathcal{N}\!=\!1$ supersymmetry that
we'll get. Because of tadpole cancellation \emph{etc.}, we'll again get an $SO$ group on the
higher-dimensional branes and a $U\!Sp$ one on the lower
dimensional branes. We'll thus have $SO(32)$ on the $D9$s (as
usual in Type I) which from the 6-d perspective will again become
a global symmetry. On the $N_c$ $D5$s we'll have gauge group
$U\!Sp(N_c)$.

Now, in 6-d a gauge field has 6 components although as usual two of these are
unphysical and so it has 4 physical degrees of freedom. Similarly, a generic
spinor will have $2^{d/2}=2^{6/2}=8$ complex degrees of freedom. Requiring this
to be physical (\emph{i.e.} obey the Dirac equation) will halve this number and
requiring it to be a Weyl spinor (which is what we'll need) will halve it
again. Thus our 6-d spinors will have 2 complex or 4 real degrees of freedom each.

The multiplets we will get are: 1 vector multiplet (one gauge
Boson plus one Weyl Fermion) in the adjoint of $U\!Sp(N_c)$, that is
$(2,2,0,2,2)$ in terms of real degrees of freedom in a helicity
scheme; 1 hypermultiplet (one Weyl Fermion and 4 real scalars) in
the antisymmetric of $U\!Sp(N_c)$ -- $(0,2,4,2,0)$ in a helicity
scheme. We'll also want (and get) a half-hypermultiplet of 6-d
$\mathcal{N}\!=\!1$ SUSY in the fundamental of $U\!Sp(N_c)$ --
\textit{i.e.} something like $(0,1,2,1,0)$.

And what about our boundary conditions? Again we are only really interested in the 6-d
theory so we'll forget about 9-9 strings from now on. 5-5 strings are clearly NN in
$0\ldots 5$ and DD in $6\ldots 9$. 5-9 strings are NN in $0\ldots 5$ and DN in $6\ldots 9$. 9-5 strings are similar.

\subsubsection{5-5 Strings}

We again have to think about the ground state energy (\textit{i.e.} our $a_{NS/R}$),
but for 5-5 strings all b.c.'s are NN or DD and so as before we get the usual $a$'s
that we get for the Type I string in 10-d. In fact, in a sense what we have is the
same as for the 3-3 strings in the previous $D3$-$D7$ setup, except that what were the 8-9
directions there we can now consider to be part of the $D5$-brane world-volume (with NN
b.c.'s). We might worry briefly that we had extra signs when doing the orientifold
projection in this sector because of the orbifold action and the sign difference between
NN and DD b.c.'s, but if we look at our table of $D3$-$D7$ 3-3 strings it is clear that
this won't make any difference -- essentially we pick up a sign because of the switch
to NN b.c.'s, but then also another one due to the fact that we move these directions
`inside' the orbifold. Because these `8-9' directions are now really the 4-5 directions
of the $D5$-brane, what were spacetime scalars before will now contribute to the spacetime
vector states instead. This is all much clearer if we draw a table\footnote{Under $SO(9,1)\rightarrow SO(5,1)\times SO(4)$ 
the spinor degrees of freedom arrange as $2^{(5-1)/2}=4$ under the $SO(5,1)$ and $2^{4/2}=4$ under the $SO(4)$.}:
\begin{center}
$D5$-$D9$ Massless 5-5 Strings
\begin{displaymath}
\begin{array}{|l|c|c|}
\hline \textrm{Directions} & 0-5 & 6-9\\
\hline \textrm{b.cs} & \textrm{NN} & \textrm{DD}\\
\hline \textrm{NS} & \alpha_{-n}/\psi_{-r} & \alpha_{-n}/\psi_{-r}\\
 a_{NS}=1/2 & \psi_{-1/2}^{\mu}|0\rangle &
\psi_{-1/2}^{M}|0\rangle\\
 \textrm{\#} & 4 & 4\\
\textrm{$U\!Sp(N_c)$ rep} & N_c(N_c+1)/2 & N_c(N_c-1)/2\\\hline
\textrm{R} &
\alpha_{-n}/\psi_{-r} & \alpha_{-n}/\psi_{-r}\\
 a_R=0 & \psi_0^{\mu}|0\rangle & \psi_0^{M}|0\rangle\\
 \# & 4 & 4\\
 \textrm{$U\!Sp(N_c)$ rep} & N_c(N_c+1)/2 &
N_c(N_c-1)/2\\
\hline
\end{array}
\end{displaymath}
\end{center}
As can be seen we have the right
number of degrees of freedom to make an adjoint vector multiplet in 6-d and an antisymmetric
hypermultiplet.

\subsubsection{5-9 Strings}

In actual fact we hardly have to do any work here either because of our previous observations. Again we can package our old `8-9' sector into the `4-5' sector
here. Because NN and DD b.c.'s provide the same contributions to the zero-point energies
we again have that $a_{NS}=a_{R}=0$ and we can draw up another table:
\begin{center}
$D5$-$D9$ Massless 5-9 + 9-5 Strings
\begin{displaymath}
\begin{array}{|l|c|c|}
\hline \textrm{Directions} & 0-5 & 6-9\\
\hline \textrm{b.cs} & \textrm{NN} & \textrm{DN}\\
\hline \textrm{NS} & \alpha_{-n}/\psi_{-r} & \alpha_{-r}/\psi_{-n}\\
a_{NS}=0 & - & \psi_0^{M}|0\rangle\\
\# & - & 2\\
\textrm{$SO(32)$ rep} & - & \mathbf{32}\\
\hline \textrm{R} & \alpha_{-n}/\psi_{-n} & \alpha_{-r}/\psi_{-r}\\
a_R=0 & \psi_0^\mu|0\rangle & - \\
 \# & 2 & -\\
\textrm{$SO(32)$ rep} & \mathbf{32} & -\\
\hline
\end{array}
\end{displaymath}
\end{center}
The same story applies here for counting 5-9 strings only without
the orientifold so the above table describes all the states we
will get in this subsector. This gives us one half-hypermultiplet
of 6-d $\mathcal{N}\!=\!1$ SUSY in the
$(\mathbf{N_c},\mathbf{32})$ of $U\!Sp(N_c)\times SO(32)$. A
half-hypermultiplet contains 2 real scalars and one Weyl Fermion
satisfying a reality condition, and only exists in pseudo-real
representations of the gauge group. See \cite{Uranga} for more.

\newpage

\section{Spinors}\label{spinorappendix}
In physics we often come across 3 types of spinors. Dirac spinors, Weyl spinors and Majorana spinors.
These different types are not mutually exclusive.

\begin{itemize}
\item Dirac spinors are the ones we are probably most familiar with from the Dirac equation and QED.
Dirac spinors have $2^{d/2}$ \textit{complex} components in $d$-dimensions.

\item Weyl spinors are chiral spinors and exist in even-dimensional space-times. They can be
defined by using:
\begin{eqnarray}
\label{weyl}
{\gamma}_{d+1}\ = \ \prod_{i=0}^{d-1}{\gamma}^{i}\ , \quad d\, \in\, 2\mathbb{Z}\ .
\end{eqnarray}
For example in 2-dimensions we have ${\gamma}_3={\gamma}^0{\gamma}^1$ as in (\ref{gammas}), and in 4-dimensions
we have the usual ${\gamma}_5={\gamma}^0{\gamma}^1{\gamma}^2{\gamma}^3$. These can be used to make
chirality projectors
\begin{eqnarray}
\label{projectors}
P^{\pm}=\frac{1}{2}\left(1\pm {\gamma}_{d+1}\right)\ ,
\end{eqnarray}
which are idempotent operators whose eigenstates with eigenvalues $\pm1$ form Weyl
representations of the Lorentz algebra. Weyl spinors have $2^{d/2-1}$ \textit{complex} components
in $d$-dimensions.

\item Majorana spinors, as mentioned previously in Chapter 8 are \textit{real} spinors. In $d=2$ mod 8 dimensions
(\emph{i.e.} 2, 10, 18, 26\ldots dimensions), spinors can be \textit{both} Majorana \textit{and} Weyl.
Majorana-Weyl spinors have $2^{d/2-1}$ \textit{real} components in $d$-dimensions. In $d=4$ mod 8
dimensions (\emph{i.e.} 4, 12, 20, 28\ldots dimensions), however, spinors can be \textit{either} Majorana \textit{or}
Weyl.
\end{itemize}

We can see from these results that the number of components of a Majorana-Weyl spinor is $2^{d/2-1}$.
However, once we have imposed the conditions set by the Dirac equation this only leaves us with half of
these, \emph{i.e.} $2^{d/2-2}$, as independent components. So a Majorana-Weyl spinor in 10-dimensions has
$2^3=8$ independent components. If we now recall that the vector potential $A_{\mu}$ from Yang-Mills
has physical components $A_{i}$ ($i=1,\ldots,d-2$), then in $d=10$ dimensions this has 8 physical
components. This is a hint at supersymmetry, as this symmetry requires that the number of physical
components of Bosons and Fermions in the theory are equal to each other.

\newpage

\section{Useful Mathematical Formul\ae}
\label{mathappendix}
\begin{itemize}
\item Standard Sums of Natural Numbers:
\begin{eqnarray*}
\sum_{p=1}^n\, 1 &=& n\ ,\\
\sum_{p=1}^n\, p &=& \frac{1}{2}n(n+1)\ ,\\
\sum_{p=1}^n\, p^2 &=& \frac{1}{6}n(n+1)(2n+1)\ .
\end{eqnarray*}

\item Continuous Delta Functions can be defined by:
\begin{eqnarray*}
\delta(x-x') =
\frac{1}{2\pi}\int_{-\infty}^{\infty}\!dp\,e^{\pm{ip(x-x')}}\ ,
\end{eqnarray*}
for $p$ continuous, and
\begin{eqnarray*}
\delta(\sigma - {\sigma}') = \frac{1}{2\pi}\sum_{n =
-\infty}^{\infty}e^{\pm{in(\sigma - {\sigma}')}}\ ,
\end{eqnarray*}
for $n$ discrete.

\item Orthogonality Relations:
\begin{eqnarray*}
\int_{x_{0}}^{x_{0} +
L}\!dx\sin\left({\frac{2\pi{rx}}{L}}\right)\cos\left({\frac{2\pi{px}}{L}}\right) &=& 0 \quad
\forall\, r, p \\
\int_{x_{0}}^{x_{0} +
L}\!dx\cos\left({\frac{2\pi{rx}}{L}}\right)\cos\left({\frac{2\pi{px}}{L}}\right) &=& \left\{
\begin{array}{ll} L & r=p=0 \\ {1 \over 2}L & r=p>0 \\ 0 & r\neq p
\end{array} \right. \\
\int_{x_{0}}^{x_{0} +
L}\!dx\sin\left({\frac{2\pi{rx}}{L}}\right)\sin\left({\frac{2\pi{px}}{L}}\right) &=& \left\{
\begin{array}{ll} 0 & r=p=0 \\ {1 \over 2}L & r=p>0 \\ 0 & r\neq p
\end{array} \right. \\
\int_{x_0}^{x_0+L}\!dx\exp\left({\frac{i2\pi rx}{L}}\right)\exp\left({\frac{i2\pi px}{L}}\right)
&=& \left\{\begin{array}{ll} L & r=p=0 \\ 0 & r=p>0 \\ 0 & r\neq p\end{array} \right.\\
\int_{x_0}^{x_0+L}\!dx\exp\left({\frac{i2\pi rx}{L}}\right)\exp\left({\frac{-i2\pi px}{L}}\right)
&=& \left\{\begin{array}{ll} L & r=p=0 \\ L & r=p>0 \\ 0 & r\neq p\end{array} \right.\\
\\ &=& L\delta_{r-p}
\end{eqnarray*}

\item  For $A$, $B$, $C$ Bosonic and $P$, $Q$, $R$ Fermionic:
\newline Commutators:
\begin{eqnarray*}
\left[AB,C\right] &=& A\left[B,C\right] + \left[A,C\right]B \\
\left[PQ,R\right] &=& P\left\{Q,R\right\} - \left\{P,R\right\}Q \\
\left[AP,B\right] &=& A\left[P,B\right] + \left[A,B\right]P \\
\left\{AP,Q\right\} &=& A\left\{P,Q\right\} - \left[A,Q\right]P \\
\end{eqnarray*}
Jacobi Identity:
\begin{eqnarray*}
\left[A,\left[B,C\right]\right] + \left[B,\left[C,A\right]\right]
+ \left[C,\left[A,B\right]\right] &=& 0 \\
\left[P,\left\{Q,R\right\}\right] +
\left[Q,\left\{R,P\right\}\right] +
\left[R,\left\{P,Q\right\}\right] &=& 0 \\
\left[A,\left[B,P\right]\right] + \left[B,\left[P,A\right]\right]
+ \left[P,\left[A,B\right]\right] &=& 0 \\
\left[\left\{P,Q\right\},A\right] +
\left\{\left[A,P\right],Q\right\} +
\left\{\left[A,Q\right],P\right\} &=& 0
\end{eqnarray*}

\item Gamma Function:
\begin{eqnarray*}
\Gamma(z+1) = z! = \int_{0}^{\infty}e^{-t}t^{z}dt \quad
\left(\Re(z) > -1\right)
\end{eqnarray*}
$\Gamma(z)$ has poles for $z \in -{\mathbb{Z}^{+}}, 0$. It has no
zeroes. We also have the useful relation:
\begin{eqnarray*}
\Gamma(z-1) = \frac{\Gamma(z)}{(z-1)}
\end{eqnarray*}

\item Beta Function:
\begin{eqnarray*}
B(x,y)=\int_0^1 t^{x-1}(1-t)^{y-1}dt \quad \left(\Re(x),\Re(y) > 0\right)
\end{eqnarray*}
In terms of the gamma function
\begin{eqnarray*}
B(x,y)=\frac{\Gamma(x)\Gamma(y)}{\Gamma(x+y)}
\end{eqnarray*}

\item Saddle Point Method:
\begin{eqnarray*}
\int_C dz\,g(z)e^{sf(z)}\approx \sqrt{\frac{2\pi}{s|f''(z_0)|}}g(z_0)
e^{sf(z_0)}e^{i\alpha}\quad\textrm{as}\,\,s\rightarrow \infty\ ,
\end{eqnarray*}
where there is a saddle point at $z=z_0$ enclosed by the contour $C$ and 
\newline $\alpha=(\pi-\arg{f''(z_0)})/2$.
\end{itemize}


\newpage

\begin{thebibliography}{99}

%\cite{Green:1987sp}
\bibitem{GSW1}
  M.~B.~Green, J.~H.~Schwarz and E.~Witten,
  {\it Superstring theory. Vol. 1: Introduction,}
%\href{http://www.slac.stanford.edu/spires/find/hep/www?irn=1755021}{SPIRES entry}
Cambridge, UK: Univ. Pr. (1987) 469 p. (Cambridge Monographs On Mathematical Physics)

%\cite{Green:1987mn}
\bibitem{GSW2}
  M.~B.~Green, J.~H.~Schwarz and E.~Witten,
  {\it Superstring theory. Vol. 2: Loop amplitudes, anomalies and phenomenology,}
%\href{http://www.slac.stanford.edu/spires/find/hep/www?irn=1782614}{SPIRES entry}
Cambridge, UK: Univ. Pr. (1987) 596 p. (Cambridge Monographs On Mathematical Physics)

%\cite{Lust:1989tj}
\bibitem{Lust}
  D.~Lust and S.~Theisen,
  {\it Lectures on string theory,}\newline
  Lect.\ Notes Phys.\  {\bf 346} (1989) 1.
  %%CITATION = LNPHA,346,1;%%

%\cite{Polchinski:1998rq}
\bibitem{Polchinski1}
  J.~Polchinski,
  {\it String theory. Vol. 1: An introduction to the Bosonic string,}\newline
%\href{http://www.slac.stanford.edu/spires/find/hep/www?irn=4634799}{SPIRES entry}
Cambridge, UK: Univ. Pr. (1998) 402 p. (Cambridge Monographs On Mathematical Physics)

%\cite{Polchinski:1998rr}
\bibitem{Polchinski2}
  J.~Polchinski,
  {\it String theory. Vol. 2: Superstring theory and beyond,}\newline
%\href{http://www.slac.stanford.edu/spires/find/hep/www?irn=4634802}{SPIRES entry}
Cambridge, UK: Univ. Pr. (1998) 531 p. (Cambridge Monographs On Mathematical Physics)

%\cite{Zwiebach:2004tj}
\bibitem{Zwiebach}
  B.~Zwiebach,
  {\it A first course in string theory,}\newline
  Cambridge, UK: Univ. Pr. (2004) 558 p.

%\cite{Johnson:2003gi}
\bibitem{Johnson1}
  C.~V.~Johnson,
  {\it D-branes,}
  Cambridge, USA: Univ. Pr. (2003) 548 p. (Cambridge Monographs On Mathematical Physics)

%\cite{Johnson:2000ch}
\bibitem{Johnson2}
  C.~V.~Johnson,
  {\it D-brane primer,}
  [hep-th/0007170]

%\cite{Tong:2009np}
\bibitem{Tong:2009np}
  D.~Tong,
  {\it Lectures on string theory,}
  [arXiv:0908.0333 [hep-th]]

\bibitem{tHooft}
\mbox{G.~'t Hooft, {\it Introduction to string theory}, at:}
{\tt http://www.phys.uu.nl/$\sim$hooft101/} (last accessed May 2011)

%\cite{Ooguri:1996ik}
\bibitem{Ooguri}
  H.~Ooguri, Z.~Yin,
  {\it TASI lectures on perturbative string theories,}
  [hep-th/9612254]

\bibitem{hoker}
\mbox{E.~D'Hoker, {\it Point particles vs. strings}, at:}
{\tt http://www.math.ias.edu/QFT/spring/} (last accessed May 2011)

%\cite{Kiritsis:1997hj}
\bibitem{Kiritsis1}
  E.~Kiritsis,
  {\it Introduction to superstring theory,}
  Leuven, Belgium: Leuven Univ. Pr. (1998) 315 p.
  [hep-th/9709062]
  
%\cite{Kiritsis:2007zz}
\bibitem{Kiritsis2}
  E.~Kiritsis,
  {\it String theory in a nutshell,}
  Princeton, USA: Univ. Pr. (2007) 588 p.

\bibitem{Uranga}
A.~Uranga, {\it Introduction to string theory}, at:\newline
{\tt http://gesalerico.ft.uam.es/paginaspersonales/angeluranga/firstpage.html} 
(last accessed May 2011)

\bibitem{mukhi}
S.~Mukhi, {\it Mini-course on string theory}, at: \newline
{\tt http://theory.tifr.res.in/~mukhi/Physics/ministring.html}\newline (last accessed May 2011)

%\cite{Becker:2007zj}
\bibitem{Beckers}
  K.~Becker, M.~Becker, J.~H.~Schwarz,
  {\it String theory and M-theory: A modern introduction,}
  Cambridge, UK: Univ. Pr. (2007) 739 p.

%\cite{Vonk:2005yv}
\bibitem{Vonk:2005yv}
  M.~Vonk,
  {\it A Mini-course on topological strings,}
  [hep-th/0504147]
  
%\cite{Gimon:1996rq}
\bibitem{Gimon}
  E.~G.~Gimon, J.~Polchinski,
  {\it Consistency conditions for orientifolds and d manifolds,}
  Phys.\ Rev.\  D {\bf 54 } (1996)  1667.
  [hep-th/9601038]
  
%\cite{Dabholkar:1997zd}
\bibitem{Dabholkar}
  A.~Dabholkar,
  {\it Lectures on orientifolds and duality,}
  [hep-th/9804208]
  
%\cite{Giveon:1998sr}
\bibitem{Giveon}
  A.~Giveon, D.~Kutasov,
  {\it Brane dynamics and gauge theory,}
  Rev.\ Mod.\ Phys.\  {\bf 71 } (1999)  983.
  [hep-th/9802067]
  
%\cite{Gava:1999ky}
\bibitem{gava}
  E.~Gava, K.~S.~Narain and M.~H.~Sarmadi,
  {\it Instantons in $\mathcal{N} = 2$ Sp(N) superconformal gauge theories and the AdS/CFT
  correspondence,}
  Nucl.\ Phys.\ B {\bf 569}, 183 (2000)
  [arXiv:hep-th/9908125]
  %%CITATION = HEP-TH 9908125;%%
  
%\cite{Bedford:2007qj}
\bibitem{BPZ}
  J.~Bedford, C.~Papageorgakis, K.~Zoubos,
  {\it Twistor Strings with Flavour,}
  JHEP {\bf 0711 } (2007)  088.
  [arXiv:0708.1248 [hep-th]]
  
%\cite{Veneziano:1968yb}
\bibitem{Veneziano}
  G.~Veneziano,
  {\it Construction of a crossing-symmetric, Regge behaved amplitude for linearly rising trajectories,}
  Nuovo Cim.\  A {\bf 57}, 190 (1968)

\end{thebibliography}
\end{document}